\documentclass[11pt]{amsart}
\usepackage{latexsym,amssymb}
\usepackage{graphics}
\usepackage{harvard}
\usepackage{enumerate}
\usepackage{color}
\usepackage{url}
\usepackage{graphicx,subfig}
\usepackage{setspace}
\usepackage{soul}
\usepackage{multirow}
\usepackage{booktabs}
\usepackage{lscape} 
\usepackage{afterpage}
\usepackage{float}
\usepackage{caption}
\usepackage[boxed]{algorithm2e}

\newcommand\R{\mathbb{R}}

\newcommand\eps{\varepsilon}
\newcommand\E{\mathbb{E}}     

\newcommand\Gc{\mathcal{G}}

\newtheoremstyle{normal}
{2ex}               
{3ex}               
{}                  
{}                  
{\bfseries} 
{}                  
{2pt}   
{\thmname{#1}\thmnumber{ #2.} \thmnote{(#3)}}

\newtheoremstyle{italic}
{2ex}
{3ex}
{\itshape}
{}
{\bfseries} 
{}
{2pt}
{\thmname{#1}\thmnumber{ #2.} \thmnote{(#3)}}

\theoremstyle{italic}
\newtheorem{definition}{Definition}[section]

\newtheorem{theorem}[definition]{Theorem}
\newtheorem{lemma}[definition]{Lemma}

\theoremstyle{normal}
\newtheorem{remark}[definition]{Remark}

\usepackage[margin=1.1in]{geometry} 

\def\1{{\mathbf{I}}}
\def\argmin{\mathop{\rm arg\,min}}

\begin{document}
\bibliographystyle{econometrica}
\title{Panel Data Quantile Regression with Grouped Fixed Effects}

\author{Jiaying Gu}
\author{Stanislav Volgushev}
\thanks{Version:  \today .  The authors would like to thank Jacob Bien for bringing the convex clustering literature to their attention. We are also grateful to two anonymous Referees and the Associate Editor whose comments helped to considerably improve the presentation of this manuscript.
}

\begin{abstract}
This paper introduces estimation methods for grouped latent heterogeneity in panel data quantile regression. We assume that the observed individuals come from a heterogeneous population with a finite number of types. The number of types and group membership is not assumed to be known in advance and is estimated by means of a convex optimization problem. We provide conditions under which group membership is estimated consistently and establish asymptotic normality of the resulting estimators. Simulations show that the method works well in finite samples when $T$ is reasonably large. To illustrate the proposed methodology we study the effects of the adoption of Right-to-Carry concealed weapon laws on violent crime rates using panel data of 51 U.S. states from 1977 - 2010. 
\end{abstract}
\maketitle
\pagestyle{myheadings}
\markboth{\sc Panel Quantile regression with group fixed effects}{\sc Gu and Volgushev}

\section{Introduction}
It is widely accepted in applied Econometrics that individual latent effects constitute an important feature of many economic applications. When panel data are available, a common approach is to incorporate latent structures in completely nonrestrictive way, i.e. the fixed effect approach. The fixed effect approach is attractive as it imposes minimal assumptions on the structure of the latent effects and on the correlation between the latent effects and the observed covariates and hence has become a very common empirical tool (see \citeasnoun{Hsiao} for a textbook treatment). 

A major challenge of the fixed effects approach lies in the fact that it introduces a large number of parameters which grows linearly with the number of individuals. For a few specific models this can be avoided by differencing out individual effects and learning about the common parameter of interest. However, for most models, including quantile regression, this simple differencing method no longer exists. The literature contains various approaches that put additional structure on latent effects in order to reduce the number of parameters and obtain more interpretable models. One popular approach is to introduce some parametric distributional structure on the latent effects, see for example \citeasnoun{Mundlak78}, \citeasnoun{Chamberlain1982} and the correlated random effects literature. An alternative is to assume that the fixed effects have a group structure and hence only take a few distinct values which is the approach we take in this paper. 


There is ample evidence from empirical studies that it is often reasonable to consider a number of homogeneous groups (clusters) within a heterogeneous population. This discrete approach was taken by \citeasnoun{HeckmanSinger82} for duration analysis of unemployment spells of a heterogeneous population of workers. \citeasnoun{BesterHansen} argue that in many applications individuals or firms are grouped naturally by some observable covariates such as classes, schools or industry codes. It is also widely accepted in the discrete choice model literature that individual agents are classified as a number of latent types (for instance \citeasnoun{Keane} among many others). 

Estimating cluster structure has a long history in Statistics and Economics, and has generated a rich and mature literature. A general overview is given in \citeasnoun{Kaufman}. Among the many available clustering algorithms, the k-means algorithm (\citeasnoun{MacQueen}) is one of the most popular methods. It has been successfully utilized in many economic applications, for instance \citeasnoun{LinNg}, \citeasnoun{BM} and \citeasnoun{AndoBai}. Finite mixture models provide an alternative, likelihood based approach. In the latter, grouping is usually achieved by maximizing the likelihood of the observed data. \citeasnoun{Sun} builds a multinomial logistic regression model to infer the group pattern while nonparametric finite mixture models  are considered in \citeasnoun{Allman} and \citeasnoun{Kasahara} among many others.

The focus of the present paper is on quantile regression for panel data with grouped individual heterogeneity. Panel data quantile regression has recently attracted a lot of attention, and there is a rich and growing literature that proposes various approaches to dealing with individual heterogeneity in this setting. In a pioneering contribution, \citeasnoun{Koenker2004} takes the fixed effect approach and introduces individual latent effects as location shifts. These individual effects are regularized through an $\ell_1$ penalty which shrinks them towards a common value. \citeasnoun{Lamarche} proposes an optimal way to choose the corresponding penalty parameter in order to optimize the asymptotic efficiency of the common parameters of the conditional quantile function, see \citeasnoun{harding2017} for an extension of this approach. Another line of work that focuses on estimating common parameters while putting no structure on individual effects includes \citeasnoun{Kato}, \citeasnoun{galvao2015} and \citeasnoun{galvao2016}. Alternative approaches have also emerged. \citeasnoun{AD} take a random effect view of these individual latent effects. They consider a correlated random-effect model in the spirit of \citeasnoun{Chamberlain1982} where the individual effects are modeled through a linear regression of some covariates. This is further developed in \citeasnoun{AB16} and \citeasnoun{CLP} where the conditional quantile function of the unobserved heterogeneity is modelled as a function of observable covariates.\footnote{For related literature on non-separable panel data models see \citeasnoun{Evdokimov}, \citeasnoun{Chernozhukov} and the references therein.}

Our contribution, which builds upon \citeasnoun{Koenker2004}, is a linear quantile regression method that accommodates grouped fixed effects. The advantages of our proposal over existing proposals are twofold. First, grouped fixed effects maintain the merit of unrestricted correlation between the latent effects and the observables and strike a good balance between the classical fixed effects approach and the other extreme which completely ignores latent heterogeneity. Second, in contrast to \citeasnoun{Koenker2004}, where the fixed effects are treated as nuisance parameters and are regularized to achieve a more efficient estimator for the global parameter, our method allows the researcher to learn the particular group structure of the latent effects together with common parameters of interest in the model. To the best of our knowledge, panel data quantile regression with grouped fixed effects has not been considered in the literature before. The only paper that goes in this direction is \citeasnoun{su2016}. While the general framework developed in this paper does include a version of quantile regression with smoothed quantile objective function, the theoretical analysis requires the smoothing parameter to be fixed. This results in a non-vanishing bias and hence does not correspond to quantile regression in a strict sense.

We do not assume any prior knowledge of the group structure and combine the quantile regression loss function with the recently proposed convex clustering penalty of \citeasnoun{Hocking}. The convex clustering method introduces a $\ell_1$-constraint on the pair-wise difference of the individual fixed effects, which tends to push the fixed effects into clusters. The number of clusters is controlled by a penalty parameter. The resulting optimization problem remains convex and can be solved in a fast and reliable fashion. Further modifications and a theoretical analysis of convex clustering were considered in \citeasnoun{zhu2014}, \citeasnoun{TanWitten} and \citeasnoun{Radchenko}. All of those authors combine $\ell_1$ penalties with the classical $\ell_2$ loss, and only consider clustering for cross-sectional data. Their theoretical results are not directly applicable to panel data or the non-smooth quantile loss function which is the main objective in this paper (all of the available theoretical results
explicitly make use of the differentiability of the $\ell_2$ loss function in their proofs). 

Our main theoretical contribution is to show consistency of the estimated grouping for a suitable range of penalty parameters when $n$ and $T$ tend to infinity jointly. We also propose a completely data-driven information criterion that facilitates the practical implementation of the method and prove its consistency for group selection as well as asymptotic normality of the resulting parameter estimators. 

The remaining part of this paper is organized as follows. Section 2 contains a detailed description of the proposed methodology and provides details on its practical implementation. Assumptions and theoretical results are included in Section 3. Section 4 presents the convex optimization problem and its computational details. Monte Carlo simulation results are included in Section 5 where investigate the final sample behavior of the proposed methodology. In Section 6 we apply the method to an empirical application in studying the effect of the adoption of Right-to-Carry concealed weapon law on violent crime rate using a panel data of 51 U.S. states from 1977 - 2010. All proofs are collected in Section~\ref{sec:proofs} while additional simulation results and details for the empirical application are relegated to the Appendix.

\newpage

\section{Methodology}\label{sec:meth}

Assume that for individuals $i=1,...,n$ we observe repeated measures $(X_{it},Y_{it})_{t=1,...,T}$ where $X_{it}$ denote covariates and $Y_{it}$ are responses.\footnote{Here, $T$ is assumed to be the same across individuals for notational simplicity. All results that follow can be extended to individual-specific values of $T_i$ as long as the ratio $(\max_{i=1,...,n} T_i)/(\min_{i=1,...,n} T_i)$ is uniformly bounded. In this case the theory goes through without changes if all instances of $T$ are replaced by $n^{-1} \sum_{i=1,...,n} T_i$} We shall maintain the assumption that data are i.i.d. within individuals and independent across individuals. The main object of interest in this paper is the conditional $\tau$-quantile function of $Y_{i1}$ given $X_{i1}$, which we will denote by $q_{i,\tau}$. We assume that $q_{i,\tau}$ is of the form
\[
q_{i,\tau}(x) = \beta_0(\tau)^\top x + \alpha_{0i}(\tau), \quad i=1,....,n
\]
with individual fixed effects $\alpha_{0i}(\tau)$ taking only a finite number, say $K$, of different values, say $\alpha_{(01)}(\tau),...,\alpha_{(0K)}(\tau)$.\footnote{We follow \citeasnoun{Koenker2004} in treating the $\alpha_i$ as fixed parameters. An alternative approach which leads to equivalent results is to treat the $\alpha_i$ as random (with no restrictions placed on the dependence with $X_{it}$). In this case the model can be written as $Q_{Y_{it}|X_{it},\alpha_i(\tau)}(\tau) = \beta_0(\tau)^\top x + \alpha_{0i}(\tau)$; here $Q_{Y_{it}|X_{it},\alpha_i(\tau)}$ denotes the conditional quantile function of $Y_{it}$ given $(X_{it},\alpha_i(\tau))$ (see for instance \citeasnoun{Kato}, \citeasnoun{galvao2015} and \citeasnoun{galvao2016} for this interpretation). Both interpretations lead to the same asymptotic results.} We explicitly allow the group membership, and even the number of groups to be unknown and to depend on $\tau$ but will not stress this dependence in the notation for the sake of simplicity. Our main objective is to jointly estimate the number of groups, unknown group structure, and parameters $\alpha_{(01)},...,\alpha_{(0K)},\beta$ from the observations. To achieve this, we consider penalized estimators of the form
\begin{equation}\label{eq:opt}
(\hat\alpha_1,...,\hat\alpha_n,\hat\beta) := \arg\min_{\alpha_1,...,\alpha_n,\beta} \Theta(\alpha_1,...,\alpha_n,\beta)
\end{equation} 
where\footnote{As pointed out by a Referee, one could also consider combining the objective functions corresponding to several quantiles as was done in \citeasnoun{Koenker2004} and force all coefficients $\alpha_i$ to be independent of $\tau$. This would result in efficiency gains if all $\alpha_i$ are purely location-shift effects but can introduce bias otherwise. We leave this extension for future research.} 
\[
\Theta(\alpha_1,...,\alpha_n,\beta) := \sum_{i,t} \rho_\tau(Y_{it} - X_{it}^\top \beta - \alpha_i) + \sum_{i\neq j} \lambda_{ij} |\alpha_i - \alpha_j|.
\]
Here $\rho_\tau$ denotes the usual 'check function' and the weights $\lambda_{i,j}$ are allowed to depend on $n,T$ and the data; one particular choice is discussed in below. The form of the penalty is motivated by the work of \citeasnoun{Hocking}. Intuitively, large values of $\lambda_{ij}$ will push different coefficients closer together and result in clustered structure of the estimators $\hat \alpha_i$. High-level conditions on the weights $\lambda_{i,j}$ which guarantee consistency of the resulting grouping procedure are provided in Theorem~\ref{th:cluster}.  

There are various possible choices for the penalty parameters $\lambda_{i,j}$. We propose to use weights of the form 
\begin{equation} \label{eq:lambdacheck}
\check \lambda_{ij} := \lambda |\check \alpha_i - \check\alpha_j|^{-2}
\end{equation}
where $(\check\alpha_1,...,\check\alpha_n)$ are the fixed effects quantile regression estimators
\begin{equation} \label{eq:alphacheck}
(\check\alpha_1,...,\check\alpha_n,\check\beta) := \argmin_{\alpha_1,...,\alpha_n,\beta} \sum_{i,t} \rho_\tau(Y_{it} - X_{it}^\top \beta - \alpha_i)
\end{equation}
of~\citeasnoun{Kato}\footnote{As pointed out by a Referee, an alternative approach to obtain preliminary estimators for $\alpha_{i0}$ would be to run separate quantile regressions for each individual. This did not improve the performance of our procedure in the simulations that we tried.} and $\lambda$ is a tuning parameter. This form of weighting by preliminary estimators is motivated by the work of~\citeasnoun{zou2006} on adaptive lasso. Intuitively, weighting by preliminary estimated distances tends to give smaller penalties to coefficients from different groups thus reducing some of the bias that is typically present in the classical lasso.

Given the developments above, it remains to find a value for the tuning parameter $\lambda$. The high-level results in Theorem~\ref{th:cluster} together with findings in \citeasnoun{Kato} provide a theoretical range for those values (see the discussion following Theorem~\ref{th:cluster} for additional details), but this range is not directly useful in practice since only rates and not constants are provided. Moreover, despite the fact that the weights $\check \lambda_{ij} := \lambda |\check \alpha_i - \check\alpha_j|^{-2}$ lead to asymptotically unbiased estimates, bias can still be a problem in finite samples. A typical approach in the literature to reduce bias which results from lasso-type penalties is to view the lasso problem solution as a candidate model (in our case, a candidate grouping of $\alpha_i$) and re-fit based on this candidate model (see \citeasnoun{belloni2009} or \citeasnoun{su2016} among many others).  

To deal with bias issues and the choice of $\lambda$ in practice, we propose to combine the re-fitting idea with a simple information criterion which will simultaneously reduce the bias problem and provide a simple way to select a final model. A formal description of our approach is given in Algorithm~\ref{alg:IC}.


\begin{algorithm}[h]
\SetKwInOut{Input}{input}\SetKwInOut{Output}{output}
\Input{Data $(X_{it},Y_{it})$, grid of values $\lambda_1,...,\lambda_L$, quantile level of interest $\tau$}
\Output{Estimated number of groups $\hat K^{IC}$, estimated group membership $\hat I_1^{IC},...,\hat I_{\hat K}^{IC}$, estimated coefficients $\hat\alpha_k^{IC}$, $\hat \beta^{IC}$}
\For{$i\leftarrow 1$ \KwTo $n$}{
compute $\check \alpha_i$ given in~\eqref{eq:alphacheck} 
}
\For{$l\leftarrow 1$ \KwTo $L$}{
Compute
\[
(\hat\alpha_{1,\ell},...,\hat\alpha_{n,\ell},\hat\beta_\ell) := \argmin_{(\alpha_1,...,\alpha_n, \beta)} \Big\{ \sum_{i,t} \rho_\tau(Y_{it} - X_{it}^\top \beta - \alpha_i) + \lambda_\ell \sum_{i\neq j} \frac{|\alpha_i - \alpha_j|}{|\check \alpha_i - \check  \alpha_j|^2}\Big\}
\]
Let $\hat \alpha_{(1,\ell)} < ... < \hat \alpha_{(K_\ell,\ell)}$ denote the unique values of $\hat \alpha_{1,\ell},...,\hat\alpha_{n,\ell}$, and define $\hat I_{j,\ell} := \{i: \hat\alpha_i = \hat \alpha_{(j,\ell)}\}$ as the estimated groups. Compute re-fitted estimators
\[
(\widetilde\alpha_{1,\ell},...,\widetilde\alpha_{K_\ell,\ell},\widetilde\beta_\ell) := \argmin_{(\alpha_1,...,\alpha_{K_\ell}, \beta)} \sum_{k = 1}^{K_\ell} \sum_{i \in \hat I_{k,\ell}} \sum_t \rho_\tau(Y_{it} - X_{it}^\top \beta - \alpha_k). 
\]
Compute the IC criterion 
\[
IC(\ell) := \sum_{k = 1}^{K_\ell} \sum_{i \in \hat I_{k,\ell}} \sum_t \rho_\tau(Y_{it} - X_{it}^\top \widetilde\beta_\ell - \widetilde\alpha_{k,\ell}) + \hat C K_\ell p_{n,T},
\]
where the choice of $\hat C$ and $p_{n,T}$ is given in~\eqref{eq:ChatpnT}.
}
\BlankLine
Set $\hat \ell^{IC} := \argmin_{\ell=1,...,L} IC(\ell)$ and denote by $\hat K^{IC} := K_{\hat \ell^{IC}}$ the corresponding number of groups. Set $\hat I^{IC}_k := \hat I_{k,\hat \ell^{IC}}$, $\hat\alpha_k^{IC} := \widetilde\alpha_{1,\hat \ell^{IC}}, \hat\beta^{IC} := \widetilde\beta_{\hat \ell^{IC}}$.
\BlankLine
\BlankLine

\caption{Grouping via IC criterion}\label{alg:IC}
\end{algorithm}

\medskip

Theorem~\ref{th:ic} provides a formal justification of Algorithm~\ref{alg:IC} under high-level conditions on $\hat C, p_{n,T}$. In particular we prove that the group structure is estimated consistently with probability tending to one and derive the asymptotic distribution of the resulting estimators $\hat \alpha_i^{IC}, \hat \beta^{IC}$. In order to make the proposed estimation procedure fully data-driven, we need to specify a choice for the tuning parameters $\hat C$ and $p_{n,T}$. In our simulations, we found that the following choices lead to good results\footnote{The exact constant $1/10$ in the factor $p_{n,T}$ does not matter asymptotically. The value $1/10$ was found to work well for a wide range of values of $n,T$ and for various models, details are provided in the Monte Carlo section \ref{MCsec}. There we also show that the impact of the precise form of the factor in $\hat p_{n,T}$ becomes less pronounced as $T$ increases}: 
\begin{equation}\label{eq:ChatpnT}
p_{n,T} = nT^{1/4}/10, \quad \hat C := \tau(1-\tau)\hat s(\tau)
\end{equation} 
with 
\[
\hat s(\tau) := (\hat F^{-1}(\tau + h_{n,T}) - \hat F^{-1}(\tau - h_{n,T}))/(2h_{n,T})
\]    
where $\hat F(y) := \frac{1}{nT}\sum_{i,t} I\{Y_{it} - X_{it}^\top \check\beta - \check\alpha_{i}\leq y\}$ denotes the empirical cdf of the regression residuals from the fixed effects quantile regression estimator given in~\eqref{eq:alphacheck}, $\hat F^{-1}$ denotes the corresponding empirical quantile function, and $h_{n,T} \to 0$ is a bandwidth parameter (we use the Hall-Sheather rule in our simulations, see \citeasnoun{koenker2005}). 

To motivate this particular choice of constant $\hat C$, observe the following expansion, which is derived in detail in the proof of Theorem~\ref{th:ic}
\[
\sum_{i,t} \rho_\tau(Y_{it} - X_{it}^\top \check\beta - \check\alpha_{i}) - \rho_\tau(Y_{it} - X_{it}^\top \beta_0 - \alpha_{0i}) = - \sum_i \frac{\tau(1-\tau)}{2 \E[f_{Y_{i1}|X_{i1}}(q_{i,\tau}(X_{i1})|X_{i1})]}+ o_P(n).
\]
This shows that plugging in the estimated (by fixed effects quantile regression) instead of true errors underestimates the objective function evaluated at the residuals by roughly the first term on the right-hand side in the above expression. This term needs to be dominated by the penalty if we want to avoid selecting models that are too large, and so it is natural to scale the penalty by a constant which is proportional to $\sum_i 1/\E[f_{Y_{i1}|X_{i1}}(q_{i,\tau}(X_{i1})|X_{i1})]$ in order to ensure reasonable performance across different data generating processes. Under the simplifying assumption that $f_{Y_{i1}|X_{i1}}(q_{i,\tau}(X_{i1})|X_{i1}) =: f_\eps(0)$ does not depend on $i, X_{i1}$, this term equals $n/f_\eps(0)$, and under the same assumptions $\hat s$ provides a consistent estimator for the latter, see \citeasnoun{koenker2005}. Note that the sparsity term introduced here plays a similar role as the noise variance in classical information criteria such as AIC and BIC in least squares regression.


\section{Theoretical analysis}\label{sec:th}

In this section we provide a theoretical analysis of the methodology proposed in Section~\ref{sec:meth}. We begin by stating an assumption on the true (but unknown) underlying group structure. 
\begin{itemize}
\item[(C)]\label{C} For each quantile $\tau$ of interest, there exists a fixed number $K_\tau$, values $\alpha_{(01)}(\tau) < ...<\alpha_{(0K_\tau)}(\tau)$ and disjoint sets $I_1(\tau),...,I_{K_{\tau}}(\tau)$ with $\cup_k I_k(\tau) = \{1,...,n\}$, $|I_k(\tau)|/n \to \mu_k(\tau) \in (0,1)$, $\alpha_{0i}(\tau) = \alpha_{0j}(\tau) = \alpha_{(0k)}(\tau)$ for $i,j \in I_k(\tau)$. There exists $\eps_0 >0$ independent of $\tau$ with 
\[
\min_{k = 1,...,K_\tau-1}|\alpha_{(0k)}(\tau) - \alpha_{(0k+1)}(\tau)| \geq \eps_0.
\] 
\end{itemize}
Assumption (C) implies that the individual fixed effects are grouped into $K$ distinct groups and that the group centers are separated. Note that the number of groups as well as group membership is allowed to differ across quantiles. For the sake of a concise notation, the dependence of the number of groups and group centers on $\tau$ will from now on be dropped unless there is risk of confusion. Note also that we require the number of groups to be fixed (i.e. independent of $n,T$ and non-random) and exogenous, i.e. independent of the covariates $X_{it}$.  

Next we collect some technical assumptions on the data generating process. Define $Z_{it}^\top = (1, X_{it}^\top)$ and let $\mathcal{Z}$ denote the support of $Z_{it}$.

\begin{itemize}
\item[(A1)]\label{A1} Assume that $\sup_i \|Z_{it}\| \leq M < \infty$ a.s. and that 
\[
c_{\lambda}\leq \inf_i \lambda_{\min}(\E [Z_{it} Z_{it}^\top])\leq \sup_i \lambda_{\max}(\E [Z_{it} Z_{it}^\top])\leq C_{\lambda}
\] 
for some fixed constants $c_{\lambda}>0$ and $C_{\lambda} <\infty$.
\item[(A2)]\label{A2} The conditional distribution functions $F_{Y_{i1}|Z_{i1}}(y|z)$ are twice differentiable w.r.t. $y$, with the corresponding derivatives $f_{Y_{i1}|Z_{i1}}(y|z)$ and $f'_{Y_{i1}|Z_{i1}}(y|z)$. Assume that 
\[
f_{max} := \sup_i\sup_{y \in \mathbb{R},z\in \mathcal{Z}} |f_{Y_{i1}|Z_{i1}}(y|z)| < \infty,
\quad \quad  
\overline{f'} :=\sup_{y \in \mathbb{R},z\in \mathcal{Z}} |f'_{Y_{i1}|Z_{i1}}(y|z)| < \infty.
\] 
\item[(A3)]\label{A3} Denote by $\mathcal{T}$ an open neighbourhood of $\tau$. Assume that there exists a constant $f_{\min} \leq f_{max}$ such that
\[
0 < f_{\min} \leq \inf_i \inf_{\eta \in \mathcal{T}} \inf_{z \in \mathcal{Z}} f_{Y_{i1}|Z_{i1}}(q_{i,\eta}(z)|z).
\]
\end{itemize}

Assumptions (A1)-(A3) are fairly standard and routinely imposed in the quantile regression literature. Similar assumptions have been made, for instance in~\citeasnoun{Kato} [see assumptions (B1)-(B3) in that paper]. 

\subsection{Analysis of the estimators in~\eqref{eq:opt}}

To state our first main result define
\[
\Lambda_D := \sup_{i \in I_k, j \in I_{k'}, k\neq k'} \lambda_{i,j}, \quad \Lambda_S := \inf_k \inf_{i,j \in I_k} \lambda_{i,j}. 
\]
In words, $\Lambda_D$ corresponds to the largest penalty corresponding to the difference between two individual effects from different groups while $\Lambda_S$ describes the smallest penalty between two effects from the same group. Our first result provides high-level conditions on $\Lambda_S, \Lambda_D$ that guarantee asymptotically correct grouping.

\begin{theorem} \label{th:cluster} Let assumptions (A1)-(A3), (C) hold and assume that $\min(n,T) \to \infty$, $\log n = o(T)$ and 
\begin{equation}\label{c:lambda}
\frac{\Lambda_D}{\Lambda_S} = o_P(1), \quad n\Lambda_D = o_P(T^{1/2}), \quad \frac{T^{3/4}(\log n)^{3/4}}{n\Lambda_S} = o_P(1).
\end{equation} 
Denote the ordered unique values of $\hat\alpha_1,...,\hat\alpha_n$ by $\hat\alpha_{(1)}<...<\hat\alpha_{(\hat K)}$ (i.e. $\hat K$ denotes the number of distinct values taken by $\hat\alpha_1,...,\hat\alpha_n$ which we interpret as the estimated number of groups) and define the sets $\hat I_k := \{i: \hat\alpha_i = \hat\alpha_{(k)}\}, k=1,...,\hat K$. Then
\[
P\Big( \hat K = K, \hat I_k = I_k, k=1,...,K \Big) \to 1.
\]
\end{theorem}

Next we discuss the implications of this general result for the specific choice $\check \lambda_{i,j}$ given in~\eqref{eq:lambdacheck}.  
Define
\[
\check\Lambda_D := \sup_{i \in I_k, j \in I_{k'}, k\neq k'} \check\lambda_{i,j}, \quad \check\Lambda_S := \inf_k \inf_{i,j \in I_k} \check\lambda_{i,j}. 
\]
From \citeasnoun{Kato} \footnote{more precisely, from the paragraph following equation (A.14) in the latter paper; note that this result is derived under the assumption that $n$ grows at most polynomially with $T$} we obtain the bound 
\[
\sup_{i=1,...,n} |\check \alpha_i - \alpha_{0i}| = O_P((\log n)^{1/2}/T^{1/2}).
\]
Now if $i,j \in I_k$ then $\alpha_{0i} = \alpha_{0j}$ and thus
\[
1/\check\Lambda_S = \Big\{\inf_{k}\inf_{i,j \in I_k} |\check\alpha_{i} - \check\alpha_{j}|^{-2} \lambda\Big\}^{-1} = \lambda^{-1} \sup_{k}\sup_{i,j \in I_k} |\check\alpha_{i} - \check\alpha_{j}|^2 = O_P\Big(\frac{\log n}{\lambda T}\Big).
\]
Moreover, under (C) we have 
\[
\inf_{k\neq k'} \inf_{i \in I_k, j \in I_{k'}} |\alpha_{0i} - \alpha_{0j}| \geq \eps_0 > 0
\]
and thus
\[
\check \Lambda_D \leq \lambda \Big\{ \inf_{k\neq k'} \inf_{i \in I_k, j \in I_{k'}} |\check \alpha_{i} - \check\alpha_{j}|\Big\}^{-2} \leq \lambda/(\eps_0 - o_P(1))^2 = O_P(\lambda).
\]


Given this choice of weights, the conditions $\frac{\Lambda_D}{\Lambda_S} = o_P(1)$ is satisfied provided that $T/\log n \to \infty$. The other conditions in~\eqref{c:lambda} take the form
\[
T^{1/2} \gg n\lambda \gg T^{-1/4} (\log n)^{7/4}. 
\]
Assuming that $(\log n)^{7/3} = o(T)$, this provides a range of possible values for $\lambda$ which will ensure that~\eqref{c:lambda} holds. 

\subsection{Analysis of the information criterion in Algorithm~\ref{alg:IC}}
In this section we provide theoretical guarantees for the performance of the information criterion based estimators $\hat \beta^{IC}, \hat \alpha_k^{IC}, \hat{I}_k^{IC}, \hat K^{IC}$
Our main result shows that, under fairly general conditions on the penalty parameter $p_{n,T}$, the IC procedure selects the correct number of groups with probability tending to one. Moreover, the estimators $(\hat\alpha_1^{IC},...,\hat\alpha_{\hat K^{IC}}^{IC},\hat\beta^{IC})$ are shown to enjoy the 'oracle property', i.e. they have the same asymptotic distribution as estimators which are based on the true (but unknown) grouping of individuals. Before making this statement more formal, we need some additional notation. Let
\begin{equation}\label{eq:or}
(\hat\alpha_{(1)}^{(OR)},...,\hat\alpha_{(K)}^{(OR)},\hat\beta^{(OR)}) := \argmin_{(\alpha_1,...,\alpha_K, \beta)} \sum_k \sum_{i \in I_k} \sum_t \rho_\tau(Y_{it} - X_{it}^\top \beta - \alpha_k)
\end{equation}
denote the infeasible 'oracle' which uses the true group membership. The asymptotic variance of the oracle estimator is conveniently expressed in terms of the following two limits which we assume to exist\footnote{Appropriate forms of asymptotic normality of the oracle and IC estimators continue to hold without assuming that the limits exist. This assumption is made for notational convenience.}  
\begin{align*}
\Sigma_{1,\tau} &:= \tau(1-\tau) \lim_{n\to \infty} \frac{1}{n} \sum_{k=1}^K \sum_{i \in I_k} \E[\tilde Z_{ik}\tilde Z_{ik}^\top],
\\
\Sigma_{2,\tau} &:= \lim_{n\to \infty} \frac{1}{n} \sum_{k=1}^K \sum_{i \in I_k} \E[\tilde Z_{ik}\tilde Z_{ik}^\top f_{Y_{i1}|X_{i1}}(q_{i,\tau}(X_{i1})|X_{i1})],
\end{align*}
where $\tilde Z_{ik} := (e_k^\top, X_{i1}^\top)^\top$ and $e_k$ denotes the $k$'th unit vector in $\R^K$. Additionally, we need the following condition on the grid $\lambda_1,...,\lambda_L$
\begin{enumerate}
\item[(G)] For each $(n,T)$, denote the grid values by $\lambda_{1,n,T},...,\lambda_{L,n,T}$ where $L$ can depend on $n,T$. There exists a sequence $j_n$ such that $T^{-1/2} (\log n) \ll n \lambda_{j_n} \ll T^{1/2}$.
\end{enumerate}

Assumption (G) is fairly mild. It only requires that among the candidate values for $\lambda$ there exists one value so that $\check\lambda_{i,j}$ satisfies the assumptions of Theorem~\ref{th:cluster}. In practice, we recommend choosing a grid of values that results in sufficiently many different numbers of groups.

\begin{theorem}\label{th:ic} Let assumptions (A1)-(A3), (C), (G) hold and assume that $\min(n,T) \to \infty$ and $n$ grows at most polynomially in $T$ (i.e. $n = O(T^b)$ for some $b < \infty$) and $\frac{(\log T)^3(\log n)^2}{T} \to 0$. Assume that there exists $\eps > 0$ such that $\hat C > \eps$ with probability tending to one and that $nT \gg p_{n,T} \gg n, \hat C = O_P(1)$. Then $P(\hat K^{IC} = K) \to 1$ and\footnote{Strictly speaking, $\hat\alpha_k^{IC}$ is not defined if $\hat K^{IC} < K$. Since the probability of this event tends to zero, we can simply define $\hat\alpha_k^{IC} = 0$ for $\hat K^{IC} > k \geq K$.}
\begin{align*}
\sqrt{nT}\Big( (\hat\alpha_1^{IC},...,\hat\alpha_{\hat K^{IC}}^{IC},(\hat\beta^{IC})^\top) - (\alpha_{(01)},...,\alpha_{(0K)},\beta_0^\top) \Big) \stackrel{\mathcal{D}}{\longrightarrow} \mathcal{N}(0,\Sigma_{2,\tau}^{-1} \Sigma_{1,\tau} \Sigma_{2,\tau}^{-1}),
\\
\sqrt{nT}\Big( (\hat\alpha_{(1)}^{(OR)},...,\hat\alpha_{(K)}^{(OR)},(\hat\beta^{(OR)})^\top)  - (\alpha_{(01)},...,\alpha_{(0K)},\beta_0^\top) \Big) \stackrel{\mathcal{D}}{\longrightarrow} \mathcal{N}(0,\Sigma_{2,\tau}^{-1} \Sigma_{1,\tau} \Sigma_{2,\tau}^{-1}).
\end{align*}
\end{theorem}

\begin{remark} Theorem~\ref{th:ic} and Theorem~\ref{th:cluster} hold point-wise in the parameter space, and we expect that deriving a similar result uniformly in the parameter space (in particular, if cluster centers are allowed to depend on $n,T$ and if their separation is lost, see \citeasnoun{leeb2008} for such findings in the context of classical lasso penalized regression) is impossible. It is a well established fact in the Statistics and Econometrics literature that inference which is based on such 'point-wise' asymptotic results can be unreliable. Recently, several approaches to alleviate this problem and achieve uniformly valid post-regularization inference have been proposed (see, among others, \citeasnoun{belloni2014}, \citeasnoun{lockhart2014} and \citeasnoun{van2014}). Applying similar ideas to the present setting is a very important question which we leave for future research.
\end{remark}


\section{Details on the optimization problem in Algorithm 1}
To implement the proposed quantile panel data regression with group fixed effect, we need to solve the optimization problem stated in (\ref{eq:opt}). A natural normalization of the objective and the penalty function leads to 
\begin{equation}
\underset{\alpha_1, \dots, \alpha_n, \boldsymbol{\beta}}{\min} \frac{1}{nT}\sum_{i,t} \rho_\tau(Y_{it} - X_{it}^\top \boldsymbol{\beta} - \alpha_i) + \frac{\tilde \lambda}{n(n-1)}\sum_{i\neq j} \frac{|\alpha_i - \alpha_j|}{|\check{\alpha_i} - \check{\alpha_j}|^2}
\label{eq: opt1}
\end{equation}
this is equivalent to the objective (\ref{eq:opt}) except that $\tilde \lambda$ is adjusted according to $n$ and $T$ so that we can use a generic grid for $\tilde \lambda$ rather than letting the grid support  change with $(n, T)$. In practice, the grid support of $\tilde \lambda \in \{0, \tilde \lambda_1, \dots, \tilde \lambda_{\ell},\dots,  \tilde \lambda_{L}\}$ is chosen such that the number of distinct values of the solution $\{\hat \alpha_{1, \ell}, \dots, \hat \alpha_{n, \ell}\}$ for $\ell = 1, \dots, L$ takes all possible integer values in the set $\{1, 2, \dots, n\}$. This is always achievable  as long as the grid width of $\tilde \lambda$ is small enough. Since for each fixed $\tilde \lambda_{\ell}$, (\ref{eq: opt1}) is a linear programming problem which can be efficiently solved in any reliable solvers, this is not computationally expensive. 

To make this section self contained, we provide some details of the primal problem stated in (\ref{eq: opt1}) and its corresponding dual problem. Define $\lambda_{ij} := |\check{\alpha_i}  - \check{\alpha_j}|^{-2}$ and observe that we can re-write $|\alpha_i - \alpha_j| = 2\Big (\frac{1}{2} - 1\{0 \leq \alpha_i - \alpha_j\}\Big) \Big(0 - (\alpha_i - \alpha_j)\Big)$. With this notation \eqref{eq: opt1} can be equivalently expressed as follows 
\begin{align*}
&	\underset{\mathbf{u}, \mathbf{v}, \mathbf{w_1}, \mathbf{w_2}, \boldsymbol{\alpha}, \boldsymbol{\beta}}{\min} \quad  \frac{\tau}{nT} \sum_{i,t} u_{it} + \frac{(1-\tau)}{nT} \sum_{i,t} v_{it} + \frac{4 \tilde \lambda}{n(n-1)} \Big( \frac{1}{2} \sum_{j = 1}^{n(n-1)/2} w_{1j}  + \frac{1}{2}\sum_{j=1}^{n(n-1)/2}w_{2j}\Big)\\
	& \text{subject to}
	\end{align*}
\begin{align*}
	u_{it} & = \max\{Y_{it} - X_{it}^\top \boldsymbol{\beta} - \alpha_i, 0\} \\
	v_{it} & = \max \{X_{it}^\top \boldsymbol{\beta} + \alpha_i - Y_{it}, 0\}\\
	w_{1j} & = \max \{- \theta_j, 0\}\\
	w_{2j} & = \max \{\theta_j, 0\}\\
	Y_{it} & = u_{it} - v_{it} + \alpha_i + X_{it}^\top \beta\\
	0 & = w_{1j} - w_{2j} + \theta_j
	\end{align*}	
	with $\boldsymbol{\theta}$ being a vector of length $\frac{n(n-1)}{2}$ that consists entries $(\alpha_i- \alpha_j) \lambda_{ij}$ for $i < j$. We can represent $\boldsymbol{\theta}$ as $A \boldsymbol{\alpha}$ where $A$ is a $\frac{n(n-1)}{2}\times n$ matrix taking the form
	\[
	A = \begin{pmatrix} 
	\lambda_{12} & - \lambda_{12} & 0 & 0 & \dots &  0 & 0 \\
	\lambda_{13} & 0 & -\lambda_{13} & 0 &\dots &  0 & 0 \\
	&&&\dots &&&\\
	\lambda_{1n} & 0 & 0 & 0 & \dots & 0 & - \lambda_{1n}\\
	0 & \lambda_{23} & -\lambda_{23} & 0 & \dots & 0 & 0 \\
		&&&\dots &&&\\\\
		0 & 0 & 0 & 0 & \dots & \lambda_{n-1,n} & -\lambda_{n-1,n}
	\end{pmatrix}
	\]
The corresponding dual problem of (\ref{eq: opt1}) can be stated as: 
\begin{align*}
	 \underset{\boldsymbol{a_1}, \boldsymbol{a_2}}{\max} & \quad \boldsymbol{a_1}^\top Y \quad \quad  \text{subject to}\\
	X^\top \boldsymbol{a_1} &= (1-\tau) X^\top \mathbf{1}_{nT}\\
	Z^\top \boldsymbol{a_1} + \frac{4nT\tilde \lambda}{n(n-1)}A^\top \boldsymbol{a_2} & = (1-\tau) Z^\top \mathbf{1}_{nT} + \frac{2nT \tilde \lambda }{n(n-1)} A^\top \mathbf{1}_{n(n-1)/2}
		\end{align*}
	with $Z$ being the incidence matrix that identifies the $n$ individuals. The solution for $\boldsymbol{\alpha}$ and $\boldsymbol{\beta}$ in the primal problem is then the dual solutions of the dual problem. We implement the dual problem using the Mosek optimization software of \citeasnoun{mosek} through the R interface \textbf{Rmosek} of \citeasnoun{Rmosek}. We have also implemented the estimation procedure using the \textbf{quantreg} package in R and the code will be made available for public use.\footnote{Mosek is a commercial state-of-the-art convex optimization solver that provides a free academic license. We use its interior point algorithm to solve our linear programming problem. The estimation procedure implemented using the \textbf{quantreg} package calls the sfn method, which uses the Frisch-Newton algorithm and exploits the sparse algebra to compute iterates.}

\section{Monte Carlo Simulations}

\subsection{Finite Sample Performance of the Proposed Estimator}
To assess the finite sample performance of the proposed convex clustering panel quantile regression estimator, we apply the method to simulated data sets. In particular, we consider data generated from two models and two error distributions for a range of $n$ and $T$. The responses, $Y_{it}$, are generated by either a location shift model 
\begin{equation}
Y_{it} = \alpha_i + X_{it} \beta + u_{it}
\end{equation}
or a location scale shift model 
\begin{equation}
Y_{it} = \alpha_i + X_{it} \beta + (1+X_{it} \gamma) u_{it} \label{eq: lsm}
\end{equation}
where the individual latent effects $\alpha_i$ are generated from three groups taking values $\{1,2,3\}$ with equal proportions. The covariate $X_{it}$ is generated such that it has a non-zero interclass correlation coefficient. In particular, 
\[
X_{it} =\rho \alpha_i +  \gamma_i + v_{it}
\]
with $\gamma_i$ and $v_{it}$ independent and identically distributed over $i$ and $i,t$ respectively. We conduct simulation experiments with $\rho \in \{0, 0.5\}$ to investigate both cases where the fixed effect is independent or correlated with the covariate.\footnote{\citeasnoun{Koenker2004} used a similar data generating process with $\rho = 0$ for $X$ and pointed out that the interclass correlation induced by $\gamma_i$ is crucial for the penalized quantile regression fixed effect estimator to have superior performance than the unpenalized QRFE estimator. } The true parameters are $\beta = 1$ and $\gamma = 1/10$. The error terms $u_{it}$ are i.i.d. following either a standard normal distribution or a student t distribution with three degrees of freedom. Results reported are based on 2000 repetitions.

We first investigate the performance of using the information criterion for estimating the number of groups. Table \ref{tab:estnum_0.5} and Table \ref{tab:estnum_0.75} report the proportion of estimated number of groups under the two models for different combinations of $n$ and $T$ for $\tau = 0.5$ and $\tau = 0.75$, respectively. Throughout the simulations, we used an equally spaced grid of $\lambda$ values with width $1/200$ and support $[0, 0.35]$. This grid was chosen to ensure that the number of groups estimated for different $\lambda$ values covers the integers in range $[1, n]$. For the IC criteria, the sparsity function $\hat s(\tau)$ is estimated with bandwidth chosen based on the \citeasnoun{HallSheather} rule implemented in the \textbf{quantreg} package (see discussion in \citeasnoun{koenker2005}). 

The results suggest that the probability of getting the correct number of groups for $\tau = 0.5$ is slightly better than for $\tau = 0.75$. When $T \geq 30$, the estimates for the number of groups for both error distributions and for different $n$ are mostly satisfactory. The performance for $t$ error deteriorates compared to those with normal error, especially for higher quantiles. For $T = 15$ and quantiles other than the median the proposed method should be used with caution. Including correlation between individual effects and predictors does not lead to dramatic changes in the accuracy for estimating the number of groups and group membership.

Table \ref{tab:betaest_0.5}  and Table \ref{tab:betaest_0.75} summarize the finite sample properties of $\hat \beta^{IC}(\tau)$ for $\tau = 0.5$ and $\tau = 0.75$, respectively and compare with the QRFE estimator where no penalization on the individual fixed effect is used (i.e. $\lambda = 0$). 

The standard errors used for constructing confidence intervals (nominal coverage $95\%$) {are based on the \texttt{nid} option with Hall-Sheather bandwidth in the package quantreg}. Results based on the Bofinger bandwidth selection are similar and not reported here.
For the QRFE, the Bofinger bandwidth rule was used since the Hall-Sheather rule resulted in substantial under-coverage with $T=30$ for some of the models.


When covariates and fixed effects are independent, the RMSE of the PQR-FEgroup estimator, $\hat \beta(\tau)$, is smaller than that of the QRFE for all settings considered. This shows that our penalization gains efficiency for estimating $\beta$ when there is group structure in the fixed effects. The results do not change much from normal error to $t$ error and from median to higher quantiles. 

Introducing correlation between predictors and group membership leads to a bias for the grouped effect estimator, while the fixed effects estimator does not suffer from additional bias. This bias can be quite noticeable for small values of $T$, especially at the $75\%$ quantile. The bias becomes negligible as $T$ increases, so there is no contradiction to our asymptotic theory. An intuitive explanation for this behaviour is that for smaller $T$ it is difficult to get a perfect grouping, and a wrong grouping leads to bias since there is dependence between predictors and group structure.


Last, we report in Table \ref{tab:member_0.5} and Table \ref{tab:member_0.75} the proportion of perfect classification of individual effects and the average value of the percentage of correct classification together with their standard errors. Since the comparison of the estimated membership and the true membership only makes sense when $\hat K = K_0$, the estimated membership are based on $\lambda$ for which $\hat K = K_0$ (see \citeasnoun{su2016} for a similar approach). Results suggest that for $T\geq 30$ and $\tau = 0.5$, the group membership estimation is quite satisfactory. While the proportion of perfect matches is low even for $T = 30$, the average proportion of correct classification shows that those effects are typically due to very few misclassified individuals. For $T= 15$ perfect classification is almost impossible while average correct classification rates remain reasonable. Adding correlation between individual effects and covariates leads to a deterioration of the probability for achieving a perfect grouping for location-scale models, especially at higher quantiles, but does not have a strong impact on other results. Overall the simulations suggest that for small $T$, there is just not enough information available for each individual to hope for perfect classification.

\subsection{Further analysis of tuning parameters in the IC Criteria} \label{MCsec}
\begin{figure}[!h]
	\captionsetup{font=scriptsize}
	\includegraphics[scale=0.75]{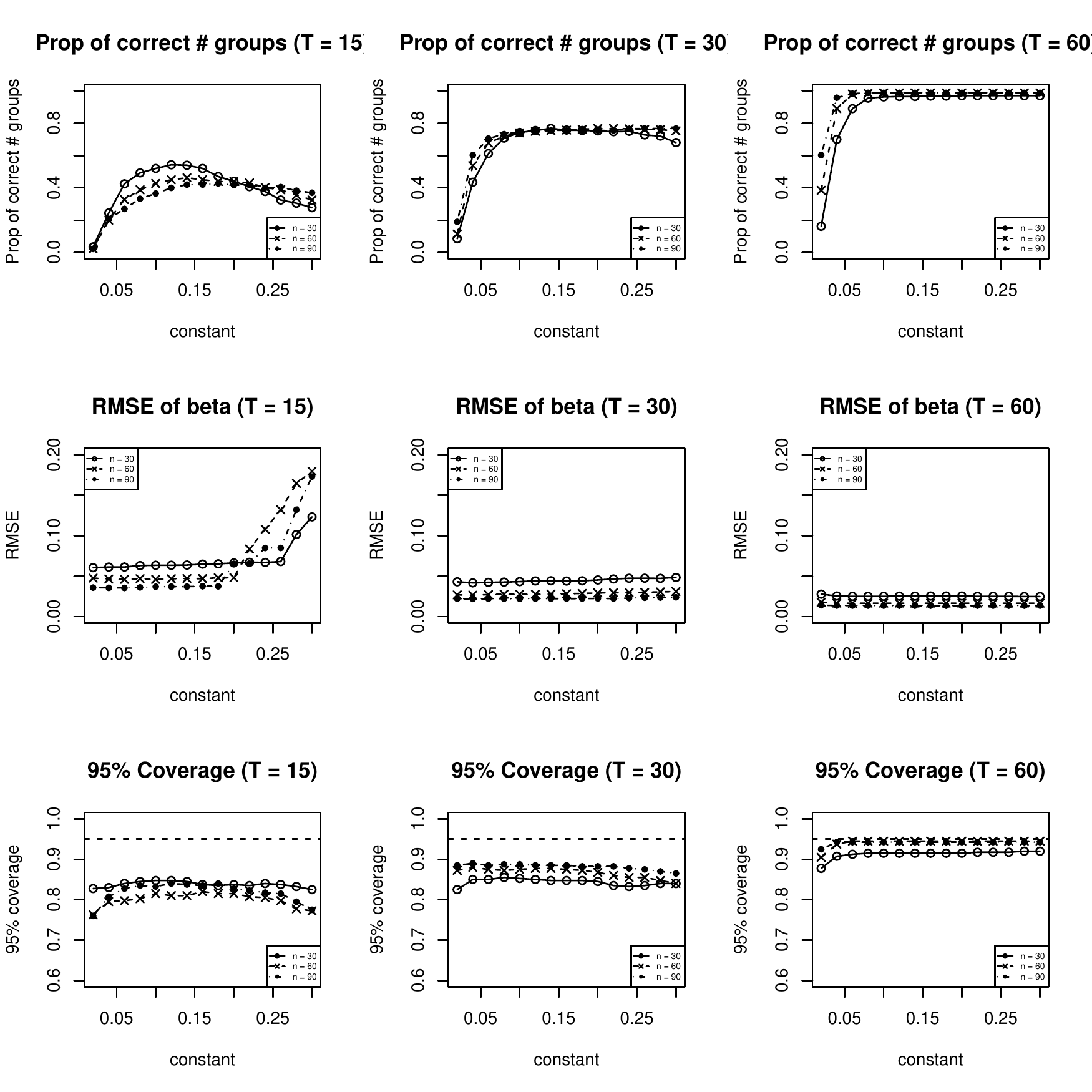}
	\caption{Use different constants in $p_{n,T}$ for the IC criteria for location-scale shift model with $t$ error on DGP1: For a equally spaced grid on $[0.01, 0.3]$ with width 0.01, the three columns represent different magnitudes of $T$ while each figures in the row overlays the curves for $n \in \{30,60,90\}$ for various performance measures. The first row plots the proportion of correctly estimated number of groups. The second row plots the RMSE of $\hat \beta^{IC}(\tau)$ where $\tau =0.5$ and the third plots the coverage rate for nominal size 5\%. Results are based on 400 repetitions. }
	\label{fig: CC1}
\end{figure}

\begin{figure}[!h]
	\captionsetup{font=scriptsize}
	\includegraphics[scale=0.75]{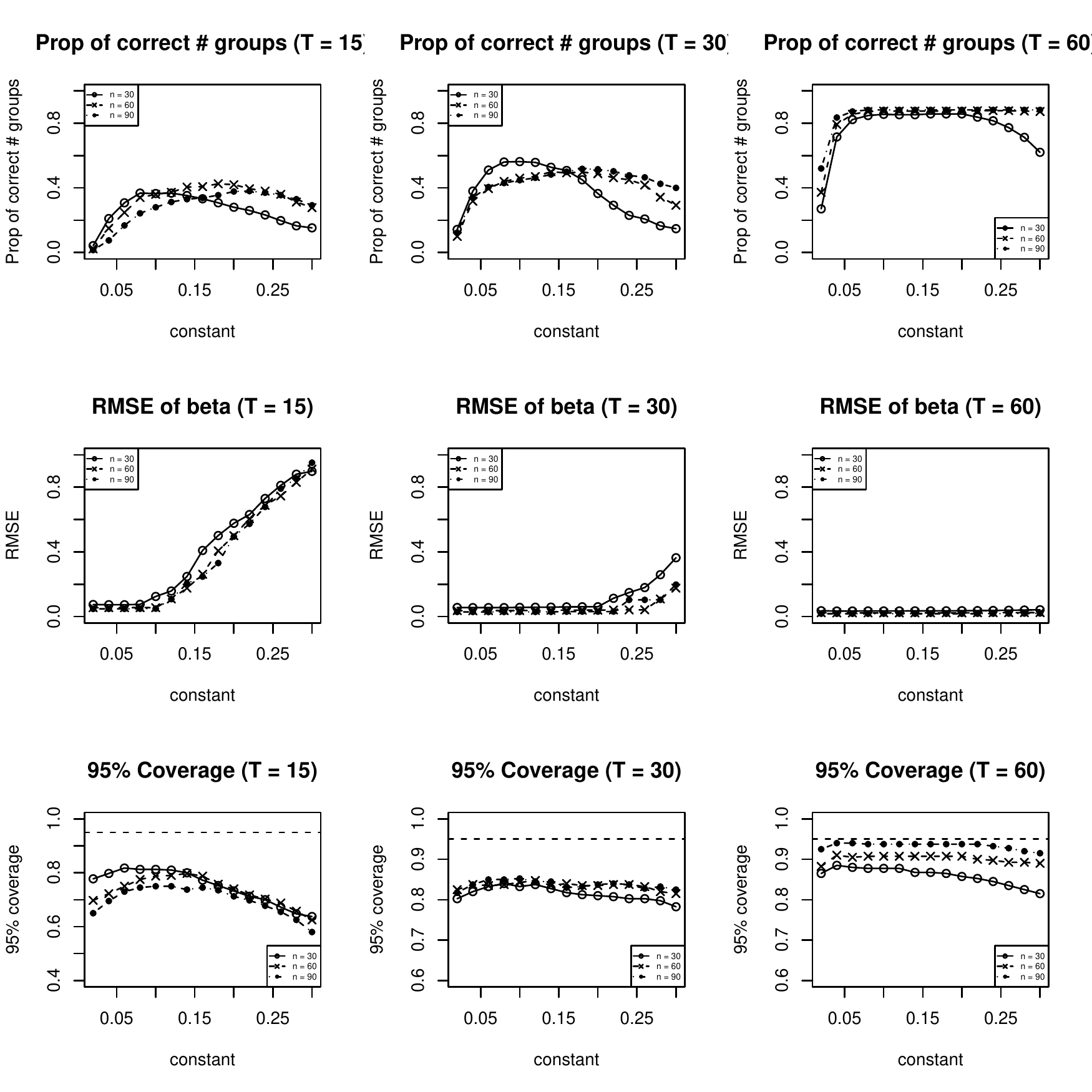}
	\caption{Use different constants in $p_{n,T}$ for the IC criteria for location-scale shift model with $t$ error on DGP1: For a equally spaced grid on $[0.01, 0.3]$ with width 0.01, the three columns represent different magnitudes of $T$ while each figures in the row overlays the curves for $n \in \{30,60,90\}$ for various performance measures. The first row plots the proportion of correctly estimated number of groups. The second row plots the RMSE of $\hat \beta^{IC}(\tau)$ where $\tau =0.75$ and the third plots the coverage rate for nominal size 5\%. Results are based on 400 repetitions. }
	\label{fig: CC2}
\end{figure}

The discussion at the end of Section~\ref{sec:meth} provides a motivation for the tuning parameters $\hat C$ in the IC criteria. Here we further investigate the impact of rescaling $p_{n,T}$ by different factors. For illustration we consider DGP1 in the previous section where data are generated based on the model (\ref{eq: lsm}). We use a grid of constants $c \in [0.01, 0.3]$ with width 0.01 and plot the associated performance of the estimated number of groups, the RMSE of $\hat \beta^{IC}(\tau)$ and the coverage rate for $p_{n,T} = cnT^{1/4}$. Figure \ref{fig: CC1} and Figure \ref{fig: CC2} contain corresponding results for the location-scale shift model with $t$ errors and $\tau=0.5,0.75$, respectively. For $T$ as small as 15, the performance is quite sensitive to the chosen constant. As predicted by the theory this dependence becomes somewhat less prominent as $T$ increases. Overall the choice $p_{n,T} = nT^{1/4}/10$ shows good performance for settings that we tried in this simulation. The patterns are similar for those with the normal error and the location-shift models and likewise for DGP2, results are reported in Figure \ref{fig: CC1_ls}- Figure \ref{fig: CC2_DGP2_ls} in the Appendix for the sake of completeness.

\section{Empirical Example} 
To further illustrate our proposed methodology, we revisit the empirical inquiry on the much-debated "More Guns Less Crime" hypothesis. \citeasnoun{LM97} provide the first empirical analysis which claims that the adoption of Right-to-Carry (RTC) laws, which allows local authorities to issue a concealed weapon permit to all applicants that are eligible, reduces crime. Ever since its publication, there has been much academic and political debate that challenges the findings. \citeasnoun{AD03} shows that the negative effect of RTC laws has no statistical significance under a more reasonable model specification and inference using both the state and county level data between 1977 - 1999 for 51 U.S. states. Their conclusion is echoed by the National Research \citeasnoun{NRC2004} (NRC) report which finds little reliable statistical support for the ``More Guns Less Crime" hypothesis. Recently, \citeasnoun{ADZ14} revisited the hypothesis using the updated panel data from 1977 - 2010. They correct several mistakes in the dataset used in earlier analysis and discuss the shortcoming of using county level crime data. We refer the readers to more details provided in \citeasnoun{ADZ14}. For our empirical analysis, we use their updated state level data kindly provided by the authors.\footnote{The data and the detailed description of the data source can be downloaded from \url{https://works.bepress.com/john_donohue/107/}.} 

The main parameter of interest is the effect of the indicator of the RTC laws (denoted as \textit{lawind}) on crime rates. In our analysis, we focus on the violent crime rate. A similar analysis is possible for other categories of offense. In addition to the law indicator, there is also information on the incarceration rate (sentenced prisoners per 100,000 residents; denoted as \textit{prisoner}) in the states in the previous year, real per capita personal income (denoted as \textit{rpcpi}) and other demographic variables on population proportions in different age-gender groups for various ethnicities. As argued in \citeasnoun{AD03} and \citeasnoun{ADZ14}, to avoid multicollinearity and accounting for the fact that 90\% of violent crimes in the U.S. are committed by male offenders, we follow their specification and control for only the proportion of African-American male in the age group 10-19, 20-29 and 30-39. Inevitably, there are many different model specifications that can be considered for this empirical inquiry and it is impossible to report all the results. Our goal is to emphasize the heterogeneous law adoption effect for states that have low violent crime rate versus those with high crime rate. We also provide evidence of clustering behavior of the state fixed effects which are incorporated to capture states' unobserved heterogeneity and show that by taking advantage of the dimension reduction of grouping the fixed effects, the parameters of interest on other control variables enjoy better statistical precision.  

Our main model specification is 
\begin{equation}
q_{i,\tau}(\log (\text{violent}_{it})) = \alpha_{i}(\tau) + \gamma_t(\tau) + \beta(\tau) \text{lawind}_{it} + \theta(\tau)^\top X_{it}
\label{eq: model1}
\end{equation}
where the additional control variables $X_{it}$ include lagged incarceration rate (\textit{prisoners}), real per capita personal income (\textit{rpcpi}) and the three demographic variables (\textit{afam1019}, \textit{afam2029}, \textit{afam3039}). We note that high incarceration rates in a given state may be a feedback towards rising violent crime, and therefore including those as control variables might lead to endogeneity issues. As a robustness check, we also report the corresponding results in the appendix for model (\ref{eq: model1}) without the incarcerating rate as a control covariate. The effects stay mostly unchanged. 

\begin{figure}[!h]
	\captionsetup{font=scriptsize}
\includegraphics[scale=0.75]{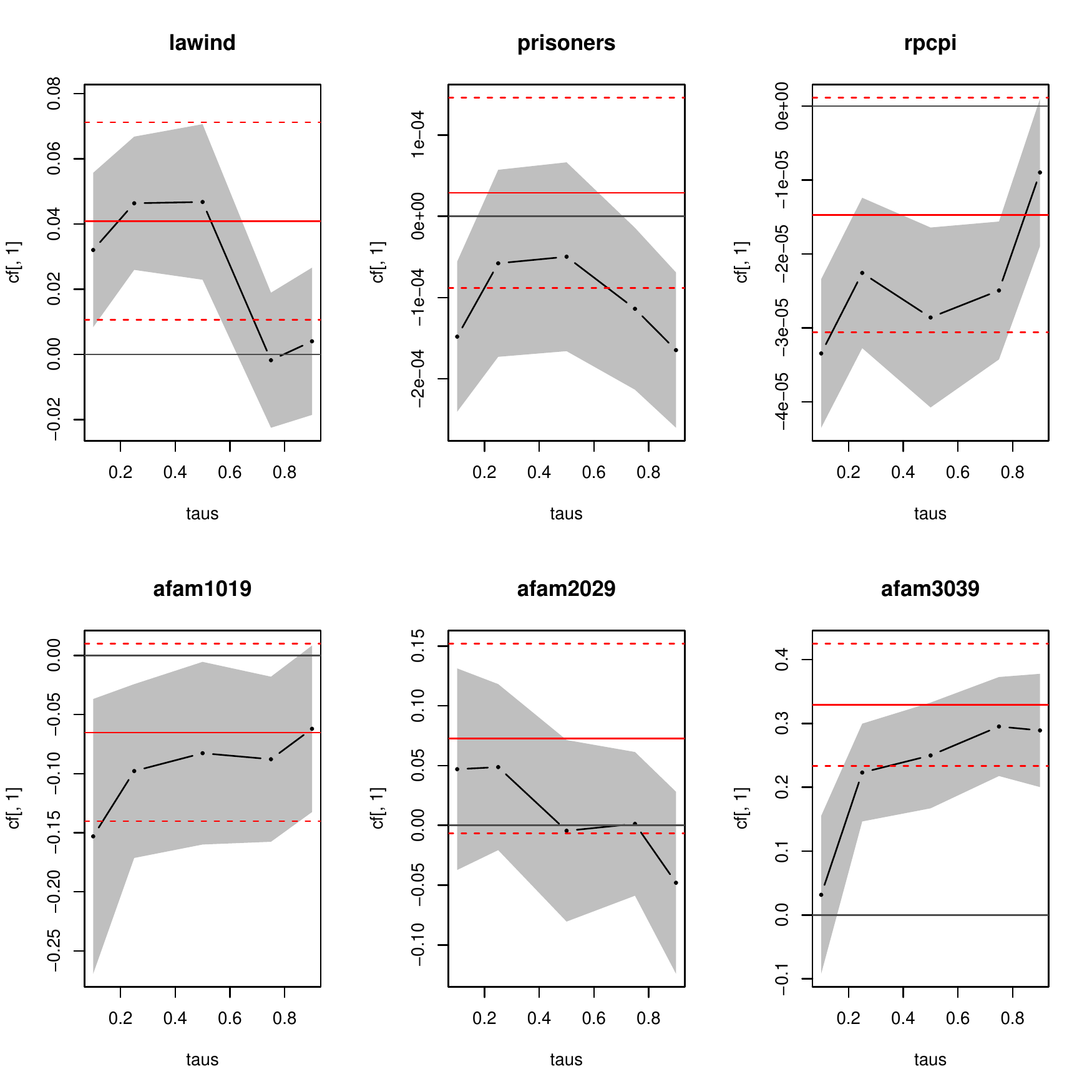}
\caption{Panel data quantile regression estimates with state fixed effects for various $\tau$ based on model specification (\ref{eq: model1}): For $\tau \in \{0.1, 0.25, 0.5, 0.75, 0.9\}$, the solid black points plot the coefficient estimates for the effects of the RTC law adoption and other control variables on the violent crime rate based on panel data of 51 U.S. states for 1977 - 2010. The shaded area is the pointwise 95\% confidence interval for which the standard errors are computed using the Hendricks-Koenker sandwich covariance matrix estimates with the Hall-Sheather bandwidth rule. The red solid line marks the fixed effect panel data mean regression estimates with the dotted red lines plots the 95\% confidence interval with robust clustered (at the states level) standard errors. }
\label{fig: QRFE}
\end{figure}

\begin{figure}[!h]
		\captionsetup{font=scriptsize}
\centering	\includegraphics[scale=0.65]{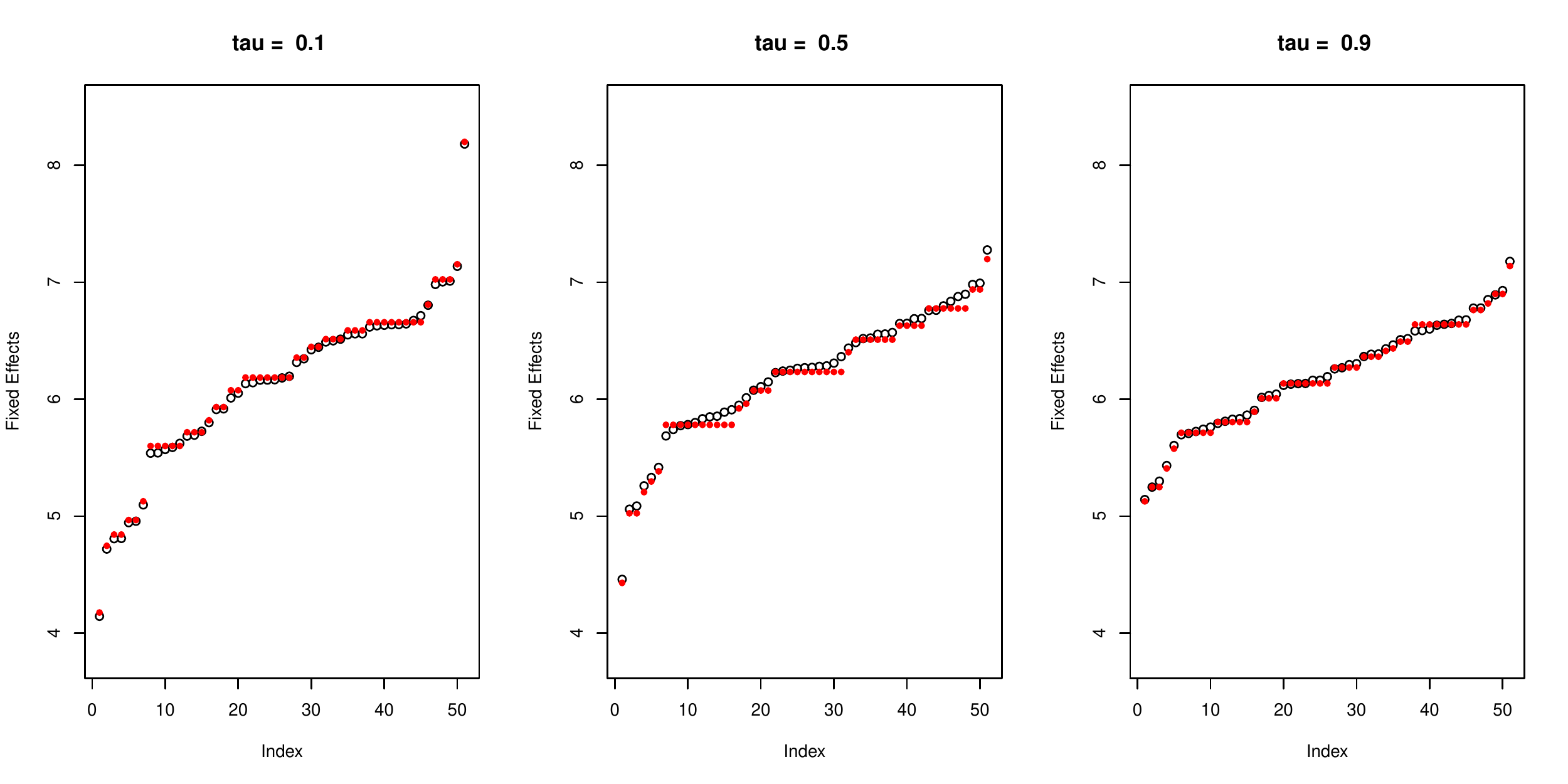}
		\caption{The estimated state fixed effects and the corresponding grouped fixed effects for various $\tau$ based on model specification (\ref{eq: model1}): For $\tau \in \{0.1,  0.5,  0.9\}$, the hollow black points plot the ordered panel data quantile regression estimates for individual state fixed effects. The red solid points mark out the estimated grouping and the corresponding group fixed effect estimates using the proposed methodology.}
			\label{fig: fixedeff}
\end{figure}

Figure \ref{fig: QRFE} reports the panel data quantile regression estimates with state fixed effects for $\tau \in \{0.1, 0.25, 0.5, 0.75, 0.9\}$ and their associated point-wise confidence intervals. Similar to the findings in \citeasnoun{ADZ14}, we see that the RTC law-adoption has a positive effects on violent crime rate. In addition this effect is significant for lower quantiles while there is no statistical significance for such an effect for states at higher quantiles of the violent crime rate. All other control variables have expected signs, with a negative effect of the lagged incarceration rate indicating that states with stricter laws have lower violent crime rates, although this effect is not very precisely estimated. Higher real per capita personal income has a negative effect on violent crime at lower quantiles; this effect is diminishing for higher quantiles of violent crime rate. The proportion of African-American population at age group 30 - 39 has a positive effect on the violent crime and this effect becomes more prominent for higher crime rate states. Figure \ref{fig: fixedeff} shows the corresponding state fixed effect estimates for different quantile levels. There is some evidence of clustered behavior of these fixed effect estimates, although these are estimates with statistical errors. Using our proposed methodology, the estimated number of groups for the five quantile levels are respectively $\{20, 20,16,17,19\}$.

\begin{figure}[!h]
		\captionsetup{font=scriptsize}
	\includegraphics[scale=0.75]{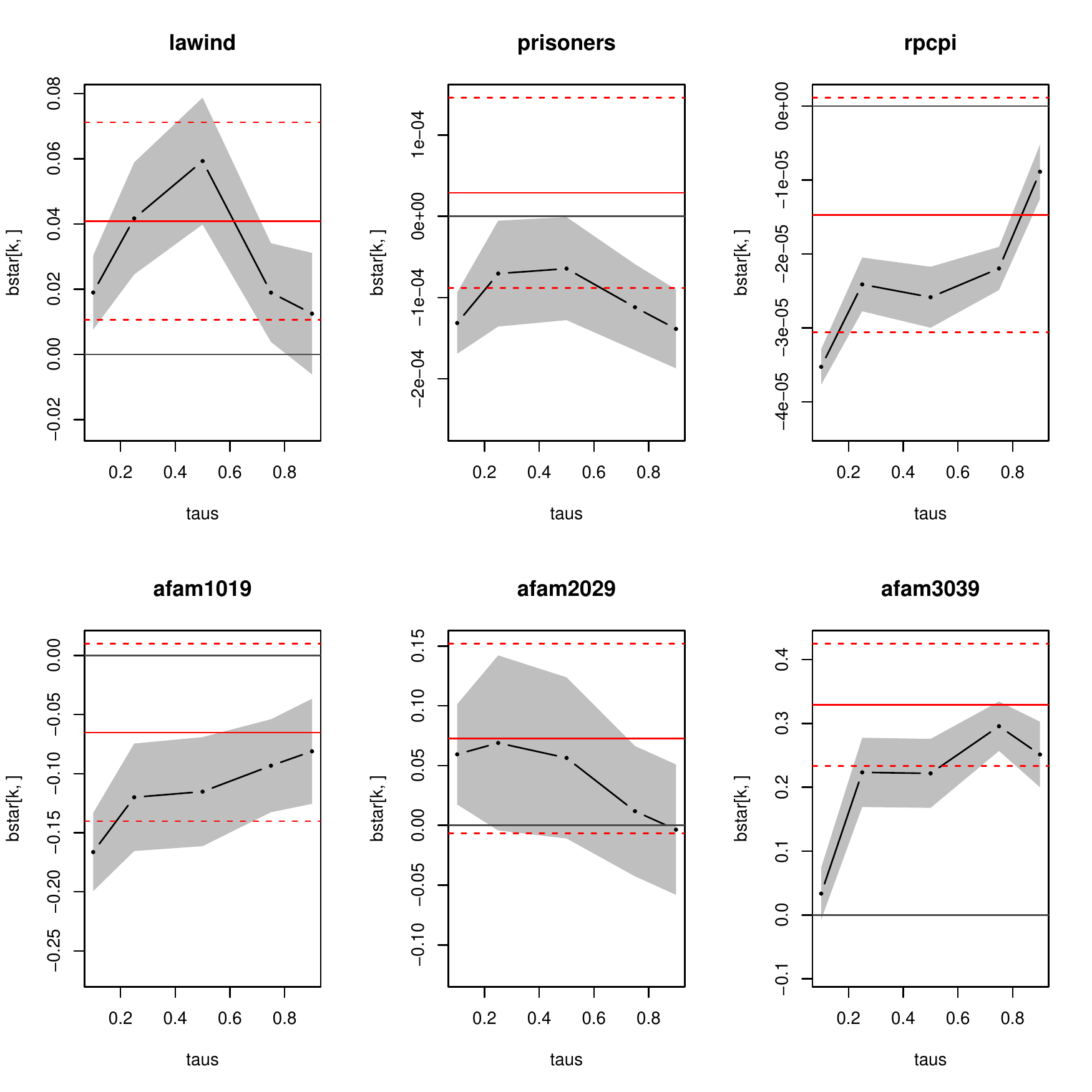}
	\caption{Panel data quantile regression estimates with grouped state fixed effects for various $\tau$ based on model specification (\ref{eq: model1}): For $\tau \in \{0.1, 0.25, 0.5, 0.75, 0.9\}$, the solid black points plot the coefficient estimates for the effects of the RTC law adoption and other control variables on the violent crime rate based on the proposed methodology with panel data of 51 U.S. states for 1977 - 2010. The shaded area are the pointwise 95\% confidence interval where the standard errors are computed using the Hendricks-Koenker sandwich covariance matrix estimates with the Hall-Sheather bandwidth rule. The red solid line marks the fixed effect panel data mean regression estimates with the dotted red lines plot the 95\% confidence interval with robust clustered (at the state level) standard errors. }
	\label{fig: QRFEgroup}
\end{figure}

Figure \ref{fig: QRFEgroup} plots the corresponding panel data quantile regression estimates with the estimated optimal grouped state fixed effects for $\tau \in \{0.1, 0.25, 0.5, 0.75, 0.9\}$ and their associated point-wise confidence intervals. The pattern of the quantile effects for all the control variables stay roughly the same compared to the fixed effect quantile regression estimates, while the variance of these estimates under the grouped fixed effects are noticeably smaller. This is similar to what we observe in the simulation section where the common parameters in quantile panel data regression with grouped fixed effects have lower variances than those of the estimates based on ``individual heterogeneity". However, we would also like to point out that the standard errors are based on refitted models with estimated group structure, which does not account for the uncertainty of model selection on the fixed effects and should be interpreted with caution. A proper inference method based on the proposed methodology accounting for such uncertainty is important and is part of our future research agenda. 

\section{Conclusions and future extensions}

The present paper suggests a simple and computationally efficient way to incorporate group fixed effects into a panel data quantile regression by means of a convex clustering penalty.
We develop theoretical results on consistent group structure estimation and discuss the asymptotic properties of the resulting joint and group-specific estimators.  
 
There are several directions that we plan to explore in the future. First, our theory focused on individual fixed effects while assuming common slope coefficients. It is equally interesting to allow for group structure in some of the slope coefficients while keeping other slope coefficients common across individuals, perhaps even allowing for individual fixed effects. This can be achieved by straightforward modifications of the penalization approach which we explored so far, but a more detailed theoretical analysis of this approach remains beyond the scope of the present paper.

Second, one can take the standpoint that in many applications there is no exact group structure. In such settings, an alternative interpretation of the penalty which we investigated is as a way of regularizing problems that have too many parameters. Such an interpretation is in the spirit of the proposals of \citeasnoun{Koenker2004} and \citeasnoun{Lamarche}, and a detailed investigation of the resulting bias-variance trade-off warrants further research.

Finally, a deeper analysis of issues that are related to uniformity of distributional approximation in the entire parameter space was not addressed here, but remains an important theoretical and practical question which we hope to address in the future.

\newpage

\bibliography{_ClusterQR}

\newpage

\section{Proofs}\label{sec:proofs}

We begin by collecting some useful facts and defining additional notation. We will repeatedly make use of Knight's identity (see~\cite{koenker2005}, p. 121)) which holds for $u \neq 0$:
\begin{equation} \label{eq:knight}
\rho_{\tau}(u-v) - \rho_{\tau}(u) = -v\psi_\tau(u) + \int_0^v \1\{u \leq s\} - \1\{u \leq 0\} ds. 
\end{equation}
Additionally, let $\gamma_{0i} := (\alpha_{0i},\beta_0^\top)^{\top}$. The symbols $a_n \lesssim b_n, a_n \gtrsim b_n$ will mean that there exists a non-random constant $C \in (0,\infty)$ which is independent of $n,T,\tau$ such that $P(a_n \leq Cb_n) = 1$ and $P(a_n \geq Cb_n) = 1$, respectively. Define $\eps_{it}^\tau := Y_{it} - Z_{it}^\top \gamma_{i0}(\tau)$ and let $F_{\eps_{it}^\tau|X_{it}}(u|X_{it}) = F_{Y_{it}|X_{it}}( Z_{it}^\top \gamma_{i0}(\tau) + u | X_{it})$ denote the conditional cdf of $\eps_{it}^\tau$ given $X_{it}$. When there is no risk of confusion, we will also write $\eps_{it}$ instead of $\eps_{it}^\tau$. Define $\psi_\tau(x) := (\1\{x\leq 0\} - \tau)$.

\subsection{Proof of Theorem~\ref{th:cluster}}

We begin by stating some useful technical results which will be proved at the end of this section.

\begin{lemma} \label{lem:rhotau1}
	For any fixed $\beta \in \R^p$ define $\eps_{it,\beta}^\tau := Y_{it} - X_{it}^\top\beta - \alpha_{i0}(\tau)$. Then we have under assumptions (A1)-(A3)
	\[
	\sum_{t=1}^T \rho_\tau(\eps_{it,\beta}^\tau - a_1) - \rho_\tau(\eps_{it,\beta}^\tau - a_2)
	= (a_2-a_1) \sum_{t=1}^T \psi_\tau(\eps_{it,\beta}^\tau) 
	+ \tilde r_{n,i}^{(1)}(a_1,a_2) + \tilde r_{n,i}^{(2)}(a_1,a_2)  
	\]
	where 
	\[
	\sup_i \tilde r_{n,i}^{(1)}(a_1,a_2) \lesssim T|a_1-a_2| \max(|a_1|,|a_2|), \quad \sup_i \tilde r_{n,i}^{(2)}(a_1,a_2) = |a_1-a_2|O_P(T^{1/2}(\log n)^{1/2}).
	\]
\end{lemma}

\begin{lemma}\label{lem:rhoulb} Under assumptions (A1)-(A3) there exist $\eps >0, \infty > c_1,c_0 > 0$ such that for all $i=1,...,n$
	\begin{equation}\label{eq:br}
	c_1 \|\gamma - \gamma_{0i}\|^2 \geq \E[\rho_\tau(Y_{it} - Z_{it}^\top \gamma)] -\E[\rho_\tau(Y_{it} - Z_{it}^\top \gamma_{0i})] \geq c_0(\|\gamma - \gamma_{0i}\|^2 \wedge \eps^2).
	\end{equation}
\end{lemma}

\begin{lemma}\label{lem:sn1} Under assumption (A1) define for fixed $B \in \R$
	\begin{equation}\label{eq:sn1}
	s_{n,1}(B) := \sup_i \sup_{|\gamma|\leq B}\Big | \sum_{t} \Big( \rho_\tau(Y_{it} - Z_{it}^\top \gamma) - \rho_\tau(Y_{it}) - \E[\rho_\tau(Y_{it} - Z_{it}^\top \gamma) - \rho_\tau(Y_{it})] \Big)\Big| 
	\end{equation} 
	We have for any fixed $B<\infty$, provided that $\min (n,T) \to \infty, \log n = o(T)$
	\begin{equation}\label{eq:sn1_b}
	s_{n,1}(B) = O_P(T^{1/2} (\log n)^{1/2}). 
	\end{equation} 
\end{lemma}

\textbf{Proof of Theorem~\ref{th:cluster}}

\noindent
\textbf{Step 1: first bounds} In this step we shall prove that
\begin{equation}\label{eq:bound1}
\|\hat \beta - \beta_0\|^2 + \sup_i |\hat \alpha_i - \alpha_{i0}|^2 = O_P(T^{-1/2}(\log n)^{1/2} + \Lambda_D n/T).
\end{equation}
Combine the results in Lemma~\ref{lem:rhoulb} and Lemma~\ref{lem:sn1} to find that any minimizer of $\Theta(\alpha_1,...,\alpha_n,\beta)$ must satisfy
\begin{align*}
c_0 T \sum_i \Big\{(\|\beta - \beta_0\|^2 + |\alpha_i - \alpha_{i0}|^2)\wedge\eps^2 \Big\} \lesssim n s_{n,1} + \Lambda_D n^2. 
\end{align*}
Let $N_\Delta := \# \{i: \|\beta - \beta_0\|^2 + |\alpha_i - \alpha_{0i}|^2 \geq \Delta\}$. Then for any $0 < \Delta < \eps^2$ 
\[
T N_\Delta \Delta = O_P( n s_{n,1} + \Lambda_D n^2),
\]
i.e. by Lemma~\ref{lem:sn1} 
\[
N_\Delta = n O_P((T^{-1/2}(\log n)^{1/2}) + \Lambda_D n/T) \Delta^{-1} 
\]
and in particular $N_\Delta = o_P(n)$ as long as $\Delta \gg T^{-1/2}(\log n)^{1/2} + \Lambda_D n/T$. Provided that $\Lambda_D n/T = o_P(1)$ we obtain
\begin{equation} \label{eq:beta1}
\|\hat \beta - \beta_0\|^2 = O_P(T^{-1/2}(\log n)^{1/2} + \Lambda_D n/T).
\end{equation}
Define $D_{n,T} := T^{-1/2}(\log n)^{1/2} + \Lambda_D n/T$. Next we will prove that, provided {$n \Lambda_D = o_P(T^{1/2}), T^{3/4}(\log n)^{3/4}/(n\Lambda_S) = o_P(1)$} also
\[
\sup_i |\hat \alpha_i - \alpha_{i0}|^2 = O_P(D_{n,T}).
\]
To this end, it suffices to prove that $\sup_i |\hat \alpha_i - \alpha_{i0}|^2 = O_P(d_{n,T})$ for any $n\Lambda_S \gg d_{n,T} \gg D_{n,T}$. Define
\[
\tilde\alpha_i = \left\{ 
\begin{array}{ccc}
\hat \alpha_i &if& |\hat \alpha_i - \alpha_{0i}|^2 \leq d_n,
\\
\alpha_{i0} + d_n^{1/2}\mbox{sgn}(\hat \alpha_i - \alpha_{i0}) &if& |\hat \alpha_i - \alpha_{0i}|^2 > d_n.
\end{array}
\right.
\] 
Define the set $E := \{i: \tilde\alpha_i = \hat\alpha_i\}$. Observe that
\begin{align*}
\Big||\tilde\alpha_i - \tilde \alpha_j| - |\hat\alpha_i - \hat \alpha_j|\Big| &\leq |\hat \alpha_i - \tilde \alpha_{i}| + |\hat \alpha_j - \tilde \alpha_{j}| & \quad \forall i,j, 
\\
|\tilde\alpha_i - \tilde \alpha_j| - |\hat\alpha_i - \hat \alpha_j| &< - |\hat \alpha_i - \alpha_{0i}| + d_{n,T}^{1/2} &\quad \exists k: i \in I_k\cap E^C, j \in I_k \cap E,
\\
|\tilde\alpha_i - \tilde \alpha_j| - |\hat\alpha_i - \hat \alpha_j| &\leq 0 &\quad \exists k: i,j \in I_k.   
\end{align*}
Thus
\begin{align*}
& \sum_{i,j} \lambda_{i,j}\Big\{|\hat \alpha_i - \hat \alpha_j| - |\tilde \alpha_i - \tilde \alpha_j| \Big\}
\\
=~& \Big(2\sum_{i \in E^C}\sum_{j \in E} + \sum_{i \in E^C}\sum_{j\in E^C} \Big)  \lambda_{i,j}\Big\{|\hat \alpha_i - \hat \alpha_j| - |\tilde \alpha_i - \tilde \alpha_j| \Big\}
\\
=~& 2 \sum_k \sum_{i \in I_k \cap E^C}\Big(\sum_{j \in I_k \cap E} + \sum_{j \in I_k^C\cap E} \Big)\lambda_{i,j}\Big\{|\hat \alpha_i - \hat \alpha_j| - |\tilde \alpha_i - \tilde \alpha_j| \Big\}
\\
& \quad + \sum_k \sum_{i \in I_k \cap E^C}\Big(\sum_{j \in I_k \cap E^C} + \sum_{j \in I_k^C\cap E^C} \Big)\lambda_{i,j}\Big\{|\hat \alpha_i - \hat \alpha_j| - |\tilde \alpha_i - \tilde \alpha_j| \Big\}
\\
\geq~& 2 \sum_k \sum_{i \in I_k \cap E^C} \Big(\Lambda_S |I_k \cap E| - n \Lambda_D \Big)\{|\hat\alpha_{i} - \alpha_{0i}| - d_{n,T}^{1/2}\} 
\\
& \quad - \sum_k \sum_{i \in I_k \cap E^C}\sum_{j \in I_k^C\cap E^C}\lambda_{i,j}\Big\{|\hat \alpha_i - \tilde \alpha_{i}| + |\hat \alpha_j - \tilde \alpha_{j}| \Big\}
\\
\geq~& 2 \sum_k \sum_{i \in I_k \cap E^C} \Big(\Lambda_S |I_k \cap E| - n \Lambda_D \Big)\{|\hat\alpha_{i} - \alpha_{0i}| - d_{n,T}^{1/2}\} 
\\
& \quad + \sum_k \sum_{i \in I_k \cap E^C} \Big\{- n\Lambda_D \{|\hat\alpha_{i} - \alpha_{0i}| - d_{n,T}^{1/2}\} - \Lambda_D \sum_{j \in I_k^C \cap E^C} \{|\hat\alpha_{j} - \alpha_{0j}| - d_{n,T}^{1/2}\}\Big\}
\\
\geq~& 2 \sum_k \sum_{i \in I_k \cap E^C} \Big(\Lambda_S |I_k \cap E| - 2 n \Lambda_D \Big)\{|\hat\alpha_{i} - \alpha_{0i}| - d_{n,T}^{1/2}\}. 
\end{align*}
Now since $N_{d_n} = o_P(n)$ and by definition of $\tilde\alpha_n$ it follows that under (C) 
\[
\max_k \Big| \frac{|I_k \cap E|}{n\mu_k} - 1 \Big| = o_P(1),
\]
and since by assumption $\Lambda_D/\Lambda_S = o_P(1)$ we obtain 
\[
\sum_{i,j} \lambda_{i,j}\Big\{|\hat \alpha_i - \hat \alpha_j| - |\tilde \alpha_i - \tilde \alpha_j| \Big\} \gtrsim n\Lambda_S \sum_{i \in E^C}  \{|\hat\alpha_{i} - \alpha_{0i}| - d_{n,T}^{1/2}\}.
\]

Next we note that for any $i$ with $|\hat \alpha_i - \alpha_{0i}|^2 \geq (2+c_1/c_0)d_{n,T}$ we have 
\begin{align*}
&\frac{1}{T}\sum_t \rho_\tau(Y_{it} - X_{it}^\top \hat \beta - \hat\alpha_i) - \rho_\tau(Y_{it} - X_{it}^\top \hat\beta - \tilde \alpha_i) 
\\
\geq~& \int \rho_\tau(y - x^\top \hat\beta - \hat \alpha_{i}) - \rho_\tau(y - x^\top \hat\beta - \tilde \alpha_{i}) dP^{Y_{i1},X_{i1}}(x,y) - 2s_{n,1}/T
\\
=~& \int \rho_\tau(y - x^\top \hat\beta - \hat \alpha_{i}) - \rho_\tau(y - x^\top \beta_0 - \alpha_{i0}) dP^{Y_{i1},X_{i1}}(x,y)
\\
& \quad \quad - \int \rho_\tau(y - x^\top \hat\beta - \tilde \alpha_{i}) - \rho_\tau(y - x^\top \beta_0 - \alpha_{i0}) dP^{Y_{i1},X_{i1}}(x,y) - 2s_{n,1}/T
\\
\geq~& c_0 (\{\|\hat \beta - \beta_0\|^2 + |\hat\alpha_i - \alpha_{0i}|^2\}\wedge\eps^2) - c_1(\|\hat \beta - \beta_0\|^2 + |\tilde\alpha_i - \alpha_{0i}|^2) - 2s_{n,1}/T
\\
>~& 0
\end{align*}
with probability tending to one by~\eqref{eq:beta1} and the definition of $d_{n,T}$. For $i$ with $|\hat \alpha_i - \alpha_{0i}|^2 < (2+c_1/c_0)d_{n,T}$ note that by Lemma~\ref{lem:rhotau1}
\begin{align*}
&
\Big| \sum_t \rho_\tau(Y_{it} - X_{it}^\top \hat \beta - \hat\alpha_i) - \rho_\tau(Y_{it} - X_{it}^\top \hat\beta - \tilde \alpha_i) \Big| 
\\
\lesssim~& |\hat \alpha_i - \tilde\alpha_i| \Big(\sum_t \psi_\tau(\eps_{it,\hat\beta}^\tau) + T d_{n,T}^{1/2} + O_P(T^{1/2}(\log n)^{1/2})\Big)
\\
\lesssim~& \{|\hat\alpha_{i} - \alpha_{0i}| - d_{n,T}^{1/2}\} \Big(T \|\hat \beta - \beta_0\| + T d_{n,T}^{1/2} + O_P(T^{1/2}(\log n)^{1/2})\Big)
\end{align*}
where the $O_P$ terms are uniform in $i$. Thus
\begin{align*}
&\sum_i\sum_t \rho_\tau(Y_{it} - X_{it}^\top \hat \beta - \hat\alpha_i) - \rho_\tau(Y_{it} - X_{it}^\top \hat\beta - \tilde \alpha_i)
\\
\gtrsim~& - \Big(T d_{n,T}^{1/2} + O_P(T^{1/2}(\log n)^{1/2})\Big)\sum_{i \in E^C}\{|\hat\alpha_{i} - \alpha_{0i}| - d_{n,T}^{1/2}\} 
\end{align*}
Summarizing we have proved that
\begin{align*}
&\Theta(\hat\alpha_1,...,\hat\alpha_n,\hat\beta) - \Theta(\tilde\alpha_1,...,\tilde\alpha_n,\hat\beta)
\\
\gtrsim~& \Big[ n\Lambda_S - \Big(T d_{n,T}^{1/2} + O_P(T^{1/2}(\log n)^{1/2})\Big)\Big]\sum_{i \in E^C}\{|\hat\alpha_{i} - \alpha_{0i}| - d_{n,T}^{1/2}\}. 
\end{align*}
Under the conditions {$n \Lambda_D = o_P(T^{1/2}), T^{3/4}(\log n)^{3/4}/(n\Lambda_S ) = o_P(1)$} the last line is strictly positive with probability tending to one unless $E^C = \emptyset$ with probability tending to one. Thus the proof of~\eqref{eq:bound1} is complete.\\

\textbf{Step 2: recovery of clusters with probability to one}


To simplify notation, assume that individual $1,...,N_1$ belongs to cluster 1, individual $N_1+1,...,N_1+N_2$ to cluster 2 and so on. Since all cluster can be handled by similar arguments we only consider the first cluster. Let $\hat \alpha_{(1)},...,\hat \alpha_{(L)}$ denote the distinct values of $\hat \alpha_1,...,\hat\alpha_{N_1}$, ordered in increasing order, and let $n_{1,k} := \#\{i: \hat \alpha_i = \hat \alpha_{(k)}\}$. Again, to simplify notation assume w.o.l.g. that $\hat \alpha_1 = ... = \hat\alpha_{n_{1,1}} = \hat \alpha_{(1)}$. To prove the result, we proceed in an iterative way. We will prove by contradiction that $L=1$, i.e. all estimators of individuals from cluster 1 take the same value. Assume that $L\geq 2$.

We will now prove by contradiction that $n_{1,1} > N_1/2$. Assume that $n_{1,1} < N_1/2$. Define $\tilde\alpha_i = \hat\alpha_{(2)}$ for $i=1,...,n_{1,1}$ and $\tilde\alpha_i = \hat \alpha_i$ for $i > n_{1,1}$. By~\eqref{eq:bound1} Lemma~\ref{lem:rhotau1} we find that
\begin{align*}
&\Big|\sum_i \sum_{t=1}^T \rho_\tau(Y_{it} - X_{it}^\top\hat\beta - \hat\alpha_i) - \rho_\tau(Y_{it} - X_{it}^\top\hat\beta - \tilde\alpha_i)\Big|
\\
\lesssim~& n_{1,1}(\hat\alpha_{(2)}-\hat\alpha_{(1)})\Big\{\Big| \sum_t \psi_{\tau}(\eps_{it,\hat\beta}^\tau) \Big| + O_P(T^{3/4}(\log n)^{1/4}) + O_P(T^{1/2}(\log n)^{1/2})\Big\}
\\
\lesssim~& n_{1,1}(\hat\alpha_{(2)}-\hat\alpha_{(1)}) O_P(T^{3/4}(\log n)^{1/4})
\end{align*}
Next, observe that by construction, under (C) and using the fact that $\sup_i |\hat \alpha_i - \alpha_{i0}| = o_P(1)$, 
\begin{align*}
|\tilde\alpha_i - \tilde\alpha_j| - |\hat\alpha_i - \hat \alpha_j| & = 
-|\hat\alpha_{(2)} - \hat\alpha_{(1)}|, & 1 \leq i \leq n_{1,1}, n_{1,1} < j \leq N_1
\\
&& \mbox{ or } 1 \leq j \leq n_{1,1}, n_{1,1} < i \leq N_1, 
\\
\Big| |\tilde\alpha_i - \tilde\alpha_j| - |\hat\alpha_i - \hat \alpha_j|\Big| &\leq |\hat\alpha_{(2)} - \hat\alpha_{(1)}|, & 1 \leq i \leq n_{1,1}, N_1 < j\mbox{ or } 1 \leq j \leq n_{1,1}, N_1 < i,
\\
|\tilde\alpha_i - \tilde\alpha_j| - |\hat\alpha_i - \hat \alpha_j| & = 0, & \mbox{else}. 
\end{align*}
From this we obtain 
\begin{align*}
&\Theta(\tilde\alpha_1,...,\tilde\alpha_n,\hat\beta) - \Theta(\hat\alpha_1,...,\hat\alpha_n,\hat\beta)
\\
\lesssim~& n_{1,1}(\hat\alpha_{(2)}-\hat\alpha_{(1)}) O_P(T^{3/4}(\log n)^{1/4}) - (\hat\alpha_{(2)}-\hat\alpha_{(1)}) \Lambda_S n_{1,1}(N_1-n_{1,1})
\\
& 
+ (\hat\alpha_{(2)}-\hat\alpha_{(1)})n_{1,1}\Lambda_D O_P(n)
\\
< 0 
\end{align*}
where the last inequality holds for sufficiently large $n,T$ since by assumption {$\Lambda_D/ \Lambda_S = o_P(1), n\Lambda_S \gg T^{3/4}(\log n)^{1/4}$} and since we assumed $n_{1,1} < N_1/2$ so that $N_1-n_{1,1} \geq N_1/2 \gtrsim n$. However, this is a contradiction to the fact that $\hat\alpha_1,...,\hat\alpha_n,\hat\beta$ minimizes $\Theta$.

In a similar fashion, one can prove that $n_{1,L} > N_1/2$. Just define $\tilde\alpha_{N_1},...,\tilde\alpha_{N_1-n_{1,L}+1} = \hat\alpha_{(L-1)}$ and proceed as above. Since $n_{1,L} + n_{1,1} \leq N_1$ and we have already proved that $n_{1,1} > N_1/2$ this leads to a contradiction with $L\geq 2$, and hence $L=1$. All other clusters can be handled in a similar fashion and that completes the proof of the second step. \hfill $\Box$


\bigskip

\textbf{Proof of Lemma~\ref{lem:rhotau1}} 
Apply Knight's identity~\eqref{eq:knight} to find that
\begin{align*}
&\sum_{t=1}^T \rho_\tau(\eps_{it,\beta}^\tau - \delta) - \rho_\tau(\eps_{it,\beta}^\tau)
\\
=~& -\delta \sum_t \psi_{\tau}(\eps_{it,\beta}^\tau) + \sum_t \int_0^\delta \E\Big[ \1\{\eps_{it,\beta}^\tau\leq s\} - \1\{\eps_{it,\beta}^\tau\leq 0\}\Big] ds
\\
&\quad +  \int_0^\delta \sum_t \Big\{ \1\{\eps_{it,\beta}^\tau\leq s\} - \1\{\eps_{it,\beta}^\tau\leq 0\} - \E\Big[\1\{\eps_{it,\beta}^\tau\leq s\} - \1\{\eps_{it,\beta}^\tau\leq 0\} \Big]\Big\} ds.
\end{align*}
Hence it follows that
\begin{align*}
&\sum_{t=1}^T \rho_\tau(\eps_{it,\beta}^\tau - a_1) - \rho_\tau(\eps_{it,\beta}^\tau - a_2)
\\
=~& (a_2-a_1) \sum_t \psi_{\tau}(\eps_{it,\beta}^\tau) +  \int_{a_2}^{a_1} \sum_t \E\Big[\1\{\eps_{it,\beta}^\tau\leq s\} - \1\{\eps_{it,\beta}^\tau\leq 0\}\Big] ds
\\
&\quad +  \int_{a_2}^{a_1} \sum_t \Big\{ \1\{\eps_{it,\beta}^\tau\leq s\} - \1\{\eps_{it,\beta}^\tau\leq 0\} - \E\Big[\1\{\eps_{it,\beta}^\tau\leq s\} - \1\{\eps_{it,\beta}^\tau\leq 0\} \Big]\Big\} ds
\\
=:~& (a_2-a_1) \sum_{t=1}^T \psi_\tau(\eps_{it,\beta}^\tau) 
+ \tilde r_{n,i}^{(1)}(a_1,a_2) + \tilde r_{n,i}^{(2)}(a_1,a_2).
\end{align*}
Now by a Taylor expansion
\begin{multline*}
\sup_i \Big| \sum_t \int_{a_2}^{a_1} \E\Big[\1\{\eps_{it,\beta}^\tau\leq s\} - \1\{\eps_{it,\beta}^\tau\leq 0\}\Big] ds \Big| 
\\
= \sup_i \Big|\int_{a_2}^{a_1} \sum_t \E[F_{Y_{i1}|X_{i1}}(\beta^{\top}X_{it} + s|X_{it}) - F_{Y_{i1}|X_{i1}}(\beta^{\top}X_{it} |X_{it})] ds \Big| 
\lesssim 
T|a_1-a_2| \max(|a_1|,|a_2|),
\end{multline*}
so the bound on $\tilde r_{n,i}^{(1)}(a_1,a_2)$ is established. Next define the classes of functions 
\begin{align*}
\Gc_1 &:= \Big\{ (y,x) \mapsto \1\{y - \beta^\top x \leq s\} - \1\{y - \beta^\top x\leq 0\} ~\Big|~ s \in \R, \beta\in \R^d \leq B\Big\},
\\
\Gc_2 &:= \Big\{ (y,x) \mapsto \1\{y - \beta^\top x \leq s\} ~\Big|~ s \in \R, \beta\in \R^d\Big\}.
\end{align*}
Note that the class of functions $\Gc_2$ has envelope function $F \equiv 1$. Thus by Lemma~2.6.15 and Theorem~2.6.7 of~\cite{vdVW} the class of functions $\Gc_2$ satisfies, for any probability measure $Q$, $N(\eps,\Gc_2,L_2(Q)) \leq K(1/\eps)^V$ for some finite constants $K,V$(here, $N(\eps,\Gc_2,L_2(Q))$ denotes the covering number, see Section 2.1 of~\cite{vdVW}). Moreover, $\Gc_1 \subseteq \{g_1-g_2|g_1,g_2\in \Gc_2\}$, and elementary computations with covering numbers show that $N(\eps,\Gc_1,L_2(Q)) \leq \tilde K(1/\eps)^{\tilde V}$ for some finite constants $\tilde V, \tilde K$. Hence we find that by Theorem~2.14.9 of~\cite{vdVW}), for any $h > 0$,
\[
P^* \Big( \sup_{g \in \Gc_1}  \frac{1}{\sqrt{T}} \Big|\sum_t g(Y_{it},X_{it}) - \E[g(Y_{it},X_{it})]\Big| \geq h \Big) \leq \Big(\frac{Dh}{\sqrt{\tilde V}}\Big)^{\tilde V} e^{-2h^2}
\]  
for some constant $D$ that depends only on $\tilde K$ (here, $P^*$ denotes outer probability). Letting $h = \sqrt{\log n}$ and applying the union bound for probabilities we obtain
\begin{align*}
&\sup_i \sup_{\beta\in\R^d, s\in\R} \Big| \sum_t \Big\{ \1\{\eps_{it,\beta}^\tau\leq s\} - \1\{\eps_{it,\beta}^\tau\leq 0\} - \E\Big[ \1\{\eps_{it,\beta}^\tau\leq s\} - \1\{\eps_{it,\beta}^\tau\leq 0\} \Big] \Big|
\\
=~&O_P(T^{1/2} (\log n)^{1/2}).
\end{align*}
Hence 
\begin{align*}
&\sup_i \sup_{\beta\in \R^p} \Big|\int_{a_2}^{a_1} \sum_t \Big\{ \1\{\eps_{it,\beta}^\tau\leq s\} - \1\{\eps_{it,\beta}^\tau\leq 0\} - \E\Big[\1\{\eps_{it,\beta}^\tau\leq s\} - \1\{\eps_{it,\beta}^\tau\leq 0\} \Big] ds \Big| 
\\
=~&
O_P(T^{1/2}(\log n)^{1/2})|a_1-a_2|. 
\end{align*}
Thus the bound on $\tilde r_{n,i}^{(2)}(a_1,a_2)$ follows and the proof is complete. 
\hfill $\Box$

\bigskip
\bigskip

\textbf{Proof of Lemma~\ref{lem:rhoulb}}
Observe that by Knight's identity~\eqref{eq:knight}
\begin{align*}
&\E\Big[\rho_{\tau}(Y_{it} - Z_{it}^\top \gamma) - \rho_{\tau}(Y_{it} - Z_{it}^\top \gamma_{0i})\Big] 
\\
=~& \E\Big[\rho_{\tau}(Y_{it} - Z_{it}^\top \gamma_{0i} - Z_{it}^\top (\gamma - \gamma_{0i})) - \rho_{\tau}(Y_{it} - Z_{it}^\top \gamma_{0i})\Big]  
\\
=~& \E\Big[-(\gamma - \gamma_{0i})^\top Z_{it}\psi_\tau(\eps_{it}) + \int_0^{(\gamma - \gamma_{0i})^\top Z_{it}} \1\{\eps_{it} \leq s\} - \1\{\eps_{it} \leq 0\} ds\Big] 
\\
=~& \E\Big[\int_0^{(\gamma - \gamma_{0i})^\top Z_{it}} F_{\eps_{it}|X_{it}}(s|X_{it}) - F_{\eps_{it}|X_{it}}(0|X_{it}) ds\Big]. 
\end{align*}
Now under assumption (A2) $|F_{\eps_{it}|X_{it}}(s|X_{it}) - F_{\eps_{it}|X_{it}}(0|X_{it})| \leq s \overline{f'}$ a.s., and thus given (A1)
\begin{align*}
\E\Big|\int_0^{(\gamma - \gamma_{0i})^\top Z_{it}} F_{\eps_{it}|X_{it}}(s|X_{it}) - F_{\eps_{it}|X_{it}}(0|X_{it}) ds\Big| \leq \frac{\overline{f'}}{2} \E\Big[ ((\gamma - \gamma_{0i})^\top Z_{it})^2 \Big] \leq \frac{M^2 \overline{f'}}{2} \|\gamma - \gamma_{0i}\|^2.
\end{align*}
This shows the upper bound in~\eqref{eq:br}. For the lower bound, note that $s \mapsto F_{\eps_{it}|X_{it}}(s|X_{it})$ is non-decreasing almost surely. Moreover, $f_{\eps_{it}|X_{it}}(0|X_{it}) \geq f_{min}$ a.s. by (A3) and thus by (A2) and (A3) we have almost surely
\[
\inf_{|s| \leq f_{min}/2\overline{f'}} f_{\eps_{it}|X_{it}}(s|X_{it}) \geq \frac{f_{min}}{2}.
\]
Define $\delta_i := (\gamma - \gamma_{0i})\min\{1,f_{min}/(2M\overline{f'}\|\gamma - \gamma_{0i}\|)\}$. Noting that $s \mapsto F_{\eps_{it}|X_{it}}(s|X_{it})$ is non-decreasing almost surely, it follows that a.s.
\begin{multline*}
\int_0^{(\gamma - \gamma_{0i})^\top Z_{it}} F_{\eps_{it}|X_{it}}(s|X_{it}) - F_{\eps_{it}|X_{it}}(0|X_{it}) ds 
\\
\geq \int_0^{\delta_i^\top Z_{it}} F_{\eps_{it}|X_{it}}(s|X_{it}) - F_{\eps_{it}|X_{it}}(0|X_{it}) ds 
\geq \frac{f_{min}}{4} (\delta_i^\top Z_{it})^2 
\end{multline*}
where the last inequality follows since by definition $|\delta_i^\top Z_{it}| \leq f_{min}/(2\overline{f'})$ a.s. Finally, under assumption (A1), $\E[(\delta_i^\top Z_{it})^2] \geq \|\delta_i\|^2 c_\lambda$. Summarizing, we find
\[
\E\Big[\rho_{\tau}(Y_{it} - Z_{it}^\top \gamma) - \rho_{\tau}(Y_{it} - Z_{it}^\top \gamma_{0i})\Big]  \geq \frac{f_{min}c_\lambda}{4}\|\delta_i\|^2 = \frac{f_{min}c_\lambda}{4}\Big(\|\gamma - \gamma_{0i}\| \wedge \frac{f_{min}}{2M\overline{f'}} \Big)^2
\]
which proves the lower bound in~\eqref{eq:br}. Thus the proof of the Lemma is complete. \hfill $\Box$   

\bigskip
\bigskip

\textbf{Proof of Lemma~\ref{lem:sn1}} Consider the class of functions
\[
\Gc_B := \Big\{ (y,z) \mapsto g_{\gamma}(y,z) := \frac{(\rho_\tau(y - z^\top \gamma) - \rho_\tau(y))\1\{|z| \leq M\} + MB}{2MB} ~\Big|~ \|\gamma\| \leq B \Big\}.
\]
Note that by construction $0 \leq g_{\gamma}(y,z) \leq 1$ for all $\|\gamma\| \leq B$ and moreover $\sup_{y,z} |g_{\gamma}(y,z) - g_{\gamma'}(y,z)| \leq \|\gamma-\gamma'\|/(2B)$. This shows the existence of constants $V,K_B < \infty$ such that for all $i=1,...,n$ $N_{[~]}(\eps,\Gc_B,L_1(P_i)) \leq (K_B/\eps)^V$ for $0<\eps<K_B$ where $K_B$ depends on $B$ only and $P_i$ denotes the measure corresponding to $(Y_{i1},Z_{i1})$. Thus we have by Theorem 2.14.9 of~\cite{vdVW},
\[
P^* \Big( \sup_\gamma  \frac{1}{\sqrt{T}} \Big|\sum_t g_\gamma(Y_{it},Z_{it}) - \E[g_\gamma(Y_{it},Z_{it})]\Big| \geq h \Big) \leq \Big(\frac{D_Bh}{\sqrt{V}}\Big)^V e^{-2h^2}
\]  
where the constant $D_B$ depends only on $K_B$ and $P^*$ denotes outer probability. Set $h = \sqrt{\log n}$ to bound the right-hand side above by $o(n^{-1})$. Defining the events
\[
E_{i,n} := \Big\{ \sup_\gamma  \frac{1}{\sqrt{T}} \Big|\sum_t g_\gamma(Y_{it},Z_{it}) - \E[g_\gamma(Y_{it},Z_{it})]\Big| \geq \sqrt{\log n} \Big\}
\]
we obtain
\[
P^*(\cup_i E_{i,n}) \leq n \sup_i P^*(E_{i,n}) \leq n o(n^{-1}) = o(1).
\] 
Finally, note that under (A1) we have a.s. 
\[
\frac{\rho_\tau(Y_{it} - Z_{it}^\top \gamma) - \rho_\tau(Y_{it}) - \E[\rho_\tau(Y_{it} - Z_{it}^\top \gamma) - \rho_\tau(Y_{it})]}{2MB} = g_\gamma(Y_{it},Z_{it}) - \E[g_\gamma(Y_{it},Z_{it})] \quad \forall i,t.
\]
This completes the proof.
\hfill $\Box$

\bigskip

\subsection{Proof of Theorem~\ref{th:ic}}

We begin by stating a useful technical result that will be proved at the end of this section.

\begin{lemma}\label{lem:rhotau2}
	Under assumptions (A1)-(A3) 
	\begin{align*}
	&\sum_{t=1}^T \rho_\tau(Y_{it} - Z_{it}^\top(\gamma_0 + \delta)) - \rho_\tau(Y_{it} - Z_{it}^\top\gamma_0)
	\\
	& = \delta^\top \sum_{t=1}^T Z_{it}\psi_\tau(\eps_{it}) + \frac{1}{2}T \delta^\top\E[Z_{it}Z_{it}^\top f_{\eps_{i1}|X_{i1}}(0|X_{it})]\delta + r_{n,i}^{(1)}(\delta) + r_{n,i}^{(2)}(\delta)  
	\end{align*}
	where, defining $\ell_{n,T} := \max\{\log n, \log T\}$, there exists a constant $C_2$ independent of $n,T, \delta$ such that 
	\[
	\sup_i \sup_{T^{-1}\ell_{n,T}^2 \leq \|\delta\| \leq 1}\frac{|r_{n,i}^{(1)}(\delta)|}{\|\delta\|^{3/2}} = O_P(T^{1/2}\ell_{n,T}^{1/2}), \quad \sup_i |r_{n,i}^{(2)}(\delta)| \leq TC_2\|\delta\|^3. 
	\]
\end{lemma}

\textbf{Proof of Theorem~\ref{th:ic}}
The proof proceeds in several steps. First, we note that the 'oracle' estimation problem~\eqref{eq:or} corresponds to a classical, fixed-dimensional quantile regression with true parameter vector $(\alpha_{(01)},...,\alpha_{(0K)},\beta_0^\top)$ and $nT$ independent observations $(Y_{it},\tilde Z_{it})$ where $\tilde Z_{it}^\top = (e_k^\top,X_{it}^\top), i \in I_k, t=1,...,T$ where $e_k$ denotes the k'th unit vector in $\R^K$. A straightforward extension of classical proof techniques in parametric quantile regression shows that under assumptions (A1)-(A3) and (C) the oracle estimator is asymptotically normal as claimed. 

Second, we observe that by definition of the optimization problem the estimated group structure $\hat I_{1,\ell},...,\hat I_{K_\ell,\ell}$ is the same for all values of $\ell$ with $\lambda_\ell$ that give rise to the same number of groups. Since the value of $IC(\ell)$ depends only on $\hat I_{1,\ell},...,\hat I_{K_\ell,\ell}$, it suffices to minimize $IC$ over those values of $\ell$ that correspond to different numbers of groups. Denote the distinct estimated numbers of groups by $\hat K_1,...,\hat K_R$, the corresponding estimated groupings by $\hat I_{(1 \hat K_r)},...,\hat I_{(\hat K_r \hat K_r)}$, and the corresponding values of $IC$ by $IC_{\hat K_1},...,IC_{\hat K_R}$. By assumption (G) and Theorem~\ref{th:cluster}, the probability of the event
\begin{equation} \label{eq:prob1}
P\Big(\exists r: \hat K_r = K, \hat I_{(k \hat K_r)} = I_k, k=1,...,K\Big) \to 1.
\end{equation} 
Hence it suffices to prove that
\begin{equation} \label{eq:oric}
P\Big( \argmin_r IC_{K_r} = K \Big) \to 1.
\end{equation} 
Once this result is established, we directly obtain 
\[
P\Big((\hat\alpha_{1}^{IC},...,\hat\alpha_{\hat K^{IC}}^{IC},(\hat\beta^{IC})^\top) = (\hat\alpha_{(1)}^{(OR)},...,\hat\alpha_{(K)}^{(OR)},(\hat\beta^{(OR)})^\top)\Big) \to 1,
\]
and thus the asymptotic distribution of $(\hat\alpha_{1}^{IC},...,\hat\alpha_{\hat K^{IC}}^{IC},(\hat\beta^{IC})^\top)$ matches that of the oracle estimator. 

We will now prove~\eqref{eq:oric}. From Theorem 3.2 in \citeasnoun{Kato} we know that under (A1)-(A3) and the additional assumptions that $n \to \infty$ but $T$ grows at most polynomially in $n$ 
\[
\check\beta - \beta_0 = O_P((T/\log n)^{-3/4}\vee (nT)^{-1/2}).
\]
If $n \to \infty$ and $T$ grows at most polynomially in $n$ it follows that $\check\beta - \beta_0 = o_P(T^{-1/2})$. Moreover, standard quantile regression arguments show that
\[
\check\alpha_i - \alpha_{0i} = - \frac{1}{\E[f_{\eps_{it}^\tau|X_{it}}(0|X_{it})]} \frac{1}{T}\sum_t \psi_\tau(\eps_{it}^\tau) + R_{n,i}
\]
where $\sup_i |R_{n,i}| = O_p\Big (  \Big (\frac{\log T}{T}\Big )^{3/4}\Big )$. Next apply Lemma~\ref{lem:rhotau2} to find that provided $\frac{(\log T)^3 (\log n)^2}{T} \to 0$, 
\begin{align*}
&\sum_{i,t} \rho_\tau(Y_{it} - Z_{it}^\top \check\gamma_{i}) - \rho_\tau(\eps_{it}^\tau)
\\
=~& \sum_i (\check\gamma_i - \gamma_{0i})^\top \sum_t Z_{it}\psi_\tau(\eps_{it}^\tau) + \frac{T}{2} \sum_i (\check\gamma_i - \gamma_{0i})^\top \E[Z_{i1}Z_{i1}^\top f_{\eps_{i1}^\tau|X_{i1}}(0|X_{i1})](\check\gamma_i - \gamma_{0i}) + o_P(n)
\\
=~&\sum_i (\check\alpha_i - \alpha_{0i}) \sum_t \psi_\tau(\eps_{it}^\tau) + \frac{T}{2} \sum_i (\check\alpha_i - \alpha_{0i})^2 \E[f_{\eps_{i1}^\tau|X_{i1}}(0|X_{i1})] + o_P(n)
\\
=~& -\sum_i \frac{1}{2\E[f_{\eps_{i1}^\tau|X_{i1}}(0|X_{i1})]} \Big(\frac{1}{\sqrt{T}}\sum_t \psi_\tau(\eps_{it}^\tau)\Big)^2  + o_P(n)
\\
=~& -\sum_i \frac{\tau(1-\tau)}{2\E[f_{\eps_{i1}^\tau|X_{i1}}(0|X_{i1})]} + o_P(n).
\end{align*}
Next, observe that by asymptotic normality of the oracle estimator
\[
\sup_{k=1,...,K} \|\hat\gamma_{(k)}^{(OR)} - \gamma_{(0k)}\| = O_P((nT)^{-1/2})
\]
where we defined $\hat\gamma_{(k)}^{(OR)} := (\hat\alpha_{(k)}^{(OR)}, \hat\beta^{(OR)})$.
Again applying Lemma~\ref{lem:rhotau2} we obtain
\begin{align*}
&\sum_k \sum_{i \in I_k}\sum_t \rho_\tau(Y_{it} - Z_{it}^\top \hat\gamma_{(k)}^{(OR)}) - \rho_\tau(\eps_{it}^\tau)
\\
=~& \sum_k (\hat\gamma_{(k)}^{(OR)} - \gamma_{(0k)})^\top \sum_{i \in I_k}\sum_t Z_{it}\psi_\tau(\eps_{it})
+ nTO_P\Big(\sup_k \|\tilde\gamma_k - \gamma_{(0k)}\|^2\Big)
+ o_P(n)
\\
=~& o_P(n).
\end{align*}
Combining the results obtained so far we have
\begin{equation}\label{eq:ic1}
\sum_k \sum_{i \in I_k}\sum_t \rho_\tau(Y_{it} - Z_{it}^\top \hat\gamma_{(k)}^{(OR)}) - \inf_{\alpha_1,...,\alpha_n,\beta} \sum_{i,t} \rho_\tau(Y_{it} - X_{it}^\top \beta - \alpha_i) \geq - \sum_i \frac{\tau(1-\tau)}{2\E[f_{\eps_{i1}|X_{i1}}(0|X_{i1})]} + o_P(n).
\end{equation}
Next, let $V_n(L)$ denote the set of all disjoint partitions of $\{1,...,n\}$ into $L$ subsets. Observe that by~\eqref{eq:br} we have under assumption (C)
\begin{align*}
&\inf_{L < K}\inf_{J_1,...,J_L \in V_N(L) }\inf_{\alpha_1,..,\alpha_L,\beta} \Big(\sum_{\ell=1}^L \sum_{i \in J_\ell} \sum_t \E[\rho_\tau(Y_{it} - \beta^\top X_{it}- \alpha_\ell)] - \sum_i\sum_t \E[\rho_\tau(\eps_{it}^\tau)]  \Big)
\\
\geq~& \frac{T c_0}{2} \min\{N_1,...,N_K\} (\eps^2 \wedge \eps_0^2).
\end{align*}
Finally, note that by Lemma~\ref{lem:sn1}
\begin{align*}
&\sup_{|\alpha_i|\leq B, \|\beta\|\leq B} \Big|\sum_{i,t} \Big( \rho_\tau(Y_{it} - X_{it}^\top \beta - \alpha) - \rho_\tau(\eps_{it}^\tau) - \E[\rho_\tau(Y_{it} - X_{it}^\top \beta - \alpha) - \rho_\tau(\eps_{it}^\tau)] \Big)\Big|
\\ 
\leq~& ns_{n,1} = O_P(nT^{1/2}(\log n)^{1/2}).
\end{align*}
Summarizing, we find that under (C)
\begin{align}
&\inf_{L < K}\inf_{J_1,...,J_L \in V_N(L) }\inf_{\alpha_1,..,\alpha_L,\beta} \Big(\sum_{\ell=1}^L \sum_{i \in I_\ell} \sum_t \rho_\tau(Y_{it} - \beta^\top X_{it}- \alpha_\ell) - \sum_i\sum_t \rho_\tau(\eps_{it}^\tau)  \Big) \nonumber
\\
\geq~& \frac{T c_0}{2} \min\{N_1,...,N_K\} (\eps^2 \wedge \eps_0^2) 
- O_P(nT^{1/2}(\log n)^{1/2}) \nonumber
\\
\gtrsim~& nT - O_P(nT^{1/2}(\log n)^{1/2}). \label{eq:ic2}
\end{align}
The final result follows from a combination of~\eqref{eq:prob1}, \eqref{eq:ic1} and~\eqref{eq:ic2}. First, observe that for $K_r > K$ we have by~\eqref{eq:prob1}, \eqref{eq:ic1} and the assumptions on $p_{n,T}, \hat C$, with probability tending to one 
\[
IC_{K_r} - \inf_s IC_{K_s} \gtrsim p_{n,T} - n + o_P(n) \gg 0.  
\] 
It follows that, with probability tending to one, $\argmin_\ell IC(\ell) \leq K$. Moreover, for $K_r < K$ we have by~\eqref{eq:ic2} and the assumptions on $p_{n,T}, \hat C$, with probability tending to one 
\[
IC_{K_r} - \inf_s IC_{K_s} \gtrsim - K p_{n,T} + nT - O_P(nT^{1/2}(\log n)^{1/2}) \gg 0.  
\]
Hence, with probability tending to one, $K \geq \argmin_r IC_{K_r} \geq K$ and thus~\eqref{eq:oric} follows. \hfill $\Box$

\bigskip
\bigskip

\textbf{Proof of Lemma~\ref{lem:rhotau2}} By Knight's identity~\eqref{eq:knight} we have
\begin{align*}
&\rho_\tau(Y_{it} - Z_{it}^\top(\gamma_{0i} + \delta)) - \rho_\tau(Y_{it} - Z_{it}^\top\gamma_{0i}) 
\\
=~& - \delta^\top Z_{it} \psi_\tau(\eps_{it}) + \int_0^{Z_{it}^\top \delta} F_{\eps_{i1}|X_{i1}}(s|X_{it}) - F_{\eps_{i1}|X_{i1}}(0|X_{it})ds
\\
& \quad + \int_0^{Z_{it}^\top \delta} \1\{\eps_{it} \leq s\} - \1\{\eps_{it} \leq 0\} - (F_{\eps_{i1}|X_{i1}}(s|X_{it}) - F_{\eps_{i1}|X_{i1}}(0|X_{it})) ds.
\end{align*}
Define
\begin{align*}
r_{n,i}^{(1)}(\delta) &:= \sum_t \int_0^{Z_{it}^\top \delta} \1\{\eps_{it} \leq s\} - \1\{\eps_{it} \leq 0\} - (F_{\eps_{i1}|X_{i1}}(s|X_{it}) - F_{\eps_{i1}|X_{i1}}(0|X_{it})) ds
\\
&\quad\quad\quad - \frac{T}{2}\delta^\top\E[Z_{i1}Z_{i1}^\top f_{\eps_{i1}|X_{i1}}(0|X_{i1})]  + \sum_t \frac{1}{2} f_{\eps_{i1}|X_{i1}}(0|X_{it}) (Z_{it}^\top \delta)^2,
\\
r_{n,i}^{(2)}(\delta) &:= \sum_t \Big\{\int_0^{Z_{it}^\top \delta} F_{\eps_{i1}|X_{i1}}(s|X_{it}) - F_{\eps_{i1}|X_{i1}}(0|X_{it})ds - \frac{1}{2} f_{\eps_{i1}|X_{i1}}(0|X_{it}) (Z_{it}^\top \delta)^2\Big\}.
\end{align*}
By a Taylor expansion we obtain 
\begin{align*}
&\Big|\int_0^{Z_{it}^\top \delta} F_{\eps_{i1}|X_{i1}}(s|X_{it}) - F_{\eps_{i1}|X_{i1}}(s|X_{it})ds - \frac{1}{2} f_{\eps_{i1}|X_{i1}}(0|X_{it}) (Z_{it}^\top \delta)^2\Big| 
\leq (Z_{it}^\top \delta)^3 \overline{f'} \leq M^3 \overline{f'} \|\delta\|^3,
\end{align*}
and thus the bound on $r_{n,2}^{(i)}$ is established. Next we note that
\begin{align*}
\E\Big[ \int_0^{Z_{it}^\top \delta} \1\{\eps_{it} \leq s\} - \1\{\eps_{it} \leq 0\} - (F_{\eps_{i1}|X_{i1}}(s|X_{it}) - F_{\eps_{i1}|X_{i1}}(0|X_{it})) ds \Big] = 0
\end{align*}
since the conditional expectation given $Z_{it}$ equals zero almost surely and moreover
\begin{align*}
|I_{it}(\delta)| := &\Big| \int_0^{Z_{it}^\top \delta} \1\{\eps_{it} \leq s\} - \1\{\eps_{it} \leq 0\} - (F_{\eps_{i1}|X_{i1}}(s|X_{it}) - F_{\eps_{i1}|X_{i1}}(0|X_{it})) ds \Big|
\\ 
\leq &M^2\|\delta\|^2 + M\|\delta\|\1\{|\eps_{it}|\leq M\|\delta\|\},
\\
|I_{it}(\delta) - I_{it}(\delta')| \leq& M \|\delta - \delta'\|.
\end{align*}
Note that in particular for $\|\delta\|\leq 1$ we have 
\[
|I_{it}(\delta)| \leq M(M+1)\|\delta\|, \quad \E[I_{it}^{2}(\delta)] \leq 2(M^4 + 4M^3 \overline{f'})\|\delta\|^3.
\]
Define $c_{1,M} := M(M+1), c_{2,M} := 2(M^4 + 4M^3 \overline{f'})$ and apply the Bernstein inequality to show that for any $1 \geq \|\delta\| \geq T^{-1}\ell_{n,T}^2, 0 < a < \infty$
\begin{align*}
P\Big(\Big|\sum_t I_{it}(\delta)\Big| > a \ell_{n,T}^{1/2}T^{1/2}\|\delta\|^{3/2}\Big)
&\leq 
2\exp\Big(-\frac{a^2 \ell_{n,T}T\|\delta\|^3/2}{T c_{2,M} \|\delta\|^3 + a c_{1,M}\ell_{n,T}^{1/2}T^{1/2}\|\delta\|^{5/2}/3} \Big)
\\
&= 2\exp\Big(-\frac{a^2 \ell_{n,T}/2}{c_{2,M} + a c_{1,M}\ell_{n,T}^{1/2}(T\|\delta\|)^{-1/2}/3} \Big).
\end{align*}
For $0 < a < 3 \ell_{n,T}^{1/2}c_{2,M}/c_{1,M}$ the last line above is bounded by $2 (n \vee T)^{-a^2/(4 c_{2,M})}$. Denote by $G_T$ a grid of values $\delta_1,...,\delta_{|G_T|}$ such that $T^{-1} \leq \|\delta_j\| \leq 1$ for all $j \in G_T$ and 
\[
\sup_{\ell_{n,T}^2T^{-1} \leq \|\delta\|\leq 1} \inf_{\tilde\delta\in G_T} \|\delta -\tilde\delta\| = o(T^{-2}).
\]
Note that it is possible to find such a $G_T$ with $|G_T| = O(T^{2(d+1)})$. It follows that
\begin{align*}
\sup_i \sup_{\ell_{n,T}^2T^{-1} \leq \|\delta\| \leq 1}\frac{\Big|\sum_t I_{it}(\delta)\Big|}{\|\delta\|^{3/2}} 
&\leq 
\sup_i \sup_{\delta\in G_T}\frac{\Big|\sum_t I_{it}(\delta)\Big|}{\|\delta\|^{3/2}} + T^{3/2} T M o(T^{-2})
\\
&= \sup_i \sup_{\delta\in G_T}\frac{\Big|\sum_t I_{it}(\delta)\Big|}{\|\delta\|^{3/2}} + o(T^{1/2}).
\end{align*}
Finally, note that for $0 < a < 3 \ell_{n,T}^{1/2}c_{2,M}/c_{1,M}$
\begin{multline*}
P\Big( \sup_i \sup_{\delta\in G_T}\frac{\Big|\sum_t I_{it}(\delta)\Big|}{\|\delta\|^{3/2}} > a \ell_{n,T}^{1/2}T^{1/2}\Big)
\\ 
\leq n |G_T| 2 (n \vee T)^{-a^2/(4 c_{2,M})} = O(n T^{2(d+1)}) (n \vee T)^{-a^2/(4 c_{2,M})}.
\end{multline*}
Since $\ell_{n,T} \to \infty$ we can pick $a$ such that the last line above is $o(1)$, and hence  
\[
\sup_i \sup_{\delta\in G_T}\frac{\Big|\sum_t I_{it}(\delta)\Big|}{\|\delta\|^{3/2}} = O_P(\ell_{n,T}^{1/2}T^{1/2}).
\]
Finally, observe that, denoting by $\|A\|_\infty$ the maximum norm of the entries of the matrix $A$, 
\begin{align*}
& \sup_i \Big|- \frac{T}{2}\delta^\top\E[Z_{i1}Z_{i1}^\top f_{\eps_{i1}|X_{i1}}(0|X_{i1})]  + \sum_t \frac{1}{2} f_{\eps_{i1}|X_{i1}}(0|X_{it}) (Z_{it}^\top \delta)^2 \Big|
\\
=~& \sup_i \Big|\frac{1}{2} \delta^\top \Big\{\sum_t Z_{i1}Z_{i1}^\top f_{\eps_{i1}|X_{i1}}(0|X_{i1}) - \E[Z_{i1}Z_{i1}^\top f_{\eps_{i1}|X_{i1}}(0|X_{i1})] \Big\} \delta \Big|
\\
\lesssim~& \|\delta\|^2 \sup_i \Big\| \sum_t Z_{i1}Z_{i1}^\top f_{\eps_{i1}|X_{i1}}(0|X_{i1}) - \E[Z_{i1}Z_{i1}^\top f_{\eps_{i1}|X_{i1}}(0|X_{i1})]\Big\|
\\
=~& \|\delta\|^2 O_P(\sqrt{T \log n})
\end{align*}
where the last line follows by a straightforward application of the Hoeffding inequality. Thus the proof of Lemma~\ref{lem:rhotau2} is complete. \hfill $\Box$

\begin{table}[h!]
	\captionsetup{font=scriptsize}
	\begin{center}
		\begin{tabular}{rrrrrrr|rrrrr}
			\hline\hline
			\multicolumn{2}{l}{}&\multicolumn{5}{c}{Normal error}&\multicolumn{5}{c}{$t_3$ error}\\\cmidrule{3-7}\cmidrule{8-12}
			\multicolumn{1}{c}{n}&\multicolumn{1}{c}{T}&\multicolumn{1}{c}{1}&\multicolumn{1}{c}{2}&\multicolumn{1}{c}{3}&\multicolumn{1}{c}{4}&\multicolumn{1}{c}{$\geq5$}&\multicolumn{1}{c}{1}&\multicolumn{1}{c}{2}&\multicolumn{1}{c}{3}&\multicolumn{1}{c}{4}&\multicolumn{1}{c}{$\geq5$}\tabularnewline
			\hline
			\multicolumn{12}{c}{\textbf{DGP1: Independence between $\alpha_i$ and $x_{it}$.}}\tabularnewline
			\multicolumn{12}{l}{Model 1: location shift model}\tabularnewline
			\hline
			$30$&$15$&$0$&$0.074$&$0.504$&$0.324$&$0.098$&$0$&$0.168$&$0.490$&$0.266$&$0.076$\tabularnewline
			$30$&$30$&$0$&$0.002$&$0.803$&$0.167$&$0.028$&$0$&$0.007$&$0.744$&$0.202$&$0.048$\tabularnewline
			$30$&$60$&$0$&$0.000$&$0.984$&$0.016$&$0.000$&$0$&$0.000$&$0.966$&$0.032$&$0.002$\tabularnewline
			$60$&$15$&$0$&$0.056$&$0.502$&$0.315$&$0.128$&$0$&$0.122$&$0.428$&$0.319$&$0.130$\tabularnewline
			$60$&$30$&$0$&$0.000$&$0.856$&$0.127$&$0.018$&$0$&$0.002$&$0.767$&$0.186$&$0.046$\tabularnewline
			$60$&$60$&$0$&$0.000$&$0.992$&$0.008$&$0.000$&$0$&$0.000$&$0.978$&$0.021$&$0.000$\tabularnewline
			$90$&$15$&$0$&$0.040$&$0.465$&$0.339$&$0.156$&$0$&$0.113$&$0.392$&$0.322$&$0.172$\tabularnewline
			$90$&$30$&$0$&$0.000$&$0.872$&$0.114$&$0.013$&$0$&$0.000$&$0.778$&$0.182$&$0.038$\tabularnewline
			$90$&$60$&$0$&$0.000$&$0.996$&$0.004$&$0.000$&$0$&$0.000$&$0.984$&$0.016$&$0.000$\tabularnewline
			\hline
			\multicolumn{12}{l}{Model 2: location-scale shift model}\tabularnewline
			\hline
			$30$&$15$&$0$&$0.059$&$0.512$&$0.336$&$0.092$&$0$&$0.158$&$0.496$&$0.266$&$0.081$\tabularnewline
			$30$&$30$&$0$&$0.000$&$0.814$&$0.159$&$0.026$&$0$&$0.006$&$0.759$&$0.198$&$0.037$\tabularnewline
			$30$&$60$&$0$&$0.000$&$0.982$&$0.017$&$0.000$&$0$&$0.000$&$0.964$&$0.034$&$0.002$\tabularnewline
			$60$&$15$&$0$&$0.038$&$0.520$&$0.324$&$0.118$&$0$&$0.112$&$0.437$&$0.330$&$0.121$\tabularnewline
			$60$&$30$&$0$&$0.000$&$0.857$&$0.126$&$0.016$&$0$&$0.002$&$0.776$&$0.182$&$0.038$\tabularnewline
			$60$&$60$&$0$&$0.000$&$0.994$&$0.006$&$0.000$&$0$&$0.000$&$0.980$&$0.020$&$0.000$\tabularnewline
			$90$&$15$&$0$&$0.032$&$0.506$&$0.318$&$0.144$&$0$&$0.108$&$0.372$&$0.328$&$0.192$\tabularnewline
			$90$&$30$&$0$&$0.000$&$0.876$&$0.110$&$0.013$&$0$&$0.001$&$0.794$&$0.170$&$0.034$\tabularnewline
			$90$&$60$&$0$&$0.000$&$0.992$&$0.008$&$0.000$&$0$&$0.000$&$0.979$&$0.020$&$0.000$\tabularnewline
			\hline
			\multicolumn{12}{c}{\textbf{DGP2: Correlation between $\alpha_i$ and $x_{it}$.}}\tabularnewline
			\multicolumn{12}{l}{Model 1: location shift model}\tabularnewline
			\hline
			$30$&$15$&$0$&$0.080$&$0.496$&$0.320$&$0.104$&$0$&$0.173$&$0.478$&$0.268$&$0.082$\tabularnewline
			$30$&$30$&$0$&$0.003$&$0.788$&$0.178$&$0.031$&$0$&$0.012$&$0.722$&$0.229$&$0.038$\tabularnewline
			$30$&$60$&$0$&$0.000$&$0.986$&$0.014$&$0.000$&$0$&$0.000$&$0.970$&$0.028$&$0.002$\tabularnewline
			$60$&$15$&$0$&$0.062$&$0.492$&$0.314$&$0.132$&$0$&$0.128$&$0.426$&$0.315$&$0.131$\tabularnewline
			$60$&$30$&$0$&$0.000$&$0.852$&$0.135$&$0.013$&$0$&$0.002$&$0.736$&$0.217$&$0.046$\tabularnewline
			$60$&$60$&$0$&$0.000$&$0.992$&$0.008$&$0.000$&$0$&$0.000$&$0.980$&$0.020$&$0.000$\tabularnewline
			$90$&$15$&$0$&$0.045$&$0.463$&$0.332$&$0.160$&$0$&$0.116$&$0.378$&$0.327$&$0.179$\tabularnewline
			$90$&$30$&$0$&$0.000$&$0.854$&$0.133$&$0.012$&$0$&$0.002$&$0.732$&$0.220$&$0.046$\tabularnewline
			$90$&$60$&$0$&$0.000$&$0.994$&$0.006$&$0.000$&$0$&$0.000$&$0.976$&$0.024$&$0.000$\tabularnewline
			\hline
			\multicolumn{12}{l}{Model 2: location-scale shift model}\tabularnewline
			\hline
			$30$&$15$&$0$&$0.138$&$0.453$&$0.330$&$0.078$&$0$&$0.228$&$0.456$&$0.249$&$0.066$\tabularnewline
			$30$&$30$&$0$&$0.008$&$0.724$&$0.224$&$0.044$&$0$&$0.040$&$0.666$&$0.251$&$0.042$\tabularnewline
			$30$&$60$&$0$&$0.000$&$0.972$&$0.025$&$0.003$&$0$&$0.000$&$0.923$&$0.068$&$0.008$\tabularnewline
			$60$&$15$&$0$&$0.099$&$0.442$&$0.312$&$0.147$&$0$&$0.176$&$0.367$&$0.313$&$0.144$\tabularnewline
			$60$&$30$&$0$&$0.001$&$0.748$&$0.210$&$0.042$&$0$&$0.016$&$0.631$&$0.272$&$0.081$\tabularnewline
			$60$&$60$&$0$&$0.000$&$0.976$&$0.024$&$0.000$&$0$&$0.000$&$0.960$&$0.036$&$0.003$\tabularnewline
			$90$&$15$&$0$&$0.098$&$0.389$&$0.324$&$0.189$&$0$&$0.145$&$0.330$&$0.316$&$0.210$\tabularnewline
			$90$&$30$&$0$&$0.001$&$0.756$&$0.207$&$0.036$&$0$&$0.016$&$0.612$&$0.288$&$0.085$\tabularnewline
			$90$&$60$&$0$&$0.000$&$0.978$&$0.022$&$0.000$&$0$&$0.000$&$0.950$&$0.050$&$0.000$\tabularnewline
			\hline\hline
	\end{tabular}\end{center}
	\caption{Frequency of estimated number of groups as $k = 1, \dots K$. We aggregate the frequency for $K \geq 5$ since the occurrence is not very often. True $K_0 = 3$. Results are based on 2000 simulation repetitions for quantile level $\tau = 0.5$. }
	\label{tab:estnum_0.5}
\end{table}

\begin{table}[h!]	
	\captionsetup{font=scriptsize}
	\begin{center}
		\begin{tabular}{rrrrrrr|rrrrr}
			\hline\hline
			\multicolumn{2}{l}{}&\multicolumn{5}{c}{Normal error}&\multicolumn{5}{c}{$t_3$ error}\\\cmidrule{3-7}\cmidrule{8-12}
			\multicolumn{1}{c}{n}&\multicolumn{1}{c}{T}&\multicolumn{1}{c}{1}&\multicolumn{1}{c}{2}&\multicolumn{1}{c}{3}&\multicolumn{1}{c}{4}&\multicolumn{1}{c}{$\geq5$}&\multicolumn{1}{c}{1}&\multicolumn{1}{c}{2}&\multicolumn{1}{c}{3}&\multicolumn{1}{c}{4}&\multicolumn{1}{c}{$\geq5$}\tabularnewline
			\hline
			\multicolumn{12}{c}{\textbf{DGP1: Independence between $\alpha_i$ and $x_{it}$.}}\tabularnewline
			\multicolumn{12}{l}{Model 1: location shift model}\tabularnewline
			\hline
			$30$&$15$&$0$&$0.159$&$0.484$&$0.284$&$0.072$&$0.002$&$0.330$&$0.408$&$0.210$&$0.048$\tabularnewline
			$30$&$30$&$0$&$0.011$&$0.734$&$0.207$&$0.048$&$0.000$&$0.200$&$0.560$&$0.207$&$0.033$\tabularnewline
			$30$&$60$&$0$&$0.000$&$0.966$&$0.032$&$0.002$&$0.000$&$0.011$&$0.866$&$0.112$&$0.012$\tabularnewline
			$60$&$15$&$0$&$0.129$&$0.398$&$0.325$&$0.148$&$0.000$&$0.188$&$0.337$&$0.306$&$0.168$\tabularnewline
			$60$&$30$&$0$&$0.002$&$0.760$&$0.198$&$0.040$&$0.000$&$0.105$&$0.472$&$0.328$&$0.094$\tabularnewline
			$60$&$60$&$0$&$0.000$&$0.970$&$0.029$&$0.000$&$0.000$&$0.000$&$0.861$&$0.128$&$0.012$\tabularnewline
			$90$&$15$&$0$&$0.110$&$0.364$&$0.331$&$0.196$&$0.001$&$0.142$&$0.314$&$0.293$&$0.251$\tabularnewline
			$90$&$30$&$0$&$0.000$&$0.768$&$0.194$&$0.038$&$0.000$&$0.072$&$0.450$&$0.341$&$0.136$\tabularnewline
			$90$&$60$&$0$&$0.000$&$0.966$&$0.034$&$0.000$&$0.000$&$0.000$&$0.852$&$0.136$&$0.012$\tabularnewline
			\hline
			\multicolumn{12}{l}{Model 2: location-scale shift model}\tabularnewline
			\hline
			$30$&$15$&$0$&$0.158$&$0.472$&$0.300$&$0.070$&$0.002$&$0.334$&$0.390$&$0.214$&$0.060$\tabularnewline
			$30$&$30$&$0$&$0.007$&$0.754$&$0.200$&$0.039$&$0.000$&$0.179$&$0.581$&$0.216$&$0.024$\tabularnewline
			$30$&$60$&$0$&$0.000$&$0.968$&$0.031$&$0.001$&$0.000$&$0.007$&$0.868$&$0.118$&$0.008$\tabularnewline
			$60$&$15$&$0$&$0.114$&$0.404$&$0.334$&$0.147$&$0.000$&$0.178$&$0.342$&$0.306$&$0.174$\tabularnewline
			$60$&$30$&$0$&$0.002$&$0.771$&$0.188$&$0.038$&$0.000$&$0.082$&$0.474$&$0.353$&$0.091$\tabularnewline
			$60$&$60$&$0$&$0.000$&$0.974$&$0.025$&$0.001$&$0.000$&$0.000$&$0.870$&$0.123$&$0.007$\tabularnewline
			$90$&$15$&$0$&$0.093$&$0.354$&$0.346$&$0.208$&$0.000$&$0.134$&$0.292$&$0.310$&$0.264$\tabularnewline
			$90$&$30$&$0$&$0.000$&$0.766$&$0.198$&$0.036$&$0.000$&$0.058$&$0.456$&$0.342$&$0.144$\tabularnewline
			$90$&$60$&$0$&$0.000$&$0.955$&$0.044$&$0.000$&$0.000$&$0.000$&$0.854$&$0.132$&$0.014$\tabularnewline
			\hline
			\multicolumn{12}{c}{\textbf{DGP2: Correlation between $\alpha_i$ and $x_{it}$.}}\tabularnewline
			\multicolumn{12}{l}{Model 1: location shift model}\tabularnewline
			\hline
			$30$&$15$&$0$&$0.172$&$0.494$&$0.265$&$0.070$&$0.006$&$0.327$&$0.416$&$0.200$&$0.052$\tabularnewline
			$30$&$30$&$0$&$0.014$&$0.716$&$0.222$&$0.050$&$0.000$&$0.214$&$0.552$&$0.208$&$0.027$\tabularnewline
			$30$&$60$&$0$&$0.000$&$0.954$&$0.044$&$0.002$&$0.000$&$0.014$&$0.844$&$0.132$&$0.009$\tabularnewline
			$60$&$15$&$0$&$0.137$&$0.414$&$0.308$&$0.141$&$0.001$&$0.178$&$0.342$&$0.304$&$0.175$\tabularnewline
			$60$&$30$&$0$&$0.003$&$0.758$&$0.202$&$0.037$&$0.000$&$0.109$&$0.464$&$0.328$&$0.099$\tabularnewline
			$60$&$60$&$0$&$0.000$&$0.974$&$0.026$&$0.000$&$0.000$&$0.000$&$0.869$&$0.118$&$0.013$\tabularnewline
			$90$&$15$&$0$&$0.110$&$0.349$&$0.336$&$0.205$&$0.002$&$0.139$&$0.320$&$0.294$&$0.246$\tabularnewline
			$90$&$30$&$0$&$0.000$&$0.768$&$0.198$&$0.034$&$0.000$&$0.074$&$0.447$&$0.349$&$0.130$\tabularnewline
			$90$&$60$&$0$&$0.000$&$0.964$&$0.034$&$0.002$&$0.000$&$0.000$&$0.836$&$0.149$&$0.015$\tabularnewline
			\hline
			\multicolumn{12}{l}{Model 2: location-scale shift model}\tabularnewline
			\hline
			$30$&$15$&$0$&$0.228$&$0.446$&$0.264$&$0.062$&$0.014$&$0.326$&$0.414$&$0.203$&$0.042$\tabularnewline
			$30$&$30$&$0$&$0.038$&$0.660$&$0.257$&$0.045$&$0.000$&$0.276$&$0.502$&$0.196$&$0.027$\tabularnewline
			$30$&$60$&$0$&$0.000$&$0.932$&$0.064$&$0.004$&$0.000$&$0.039$&$0.769$&$0.180$&$0.012$\tabularnewline
			$60$&$15$&$0$&$0.176$&$0.377$&$0.304$&$0.144$&$0.004$&$0.169$&$0.338$&$0.303$&$0.185$\tabularnewline
			$60$&$30$&$0$&$0.018$&$0.652$&$0.257$&$0.073$&$0.000$&$0.157$&$0.420$&$0.311$&$0.112$\tabularnewline
			$60$&$60$&$0$&$0.000$&$0.954$&$0.044$&$0.002$&$0.000$&$0.004$&$0.775$&$0.192$&$0.030$\tabularnewline
			$90$&$15$&$0$&$0.146$&$0.309$&$0.342$&$0.204$&$0.006$&$0.108$&$0.296$&$0.312$&$0.277$\tabularnewline
			$90$&$30$&$0$&$0.010$&$0.660$&$0.258$&$0.072$&$0.000$&$0.118$&$0.373$&$0.342$&$0.167$\tabularnewline
			$90$&$60$&$0$&$0.000$&$0.948$&$0.050$&$0.002$&$0.000$&$0.002$&$0.753$&$0.212$&$0.032$\tabularnewline
			\hline\hline
	\end{tabular}\end{center}
	\caption{Frequency of estimated number of groups as $k = 1, \dots K$. We aggregate the frequency for $K \geq 5$ since the occurrence is not very often. True $K_0 = 3$. Results are based on 2000 simulation repetitions for quantile level $\tau = 0.75$. }
	\label{tab:estnum_0.75}
\end{table}

\newgeometry{a4paper,left=1in,right=1in,top=1in,bottom=1in}

\begin{landscape}
	\begin{table}[h!]
		\captionsetup{font=scriptsize}
		\begin{center}
			\resizebox{\textwidth}{!}{	\begin{tabular}{rrrrr|rrr|rrr|rrr}
					\hline\hline
					\multicolumn{2}{l}{}&\multicolumn{6}{c}{Normal error}&\multicolumn{6}{c}{$t_3$ error}\\\cmidrule{3-8}\cmidrule{9-14}
					\multicolumn{2}{l}{}&\multicolumn{3}{c}{PQR-FEgroup}&\multicolumn{3}{c}{QRFE}&\multicolumn{3}{c}{PQR-FEgroup}&\multicolumn{3}{c}{QRFE}\\\cmidrule{3-5}\cmidrule{6-8}\cmidrule{9-11}\cmidrule{12-14}
					\multicolumn{1}{c}{n}&\multicolumn{1}{c}{T}&\multicolumn{1}{c}{Bias}&\multicolumn{1}{c}{RMSE}&\multicolumn{1}{c}{Coverage}&\multicolumn{1}{c}{Bias}&\multicolumn{1}{c}{RMSE}&\multicolumn{1}{c}{Coverage}&\multicolumn{1}{c}{Bias}&\multicolumn{1}{c}{RMSE}&\multicolumn{1}{c}{Coverage}&\multicolumn{1}{c}{Bias}&\multicolumn{1}{c}{RMSE}&\multicolumn{1}{c}{Coverage}\tabularnewline
					\hline
					\multicolumn{14}{c}{\textbf{DGP1: Independence between $\alpha_i$ and $x_{it}$.}}\tabularnewline
					\multicolumn{14}{l}{Model 1: location shift model}\tabularnewline
					\hline
					$30$&$15$&$-0.001$&$0.060$&$0.792$&$-0.001$&$0.062$&$0.924$&$ 0.001$&$0.069$&$0.797$&$ 0.000$&$0.069$&$0.931$\tabularnewline
					$30$&$30$&$ 0.001$&$0.036$&$0.900$&$ 0.001$&$0.042$&$0.906$&$-0.001$&$0.042$&$0.874$&$-0.001$&$0.047$&$0.908$\tabularnewline
					$30$&$60$&$ 0.000$&$0.022$&$0.942$&$ 0.000$&$0.030$&$0.932$&$ 0.000$&$0.025$&$0.933$&$ 0.000$&$0.033$&$0.934$\tabularnewline
					$60$&$15$&$ 0.000$&$0.040$&$0.832$&$ 0.001$&$0.042$&$0.798$&$ 0.001$&$0.046$&$0.819$&$ 0.001$&$0.048$&$0.780$\tabularnewline
					$60$&$30$&$ 0.001$&$0.024$&$0.902$&$ 0.000$&$0.030$&$0.942$&$-0.001$&$0.028$&$0.884$&$-0.001$&$0.033$&$0.942$\tabularnewline
					$60$&$60$&$ 0.001$&$0.016$&$0.936$&$ 0.001$&$0.022$&$0.936$&$ 0.000$&$0.017$&$0.940$&$ 0.000$&$0.023$&$0.942$\tabularnewline
					$90$&$15$&$-0.001$&$0.033$&$0.834$&$ 0.000$&$0.035$&$0.836$&$-0.001$&$0.037$&$0.824$&$-0.001$&$0.039$&$0.828$\tabularnewline
					$90$&$30$&$ 0.000$&$0.020$&$0.914$&$ 0.001$&$0.024$&$0.955$&$ 0.001$&$0.023$&$0.886$&$ 0.000$&$0.027$&$0.968$\tabularnewline
					$90$&$60$&$ 0.000$&$0.012$&$0.942$&$ 0.000$&$0.018$&$0.905$&$ 0.000$&$0.014$&$0.937$&$ 0.000$&$0.019$&$0.913$\tabularnewline
					\hline
					\multicolumn{14}{l}{Model 2: location-scale shift model}\tabularnewline
					\hline
					$30$&$15$&$ 0.000$&$0.059$&$0.803$&$-0.001$&$0.060$&$0.923$&$ 0.001$&$0.067$&$0.800$&$ 0.000$&$0.067$&$0.932$\tabularnewline
					$30$&$30$&$ 0.002$&$0.035$&$0.893$&$ 0.001$&$0.041$&$0.894$&$ 0.000$&$0.041$&$0.870$&$ 0.000$&$0.047$&$0.890$\tabularnewline
					$30$&$60$&$ 0.000$&$0.022$&$0.936$&$ 0.000$&$0.030$&$0.924$&$ 0.000$&$0.025$&$0.929$&$ 0.000$&$0.032$&$0.935$\tabularnewline
					$60$&$15$&$ 0.001$&$0.039$&$0.834$&$ 0.002$&$0.041$&$0.792$&$ 0.001$&$0.045$&$0.820$&$ 0.001$&$0.047$&$0.774$\tabularnewline
					$60$&$30$&$ 0.001$&$0.023$&$0.903$&$ 0.000$&$0.029$&$0.940$&$ 0.000$&$0.028$&$0.882$&$ 0.000$&$0.032$&$0.946$\tabularnewline
					$60$&$60$&$ 0.001$&$0.015$&$0.932$&$ 0.001$&$0.021$&$0.935$&$ 0.000$&$0.017$&$0.932$&$ 0.000$&$0.022$&$0.938$\tabularnewline
					$90$&$15$&$ 0.000$&$0.032$&$0.841$&$ 0.000$&$0.034$&$0.824$&$-0.001$&$0.037$&$0.818$&$-0.001$&$0.038$&$0.830$\tabularnewline
					$90$&$30$&$ 0.000$&$0.019$&$0.913$&$ 0.000$&$0.024$&$0.952$&$ 0.002$&$0.022$&$0.884$&$ 0.001$&$0.026$&$0.966$\tabularnewline
					$90$&$60$&$ 0.000$&$0.012$&$0.940$&$ 0.000$&$0.017$&$0.904$&$ 0.001$&$0.014$&$0.935$&$ 0.000$&$0.019$&$0.906$\tabularnewline
					\hline\hline
					\multicolumn{14}{c}{\textbf{DGP2: Correlation between $\alpha_i$ and $x_{it}$.}}\tabularnewline
					\multicolumn{14}{l}{Model 1: location shift model}\tabularnewline
					\hline
					$30$&$15$&$0.016$&$0.063$&$0.782$&$-0.001$&$0.062$&$0.924$&$0.023$&$0.073$&$0.762$&$ 0.000$&$0.069$&$0.931$\tabularnewline
					$30$&$30$&$0.007$&$0.037$&$0.889$&$ 0.001$&$0.042$&$0.906$&$0.007$&$0.044$&$0.868$&$-0.001$&$0.047$&$0.908$\tabularnewline
					$30$&$60$&$0.001$&$0.022$&$0.944$&$ 0.000$&$0.030$&$0.932$&$0.001$&$0.025$&$0.928$&$ 0.000$&$0.033$&$0.934$\tabularnewline
					$60$&$15$&$0.015$&$0.044$&$0.783$&$ 0.001$&$0.042$&$0.798$&$0.021$&$0.052$&$0.768$&$ 0.001$&$0.048$&$0.780$\tabularnewline
					$60$&$30$&$0.005$&$0.026$&$0.889$&$ 0.000$&$0.030$&$0.942$&$0.006$&$0.029$&$0.872$&$-0.001$&$0.033$&$0.942$\tabularnewline
					$60$&$60$&$0.001$&$0.016$&$0.934$&$ 0.001$&$0.022$&$0.936$&$0.001$&$0.017$&$0.938$&$ 0.000$&$0.023$&$0.942$\tabularnewline
					$90$&$15$&$0.014$&$0.037$&$0.774$&$ 0.000$&$0.035$&$0.836$&$0.019$&$0.043$&$0.756$&$-0.001$&$0.039$&$0.828$\tabularnewline
					$90$&$30$&$0.004$&$0.020$&$0.894$&$ 0.001$&$0.024$&$0.955$&$0.007$&$0.024$&$0.871$&$ 0.000$&$0.027$&$0.968$\tabularnewline
					$90$&$60$&$0.000$&$0.013$&$0.941$&$ 0.000$&$0.018$&$0.905$&$0.001$&$0.014$&$0.934$&$ 0.000$&$0.019$&$0.913$\tabularnewline
					\multicolumn{14}{l}{Model 2: location-scale shift model}\tabularnewline
					\hline
					$30$&$15$&$0.020$&$0.071$&$0.752$&$-0.001$&$0.066$&$0.926$&$0.027$&$0.080$&$0.751$&$ 0.000$&$0.074$&$0.933$\tabularnewline
					$30$&$30$&$0.010$&$0.042$&$0.861$&$ 0.001$&$0.045$&$0.898$&$0.011$&$0.051$&$0.828$&$-0.001$&$0.051$&$0.892$\tabularnewline
					$30$&$60$&$0.002$&$0.025$&$0.924$&$ 0.000$&$0.033$&$0.926$&$0.002$&$0.029$&$0.912$&$ 0.000$&$0.035$&$0.938$\tabularnewline
					$60$&$15$&$0.020$&$0.050$&$0.750$&$ 0.002$&$0.045$&$0.792$&$0.026$&$0.058$&$0.729$&$ 0.001$&$0.051$&$0.775$\tabularnewline
					$60$&$30$&$0.007$&$0.029$&$0.860$&$ 0.000$&$0.032$&$0.940$&$0.009$&$0.034$&$0.832$&$ 0.000$&$0.035$&$0.943$\tabularnewline
					$60$&$60$&$0.002$&$0.018$&$0.924$&$ 0.001$&$0.023$&$0.934$&$0.002$&$0.019$&$0.920$&$ 0.000$&$0.025$&$0.938$\tabularnewline
					$90$&$15$&$0.018$&$0.041$&$0.752$&$ 0.000$&$0.037$&$0.824$&$0.024$&$0.049$&$0.706$&$-0.001$&$0.042$&$0.831$\tabularnewline
					$90$&$30$&$0.006$&$0.024$&$0.861$&$ 0.001$&$0.026$&$0.954$&$0.009$&$0.028$&$0.831$&$ 0.001$&$0.029$&$0.966$\tabularnewline
					$90$&$60$&$0.001$&$0.014$&$0.932$&$ 0.000$&$0.019$&$0.905$&$0.002$&$0.016$&$0.923$&$ 0.000$&$0.020$&$0.906$\tabularnewline
					\hline\hline
					
		\end{tabular}}\end{center}
		\caption{Comparison of bias and root mean squared error of $\hat \beta(\tau)$ based on the group fixed effect quantile regression (PQR-FEgroup)and the fixed effect quantile regression estimator (QRFE). Results are based on 2000 simulation repetitions for quantile level $\tau = 0.5$. DGP1 assumes that $x_{it}$ is independent of the fixed effect $\alpha_i$. DGP2 assumes that $x_{it} = 0.5 \alpha_i + \gamma_i + v_{it}$. }
		\label{tab:betaest_0.5}
	\end{table}
\end{landscape}

\begin{landscape}
	\begin{table}[h]
		\captionsetup{font=scriptsize}
		\begin{center}
			\resizebox{\textwidth}{!}{	\begin{tabular}{rrrrr|rrr|rrr|rrr}
					\hline\hline
					\multicolumn{2}{l}{}&\multicolumn{6}{c}{Normal error}&\multicolumn{6}{c}{$t_3$ error}\\\cmidrule{3-8}\cmidrule{9-14}
					\multicolumn{2}{l}{}&\multicolumn{3}{c}{PQR-FEgroup}&\multicolumn{3}{c}{QRFE}&\multicolumn{3}{c}{PQR-FEgroup}&\multicolumn{3}{c}{QRFE}\\\cmidrule{3-5}\cmidrule{6-8}\cmidrule{9-11}\cmidrule{12-14}
					\multicolumn{1}{c}{n}&\multicolumn{1}{c}{T}&\multicolumn{1}{c}{Bias}&\multicolumn{1}{c}{RMSE}&\multicolumn{1}{c}{Coverage}&\multicolumn{1}{c}{Bias}&\multicolumn{1}{c}{RMSE}&\multicolumn{1}{c}{Coverage}&\multicolumn{1}{c}{Bias}&\multicolumn{1}{c}{RMSE}&\multicolumn{1}{c}{Coverage}&\multicolumn{1}{c}{Bias}&\multicolumn{1}{c}{RMSE}&\multicolumn{1}{c}{Coverage}\tabularnewline
					\hline
					\multicolumn{14}{c}{\textbf{DGP1: Independence between $\alpha_i$ and $x_{it}$.}}\tabularnewline
					\multicolumn{14}{l}{Model 1: location shift model: normal error}\tabularnewline
					\hline
					$30$&$15$&$-0.001$&$0.066$&$0.784$&$-0.001$&$0.066$&$0.921$&$ 0.003$&$0.130$&$0.812$&$ 0.000$&$0.082$&$0.944$\tabularnewline
					$30$&$30$&$ 0.000$&$0.041$&$0.856$&$-0.001$&$0.045$&$0.895$&$-0.001$&$0.057$&$0.816$&$-0.001$&$0.057$&$0.890$\tabularnewline
					$30$&$60$&$ 0.000$&$0.025$&$0.930$&$ 0.000$&$0.033$&$0.922$&$ 0.000$&$0.035$&$0.898$&$-0.001$&$0.040$&$0.938$\tabularnewline
					$60$&$15$&$ 0.001$&$0.046$&$0.790$&$ 0.000$&$0.047$&$0.787$&$ 0.001$&$0.073$&$0.784$&$-0.001$&$0.059$&$0.786$\tabularnewline
					$60$&$30$&$ 0.000$&$0.029$&$0.874$&$ 0.000$&$0.033$&$0.939$&$-0.001$&$0.038$&$0.846$&$ 0.000$&$0.039$&$0.945$\tabularnewline
					$60$&$60$&$ 0.000$&$0.017$&$0.932$&$ 0.000$&$0.023$&$0.890$&$ 0.000$&$0.023$&$0.923$&$ 0.001$&$0.028$&$0.902$\tabularnewline
					$90$&$15$&$-0.001$&$0.037$&$0.812$&$ 0.000$&$0.038$&$0.827$&$ 0.002$&$0.080$&$0.790$&$ 0.001$&$0.048$&$0.825$\tabularnewline
					$90$&$30$&$ 0.000$&$0.023$&$0.884$&$ 0.001$&$0.026$&$0.957$&$ 0.001$&$0.030$&$0.850$&$ 0.001$&$0.033$&$0.963$\tabularnewline
					$90$&$60$&$ 0.000$&$0.014$&$0.936$&$ 0.000$&$0.018$&$0.930$&$ 0.000$&$0.018$&$0.917$&$ 0.000$&$0.023$&$0.936$\tabularnewline
					\hline
					\multicolumn{14}{l}{Model 2: location-scale shift model: normal error}\tabularnewline
					\hline
					$30$&$15$&$-0.009$&$0.065$&$0.772$&$-0.007$&$0.064$&$0.924$&$-0.009$&$0.126$&$0.796$&$-0.009$&$0.081$&$0.942$\tabularnewline
					$30$&$30$&$-0.001$&$0.040$&$0.852$&$-0.003$&$0.045$&$0.886$&$-0.005$&$0.056$&$0.814$&$-0.005$&$0.056$&$0.886$\tabularnewline
					$30$&$60$&$ 0.000$&$0.025$&$0.928$&$-0.001$&$0.032$&$0.924$&$-0.001$&$0.034$&$0.896$&$-0.002$&$0.039$&$0.933$\tabularnewline
					$60$&$15$&$-0.006$&$0.045$&$0.792$&$-0.005$&$0.046$&$0.794$&$-0.012$&$0.057$&$0.773$&$-0.009$&$0.058$&$0.784$\tabularnewline
					$60$&$30$&$-0.001$&$0.027$&$0.884$&$-0.002$&$0.032$&$0.926$&$-0.005$&$0.037$&$0.842$&$-0.004$&$0.038$&$0.936$\tabularnewline
					$60$&$60$&$ 0.000$&$0.017$&$0.930$&$-0.001$&$0.023$&$0.884$&$-0.001$&$0.022$&$0.906$&$-0.001$&$0.028$&$0.892$\tabularnewline
					$90$&$15$&$-0.007$&$0.037$&$0.781$&$-0.005$&$0.037$&$0.831$&$-0.012$&$0.048$&$0.776$&$-0.007$&$0.047$&$0.824$\tabularnewline
					$90$&$30$&$-0.001$&$0.022$&$0.884$&$-0.002$&$0.026$&$0.946$&$-0.003$&$0.029$&$0.854$&$-0.003$&$0.032$&$0.958$\tabularnewline
					$90$&$60$&$ 0.000$&$0.014$&$0.936$&$-0.001$&$0.018$&$0.926$&$-0.001$&$0.018$&$0.918$&$-0.002$&$0.023$&$0.934$\tabularnewline
					\hline\hline
					\multicolumn{14}{c}{\textbf{DGP2: Correlation between $\alpha_i$ and $x_{it}$.}}\tabularnewline
					\multicolumn{14}{l}{Model 1: location shift model}\tabularnewline
					\hline
					$30$&$15$&$0.021$&$0.071$&$0.756$&$-0.001$&$0.066$&$0.921$&$0.044$&$0.170$&$0.757$&$ 0.000$&$0.082$&$0.944$\tabularnewline
					$30$&$30$&$0.008$&$0.042$&$0.853$&$-0.001$&$0.045$&$0.895$&$0.019$&$0.062$&$0.781$&$-0.001$&$0.057$&$0.890$\tabularnewline
					$30$&$60$&$0.001$&$0.026$&$0.925$&$ 0.000$&$0.033$&$0.922$&$0.005$&$0.036$&$0.883$&$-0.001$&$0.040$&$0.938$\tabularnewline
					$60$&$15$&$0.021$&$0.052$&$0.746$&$ 0.000$&$0.047$&$0.787$&$0.033$&$0.086$&$0.708$&$-0.001$&$0.059$&$0.786$\tabularnewline
					$60$&$30$&$0.006$&$0.030$&$0.854$&$ 0.000$&$0.033$&$0.939$&$0.015$&$0.043$&$0.788$&$ 0.000$&$0.039$&$0.945$\tabularnewline
					$60$&$60$&$0.001$&$0.018$&$0.934$&$ 0.000$&$0.023$&$0.890$&$0.003$&$0.023$&$0.909$&$ 0.001$&$0.028$&$0.902$\tabularnewline
					$90$&$15$&$0.019$&$0.043$&$0.726$&$ 0.000$&$0.038$&$0.827$&$0.035$&$0.092$&$0.673$&$ 0.001$&$0.048$&$0.825$\tabularnewline
					$90$&$30$&$0.006$&$0.024$&$0.868$&$ 0.001$&$0.026$&$0.957$&$0.015$&$0.036$&$0.787$&$ 0.001$&$0.033$&$0.963$\tabularnewline
					$90$&$60$&$0.001$&$0.014$&$0.935$&$ 0.000$&$0.018$&$0.930$&$0.003$&$0.019$&$0.912$&$ 0.000$&$0.023$&$0.936$\tabularnewline
					\hline
					\multicolumn{14}{l}{Model 2: location-scale shift model}\tabularnewline
					\hline
					$30$&$15$&$0.017$&$0.074$&$0.752$&$-0.007$&$0.071$&$0.924$&$0.052$&$0.228$&$0.760$&$-0.009$&$0.089$&$0.942$\tabularnewline
					$30$&$30$&$0.009$&$0.048$&$0.830$&$-0.003$&$0.049$&$0.889$&$0.019$&$0.067$&$0.778$&$-0.005$&$0.062$&$0.888$\tabularnewline
					$30$&$60$&$0.002$&$0.029$&$0.912$&$-0.001$&$0.035$&$0.922$&$0.007$&$0.041$&$0.871$&$-0.002$&$0.043$&$0.935$\tabularnewline
					$60$&$15$&$0.018$&$0.054$&$0.747$&$-0.005$&$0.051$&$0.794$&$0.031$&$0.135$&$0.739$&$-0.009$&$0.064$&$0.784$\tabularnewline
					$60$&$30$&$0.007$&$0.034$&$0.834$&$-0.002$&$0.035$&$0.928$&$0.015$&$0.047$&$0.785$&$-0.005$&$0.042$&$0.936$\tabularnewline
					$60$&$60$&$0.002$&$0.020$&$0.914$&$-0.001$&$0.025$&$0.886$&$0.004$&$0.027$&$0.874$&$-0.001$&$0.031$&$0.896$\tabularnewline
					$90$&$15$&$0.016$&$0.045$&$0.744$&$-0.005$&$0.041$&$0.830$&$0.034$&$0.145$&$0.718$&$-0.007$&$0.052$&$0.828$\tabularnewline
					$90$&$30$&$0.007$&$0.027$&$0.840$&$-0.002$&$0.028$&$0.946$&$0.015$&$0.039$&$0.762$&$-0.003$&$0.036$&$0.960$\tabularnewline
					$90$&$60$&$0.002$&$0.016$&$0.916$&$-0.001$&$0.020$&$0.926$&$0.003$&$0.021$&$0.896$&$-0.002$&$0.025$&$0.932$\tabularnewline
					\hline\hline
		\end{tabular}}\end{center}
		\caption{Comparison of bias and root mean squared error of $\hat \beta(\tau)$ based on the group fixed effect quantile regression (PQR-FEgroup) and the fixed effect quantile regression estimator (QRFE). Results are based on 2000 simulation repetitions for quantile level $\tau = 0.75$. DGP1 assumes that $x_{it}$ is independent of the fixed effect $\alpha_i$. DGP2 assumes that $x_{it} = 0.5 \alpha_i + \gamma_i + v_{it}$.  }
		\label{tab:betaest_0.75}
	\end{table}
\end{landscape}

\begin{table}[H]
	\captionsetup{font=scriptsize}
	\begin{center}
		\begin{tabular}{rrccc|ccc}
			\hline\hline
			\multicolumn{2}{l}{}&\multicolumn{3}{c}{Normal error}&\multicolumn{3}{c}{$t_3$ error}\\\cmidrule{3-5}\cmidrule{6-8}
			\multicolumn{1}{c}{n}&\multicolumn{1}{c}{T}&\multicolumn{1}{c}{Perfect Match}&\multicolumn{1}{c}{Avg Match}&\multicolumn{1}{c}{Std Error}&\multicolumn{1}{c}{Perfect Match}&\multicolumn{1}{c}{Avg Match}&\multicolumn{1}{c}{Std Error}\tabularnewline
			\hline
			\multicolumn{8}{c}{\textbf{DGP1: Independence between $\alpha_i$ and $x_{it}$.}}\tabularnewline
			\multicolumn{8}{l}{Model 1: location shift model}\tabularnewline
			\hline
			$30$&$15$&$0.025$&$0.659$&$0.285$&$0.008$&$0.636$&$0.267$\tabularnewline
			$30$&$30$&$0.384$&$0.877$&$0.222$&$0.223$&$0.820$&$0.262$\tabularnewline
			$30$&$60$&$0.918$&$0.989$&$0.069$&$0.822$&$0.976$&$0.105$\tabularnewline
			$60$&$15$&$0.003$&$0.651$&$0.307$&$0.000$&$0.590$&$0.297$\tabularnewline
			$60$&$30$&$0.225$&$0.900$&$0.202$&$0.088$&$0.841$&$0.256$\tabularnewline
			$60$&$60$&$0.884$&$0.992$&$0.064$&$0.735$&$0.985$&$0.077$\tabularnewline
			$90$&$15$&$0.000$&$0.646$&$0.317$&$0.000$&$0.562$&$0.320$\tabularnewline
			$90$&$30$&$0.119$&$0.912$&$0.187$&$0.024$&$0.839$&$0.270$\tabularnewline
			$90$&$60$&$0.856$&$0.995$&$0.044$&$0.654$&$0.985$&$0.080$\tabularnewline
			\hline
			\multicolumn{8}{l}{Model 2: location-scale shift model}\tabularnewline
			\hline
			$30$&$15$&$0.028$&$0.670$&$0.285$&$0.012$&$0.638$&$0.274$\tabularnewline
			$30$&$30$&$0.394$&$0.880$&$0.221$&$0.225$&$0.836$&$0.245$\tabularnewline
			$30$&$60$&$0.906$&$0.988$&$0.072$&$0.800$&$0.975$&$0.106$\tabularnewline
			$60$&$15$&$0.002$&$0.663$&$0.309$&$0.000$&$0.595$&$0.305$\tabularnewline
			$60$&$30$&$0.218$&$0.903$&$0.201$&$0.084$&$0.850$&$0.247$\tabularnewline
			$60$&$60$&$0.856$&$0.994$&$0.047$&$0.708$&$0.983$&$0.085$\tabularnewline
			$90$&$15$&$0.000$&$0.662$&$0.323$&$0.000$&$0.550$&$0.331$\tabularnewline
			$90$&$30$&$0.118$&$0.908$&$0.198$&$0.026$&$0.851$&$0.261$\tabularnewline
			$90$&$60$&$0.827$&$0.994$&$0.051$&$0.634$&$0.984$&$0.084$\tabularnewline
			\hline		
			\multicolumn{8}{c}{\textbf{DGP2: Correlation between $\alpha_i$ and $x_{it}$.}}\tabularnewline
			\multicolumn{8}{l}{Model 1: location shift model}\tabularnewline
			\hline
			$30$&$15$&$0.024$&$0.655$&$0.289$&$0.008$&$0.624$&$0.270$\tabularnewline
			$30$&$30$&$0.365$&$0.865$&$0.234$&$0.199$&$0.817$&$0.257$\tabularnewline
			$30$&$60$&$0.923$&$0.991$&$0.061$&$0.824$&$0.978$&$0.101$\tabularnewline
			$60$&$15$&$0.002$&$0.650$&$0.305$&$0.000$&$0.590$&$0.297$\tabularnewline
			$60$&$30$&$0.202$&$0.895$&$0.212$&$0.076$&$0.829$&$0.267$\tabularnewline
			$60$&$60$&$0.884$&$0.993$&$0.052$&$0.738$&$0.985$&$0.076$\tabularnewline
			$90$&$15$&$0.000$&$0.643$&$0.318$&$0.000$&$0.554$&$0.320$\tabularnewline
			$90$&$30$&$0.104$&$0.900$&$0.208$&$0.025$&$0.817$&$0.285$\tabularnewline
			$90$&$60$&$0.850$&$0.994$&$0.053$&$0.646$&$0.984$&$0.078$\tabularnewline
			\hline
			\multicolumn{8}{l}{Model 2: location-scale shift model}\tabularnewline
			\hline
			$30$&$15$&$0.006$&$0.636$&$0.263$&$0.002$&$0.608$&$0.248$\tabularnewline
			$30$&$30$&$0.215$&$0.826$&$0.250$&$0.119$&$0.780$&$0.260$\tabularnewline
			$30$&$60$&$0.818$&$0.979$&$0.096$&$0.649$&$0.949$&$0.146$\tabularnewline
			$60$&$15$&$0.000$&$0.612$&$0.293$&$0.000$&$0.545$&$0.280$\tabularnewline
			$60$&$30$&$0.083$&$0.837$&$0.254$&$0.026$&$0.764$&$0.294$\tabularnewline
			$60$&$60$&$0.715$&$0.981$&$0.091$&$0.516$&$0.968$&$0.117$\tabularnewline
			$90$&$15$&$0.000$&$0.585$&$0.312$&$0.000$&$0.514$&$0.296$\tabularnewline
			$90$&$30$&$0.023$&$0.832$&$0.268$&$0.006$&$0.744$&$0.322$\tabularnewline
			$90$&$60$&$0.648$&$0.983$&$0.085$&$0.398$&$0.962$&$0.133$\tabularnewline
			\hline\hline
	\end{tabular}\end{center}
	\caption{Membership estimation for $\tau = 0.5$ for two different error distributions: Perfect Match states the percentage of perfect membership estimation out of the 2000 repetitions. Average match reports the mean of the percentage of correct membership estimation and the standard error reports the associated standard deviation.}
	\label{tab:member_0.5}
\end{table}

\begin{table}[H]
	\captionsetup{font=scriptsize}
	\begin{center}
		\begin{tabular}{rrccc|ccc}
			\hline\hline
			\multicolumn{2}{l}{}&\multicolumn{3}{c}{Normal error}&\multicolumn{3}{c}{$t_3$ error}\\\cmidrule{3-5}\cmidrule{6-8}
			\multicolumn{1}{c}{n}&\multicolumn{1}{c}{T}&\multicolumn{1}{c}{Perfect Match}&\multicolumn{1}{c}{Avg Match}&\multicolumn{1}{c}{Std Error}&\multicolumn{1}{c}{Perfect Match}&\multicolumn{1}{c}{Avg Match}&\multicolumn{1}{c}{Std Error}\tabularnewline
			\hline
			\multicolumn{8}{c}{\textbf{DGP1: Independence between $\alpha_i$ and $x_{it}$.}}\tabularnewline
			\multicolumn{8}{l}{Model 1: location shift model}\tabularnewline
			\hline
			$30$&$15$&$0.012$&$0.666$&$0.240$&$0.000$&$0.574$&$0.181$\tabularnewline
			$30$&$30$&$0.242$&$0.849$&$0.220$&$0.034$&$0.737$&$0.228$\tabularnewline
			$30$&$60$&$0.846$&$0.982$&$0.082$&$0.471$&$0.923$&$0.160$\tabularnewline
			$60$&$15$&$0.000$&$0.634$&$0.258$&$0.000$&$0.558$&$0.197$\tabularnewline
			$60$&$30$&$0.098$&$0.873$&$0.198$&$0.002$&$0.732$&$0.248$\tabularnewline
			$60$&$60$&$0.754$&$0.984$&$0.072$&$0.260$&$0.934$&$0.142$\tabularnewline
			$90$&$15$&$0.000$&$0.633$&$0.267$&$0.000$&$0.530$&$0.210$\tabularnewline
			$90$&$30$&$0.041$&$0.880$&$0.201$&$0.000$&$0.734$&$0.253$\tabularnewline
			$90$&$60$&$0.690$&$0.984$&$0.071$&$0.182$&$0.939$&$0.132$\tabularnewline
			\hline
			\multicolumn{8}{l}{Model 2: location-scale shift model}\tabularnewline
			\hline
			$30$&$15$&$0.014$&$0.668$&$0.240$&$0.000$&$0.573$&$0.186$\tabularnewline
			$30$&$30$&$0.248$&$0.858$&$0.215$&$0.040$&$0.751$&$0.225$\tabularnewline
			$30$&$60$&$0.824$&$0.980$&$0.086$&$0.451$&$0.926$&$0.158$\tabularnewline
			$60$&$15$&$0.000$&$0.640$&$0.260$&$0.000$&$0.566$&$0.205$\tabularnewline
			$60$&$30$&$0.100$&$0.875$&$0.204$&$0.002$&$0.747$&$0.243$\tabularnewline
			$60$&$60$&$0.729$&$0.985$&$0.069$&$0.270$&$0.938$&$0.136$\tabularnewline
			$90$&$15$&$0.000$&$0.630$&$0.278$&$0.000$&$0.531$&$0.219$\tabularnewline
			$90$&$30$&$0.045$&$0.885$&$0.195$&$0.001$&$0.741$&$0.261$\tabularnewline
			$90$&$60$&$0.646$&$0.978$&$0.084$&$0.164$&$0.938$&$0.134$\tabularnewline
			\hline		
			\multicolumn{8}{c}{\textbf{DGP2: Correlation between $\alpha_i$ and $x_{it}$.}}\tabularnewline
			\multicolumn{8}{l}{Model 1: location shift model}\tabularnewline
			\hline
			$30$&$15$&$0.012$&$0.674$&$0.187$&$0.000$&$0.584$&$0.158$\tabularnewline
			$30$&$30$&$0.255$&$0.873$&$0.163$&$0.052$&$0.758$&$0.190$\tabularnewline
			$30$&$60$&$0.772$&$0.977$&$0.078$&$0.368$&$0.908$&$0.148$\tabularnewline
			$60$&$15$&$0.000$&$0.727$&$0.189$&$0.000$&$0.523$&$0.155$\tabularnewline
			$60$&$30$&$0.188$&$0.921$&$0.124$&$0.008$&$0.699$&$0.212$\tabularnewline
			$60$&$60$&$0.855$&$0.994$&$0.032$&$0.258$&$0.911$&$0.160$\tabularnewline
			$90$&$15$&$0.000$&$0.624$&$0.207$&$0.000$&$0.482$&$0.153$\tabularnewline
			$90$&$30$&$0.048$&$0.868$&$0.177$&$0.002$&$0.667$&$0.229$\tabularnewline
			$90$&$60$&$0.665$&$0.989$&$0.044$&$0.172$&$0.924$&$0.157$\tabularnewline
			\hline
			\multicolumn{8}{l}{Model 2: location-scale shift model}\tabularnewline
			\hline
			$30$&$15$&$0.002$&$0.618$&$0.161$&$0.000$&$0.538$&$0.141$\tabularnewline
			$30$&$30$&$0.072$&$0.776$&$0.183$&$0.008$&$0.665$&$0.171$\tabularnewline
			$30$&$60$&$0.468$&$0.945$&$0.103$&$0.140$&$0.819$&$0.187$\tabularnewline
			$60$&$15$&$0.000$&$0.556$&$0.162$&$0.000$&$0.476$&$0.134$\tabularnewline
			$60$&$30$&$0.000$&$0.761$&$0.188$&$0.000$&$0.591$&$0.179$\tabularnewline
			$60$&$60$&$0.355$&$0.954$&$0.094$&$0.055$&$0.825$&$0.198$\tabularnewline
			$90$&$15$&$0.000$&$0.527$&$0.172$&$0.000$&$0.428$&$0.113$\tabularnewline
			$90$&$30$&$0.002$&$0.748$&$0.212$&$0.000$&$0.554$&$0.191$\tabularnewline
			$90$&$60$&$0.240$&$0.953$&$0.103$&$0.010$&$0.805$&$0.211$\tabularnewline
			\hline\hline
	\end{tabular}\end{center}
	\caption{Membership estimation for $\tau = 0.75$ for two different error distributions: Perfect Match states the percentage of perfect membership estimation out of the 2000 repetitions. Average match reports the mean of the percentage of correct membership estimation and the standard error reports the associated standard deviation.}
	\label{tab:member_0.75}
\end{table}

\restoregeometry
\appendix
\section{Additional simulation results}

\subsection{More investigations on the tuning parameters in the IC criteria}
\begin{figure}[H]
	\captionsetup{font=scriptsize}
	\includegraphics[scale=0.75]{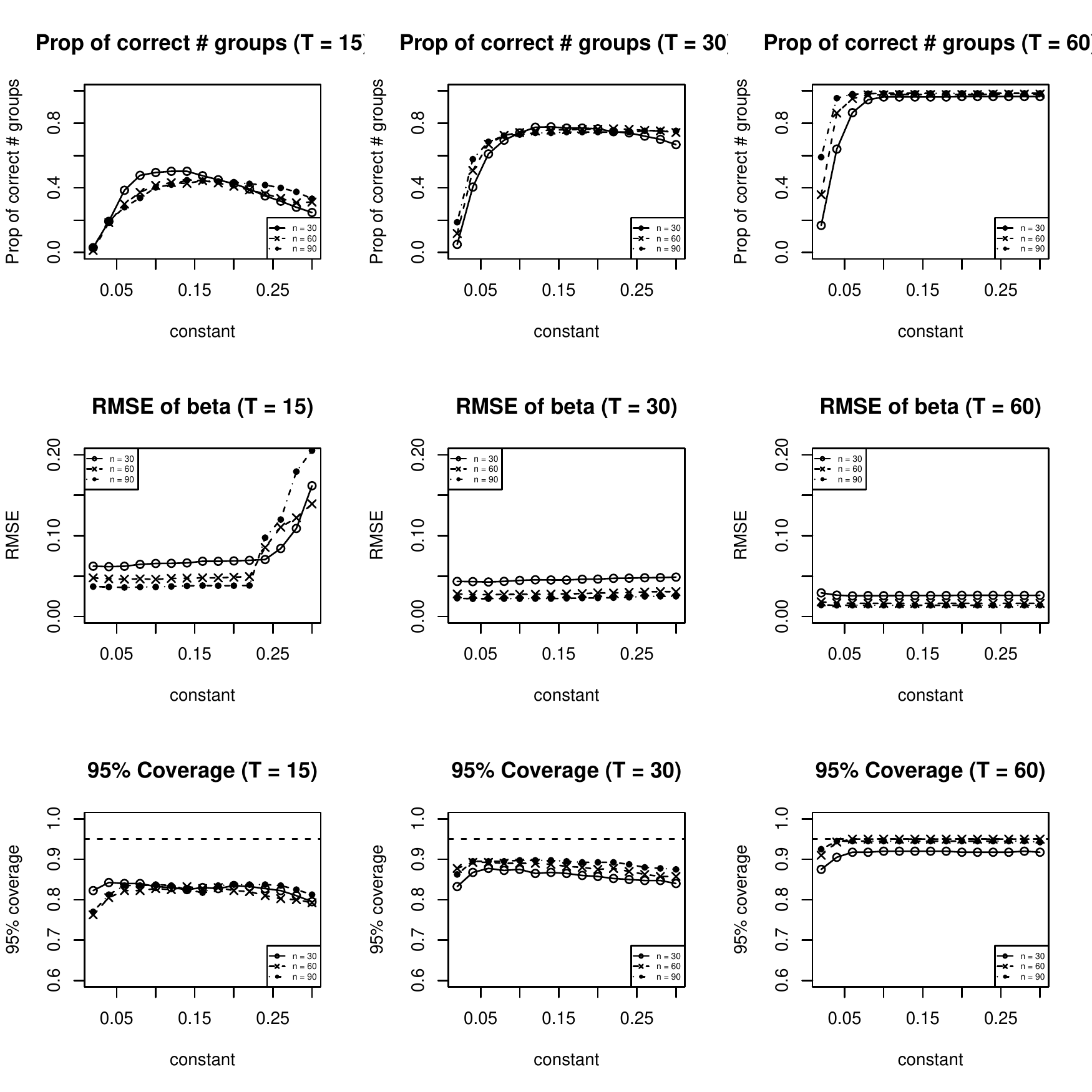}
	\caption{Use different constants in $p_{n,T}$ for the IC criteria for location shift model with $t_3$ error on DGP1: For a equally spaced grid on $[0.01, 0.3]$ with width 0.01, the three columns represent different magnitudes of $T$ while each figures in the row overlays the curves for $n \in \{30,60,90\}$ for various performance measures. The first row plots the proportion of correctly estimated number of groups. The second row plots the RMSE of $\hat \beta^{IC}(\tau)$ where $\tau =0.5$ and the third plots the coverage rate for nominal size 5\%. }
	\label{fig: CC1_ls}
\end{figure}

\begin{figure}[H]
	\captionsetup{font=scriptsize}
	\includegraphics[scale=0.75]{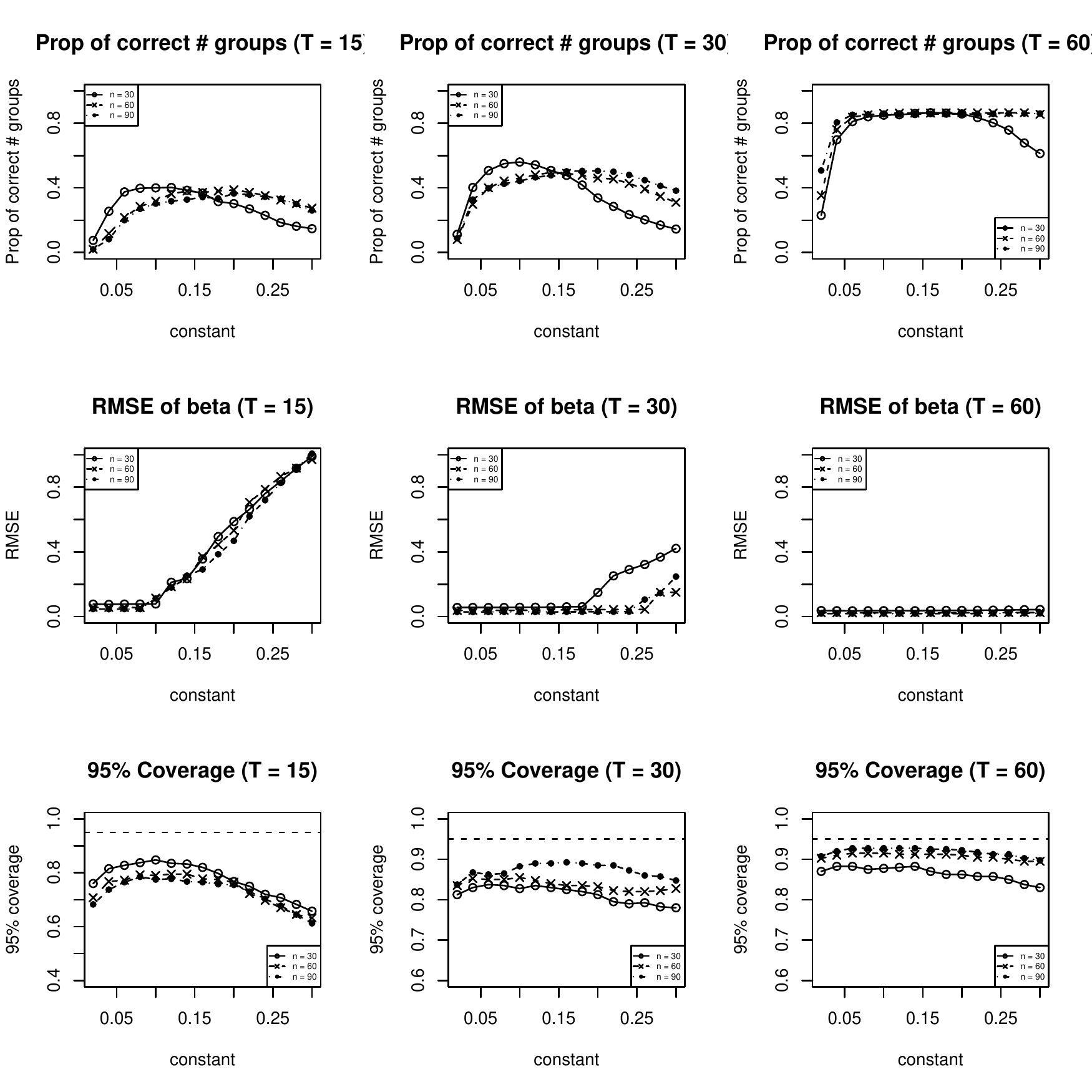}
	\caption{Use different constants in $p_{n,T}$ for the IC criteria for location shift model with $t_3$ error on DGP1: For a equally spaced grid on $[0.01, 0.3]$ with width 0.01, the three columns represent different magnitudes of $T$ while each figures in the row overlays the curves for $n \in \{30,60,90\}$ for various performance measures. The first row plots the proportion of correctly estimated number of groups. The second row plots the RMSE of $\hat \beta^{IC}(\tau)$ where $\tau =0.75$ and the third plots the coverage rate for nominal size 5\%. }
	\label{fig: CC2_ls}
\end{figure}

\begin{figure}[H]
	\captionsetup{font=scriptsize}
	\includegraphics[scale=0.75]{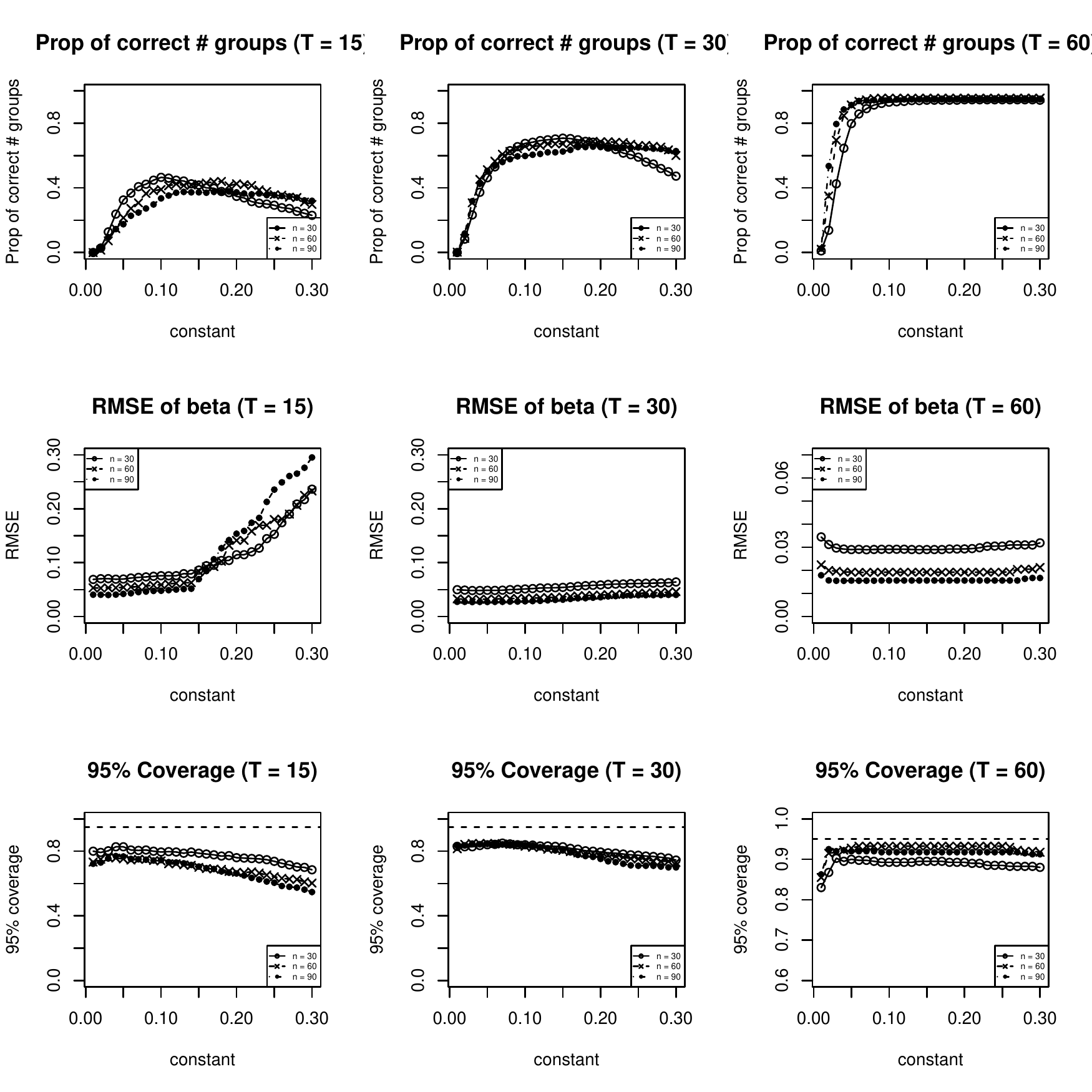}
	\caption{Use different constants in $p_{n,T}$ for the IC criteria for location-scale shift model with $t_3$ error on DGP2: For a equally spaced grid on $[0.01, 0.3]$ with width 0.01, the three columns represent different magnitudes of $T$ while each figures in the row overlays the curves for $n \in \{30,60,90\}$ for various performance measures. The first row plots the proportion of correctly estimated number of groups. The second row plots the RMSE of $\hat \beta^{IC}(\tau)$ where $\tau =0.5$ and the third plots the coverage rate for for nominal size 5\%. Results are based on 400 repetitions.}
	\label{fig: CC1_DGP2}
\end{figure}

\begin{figure}[H]
	\captionsetup{font=scriptsize}
	\includegraphics[scale=0.75]{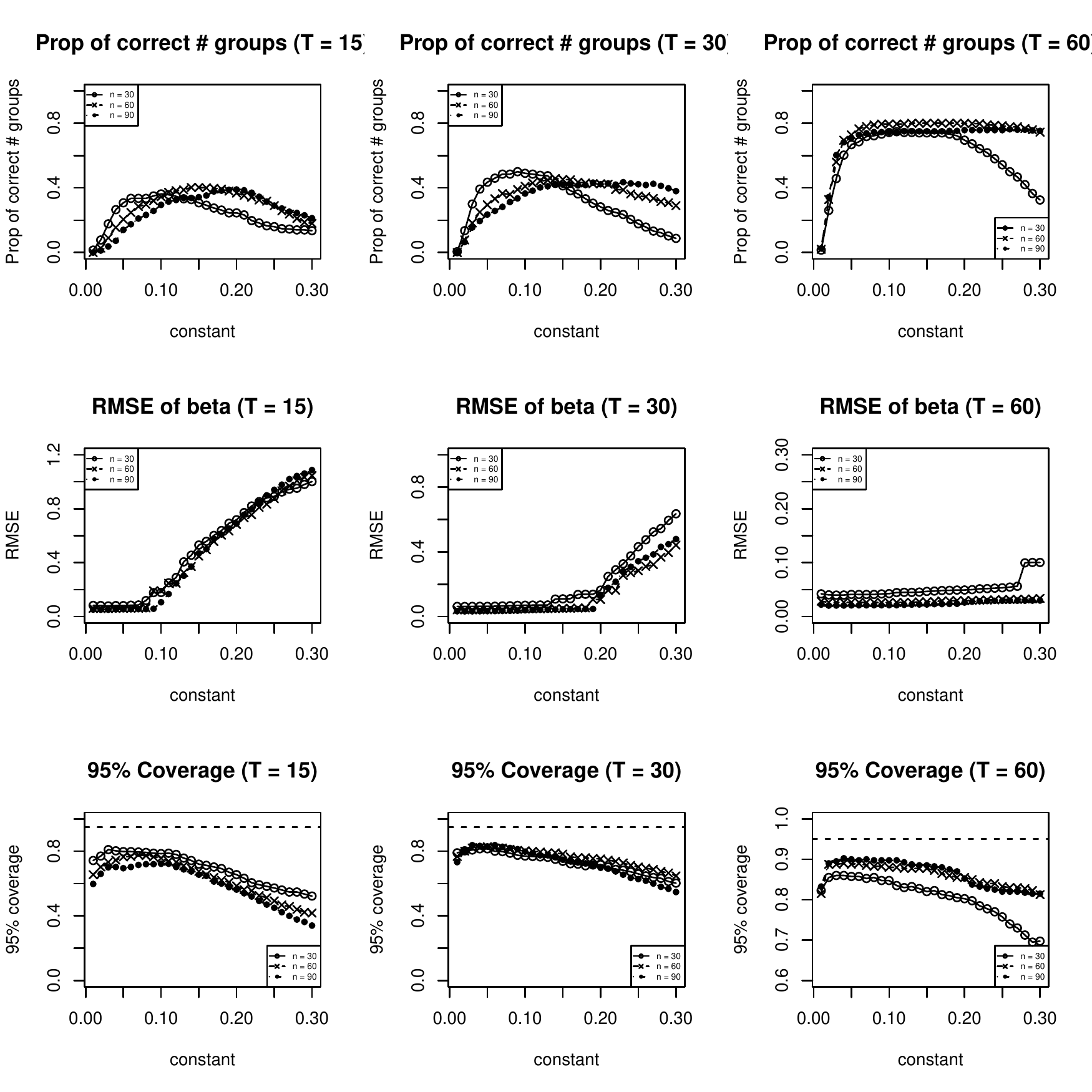}
	\caption{Use different constants in $p_{n,T}$ for the IC criteria for location-scale shift model with $t_3$ error on DGP2: For a equally spaced grid on $[0.01, 0.3]$ with width 0.01, the three columns represent different magnitudes of $T$ while each figures in the row overlays the curves for $n \in \{30,60,90\}$ for various performance measures. The first row plots the proportion of correctly estimated number of groups. The second row plots the RMSE of $\hat \beta^{IC}(\tau)$ where $\tau =0.75$ and the third plots the coverage rate for nominal size 5\%. Results are based on 400 repetitions.}
	\label{fig: CC2_DGP2}
\end{figure}

\begin{figure}[H]
	\captionsetup{font=scriptsize}
	\includegraphics[scale=0.75]{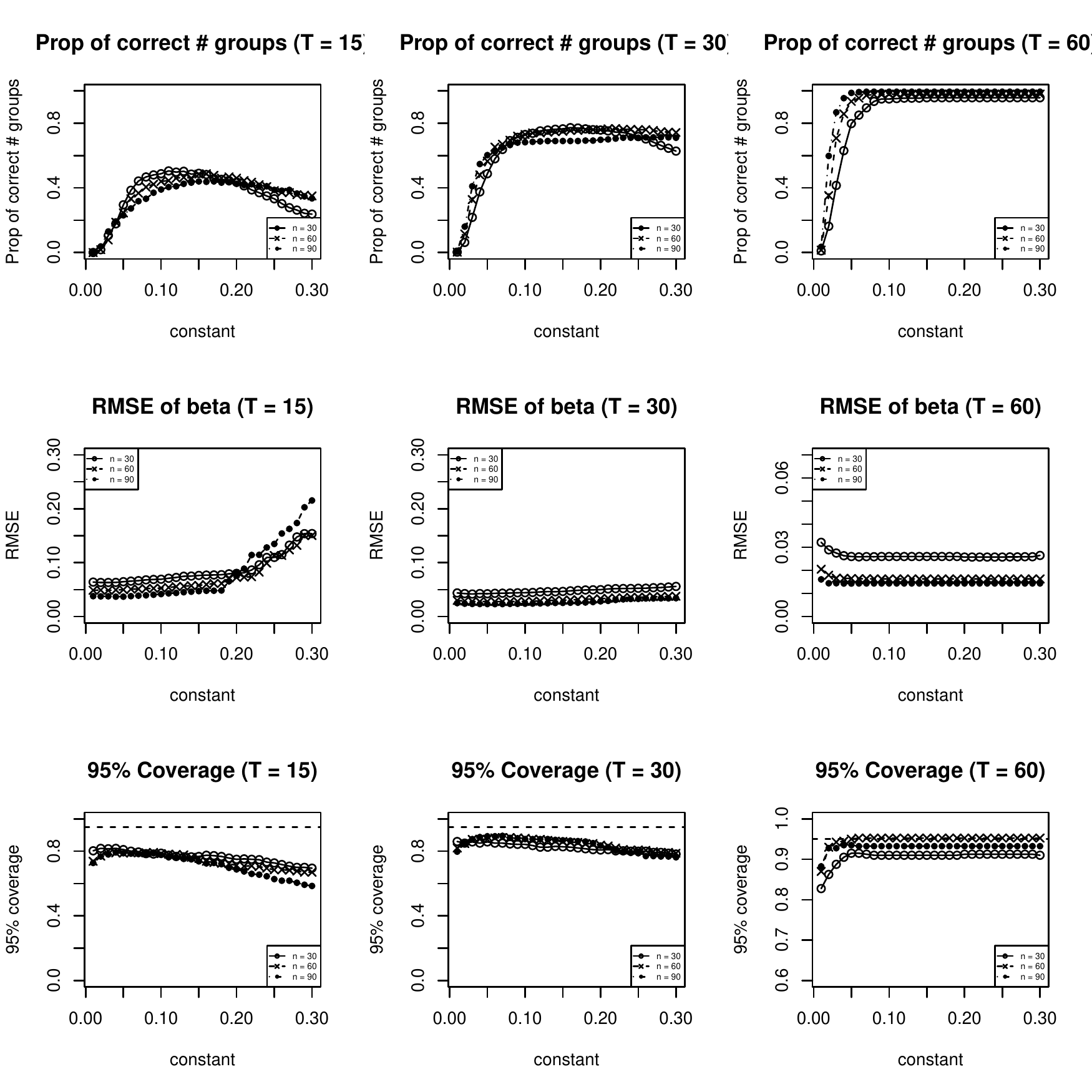}
	\caption{Use different constants in $p_{n,T}$ for the IC criteria for location shift model with $t_3$ error on DGP2: For a equally spaced grid on $[0.01, 0.3]$ with width 0.01, the three columns represent different magnitudes of $T$ while each figures in the row overlays the curves for $n \in \{30,60,90\}$ for various performance measures. The first row plots the proportion of correctly estimated number of groups. The second row plots the RMSE of $\hat \beta^{IC}(\tau)$ where $\tau =0.5$ and the third plots the coverage rate for nominal size 5\%. Results are based on 400 repetitions.}
	\label{fig: CC1_DGP2_ls}
\end{figure}

\begin{figure}[H]
	\captionsetup{font=scriptsize}
	\includegraphics[scale=0.75]{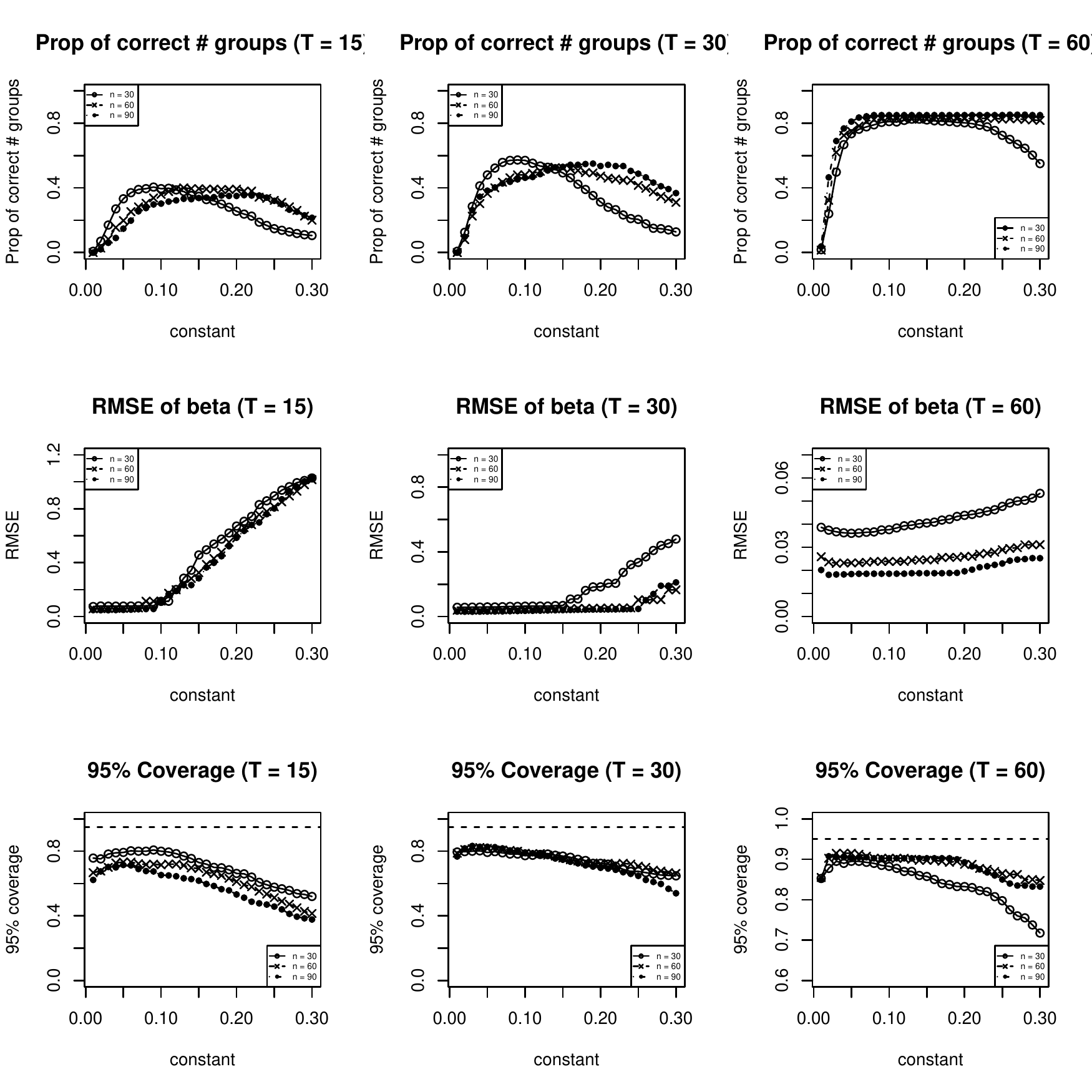}
	\caption{Use different constants in $p_{n,T}$ for the IC criteria for location shift model with $t_3$ error on DGP2: For a equally spaced grid on $[0.01, 0.3]$ with width 0.01, the three columns represent different magnitudes of $T$ while each figures in the row overlays the curves for $n \in \{30,60,90\}$ for various performance measures. The first row plots the proportion of correctly estimated number of groups. The second row plots the RMSE of $\hat \beta^{IC}(\tau)$ where $\tau =0.75$ and the third plots the coverage rate for nominal size 5\%. Results are based on 400 repetitions.}
	\label{fig: CC2_DGP2_ls}
\end{figure}

In the main manuscript, we reported the influence of the turning parameters in the IC criteria on the performance of the group and common parameters estimation for location-scale shift model with $t_3$ errors on DGP1 where the predictor $X_{it}$ and the fixed effects $\alpha_i$ are independent. Figure \ref{fig: CC1_ls} and Figure \ref{fig: CC2_ls} shows the corresponding results for location shift model with $t_3$ error on DGP1. The corresponding plots for DGP2 are collected in Figure \ref{fig: CC1_DGP2} - Figure \ref{fig: CC2_DGP2_ls}.

\section{Additional Empirical Application Analysis}
\begin{figure}[!tbp]
	\captionsetup{font=scriptsize}
	\includegraphics[scale=0.75]{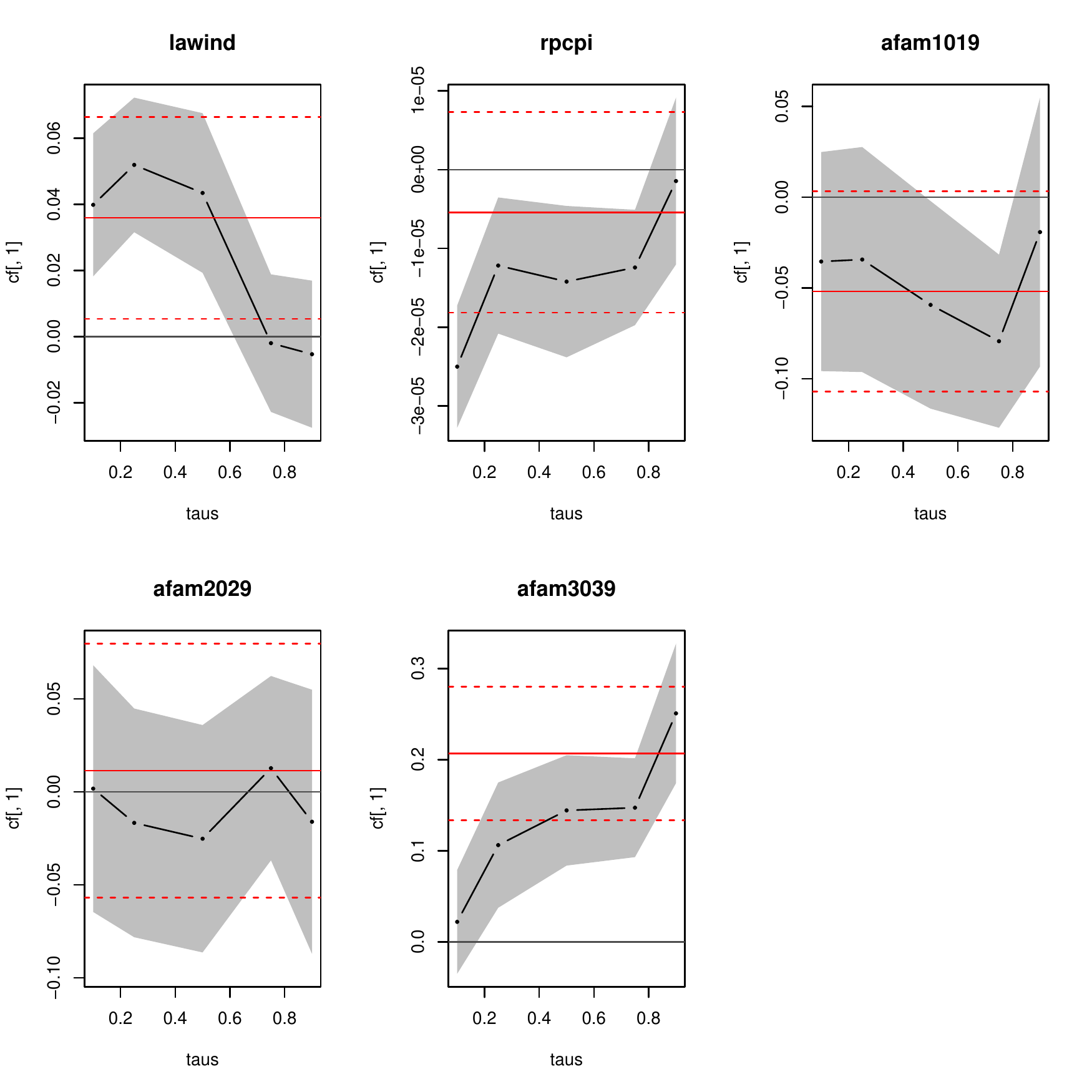}
	\caption{Panel data quantile regression estimates with state fixed effects for various $\tau$ based on model specification (\ref{eq: model1}) excluding the incarceration rate as a control variable: For $\tau \in \{0.1, 0.25, 0.5, 0.75, 0.9\}$, the solid black points plot the coefficient estimates for the effects of the RTC law adoption and other control variables on the violent crime rate based on panel data on 51 U.S. states for 1977 - 2010. The shaded area are the pointwise 95\% confidence interval where the standard errors are computed using the Hendricks-Koenker sandwich covariance matrix estimates with the Hall-Sheather bandwidth rule. The red solid line marks the fixed effect panel data mean regression estimates with the dotted red lines plot the 95\% confidence interval with robust clustered standard errors. }
	\label{fig: QRFE_small}
\end{figure}

\begin{figure}[!tbp]
	\captionsetup{font=scriptsize}
	\includegraphics[scale=0.75]{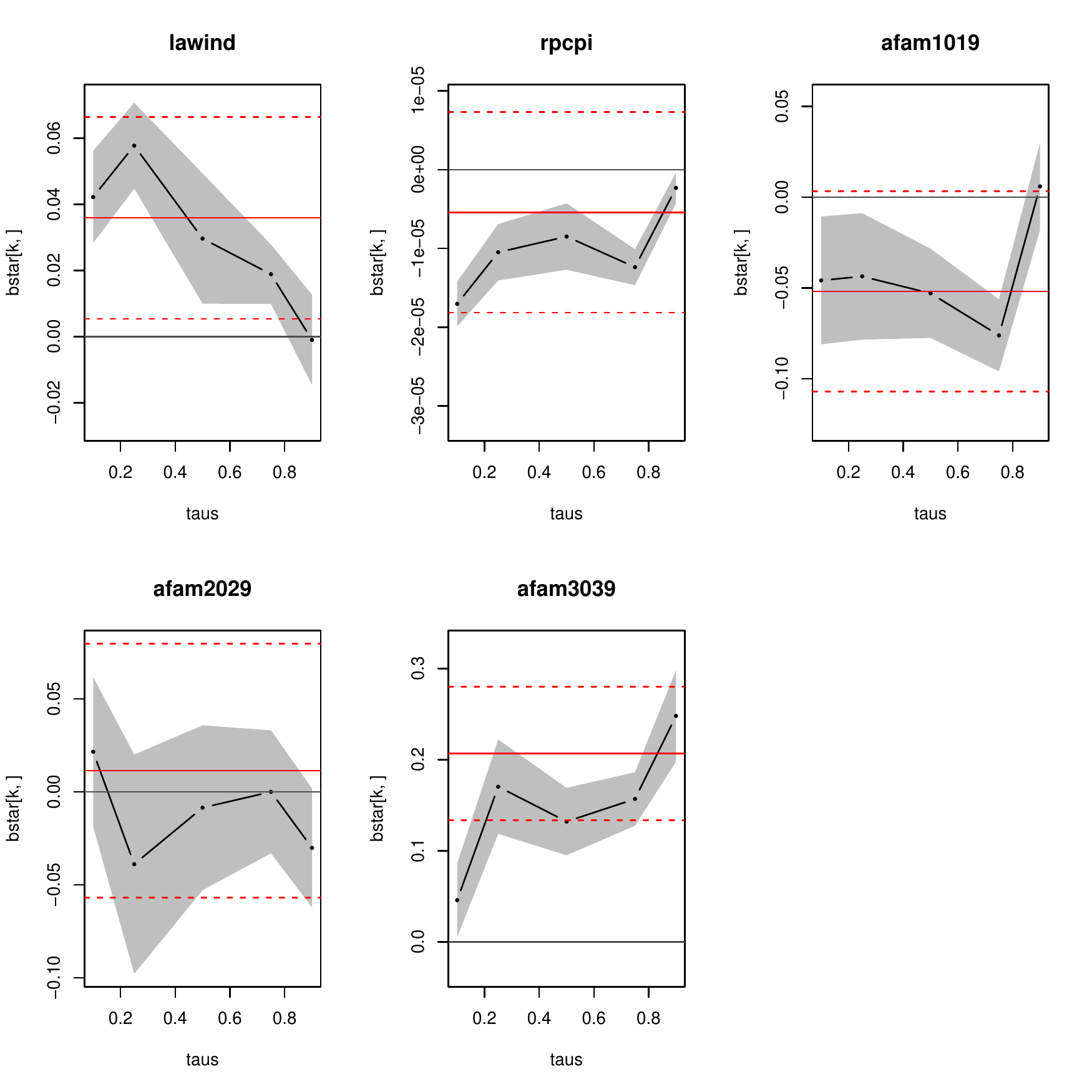}
	\caption{Panel data quantile regression estimates with grouped state fixed effects for various $\tau$ based on model specification (\ref{eq: model1}) excluding the incarceration rate as a control variable: For $\tau \in \{0.1, 0.25, 0.5, 0.75, 0.9\}$, the solid black points plot the coefficient estimates for the effects of the RTC law adoption and other control variables on the violent crime rate based on the proposed methodology with panel data on 51 U.S. states for 1977 - 2010. The shaded area are the pointwise 95\% confidence interval where the standard errors are computed using the Hendricks-Koenker sandwich covariance matrix estimates with the Hall-Sheather bandwidth rule. The red solid line marks the fixed effect panel data mean regression estimates with the dotted red lines plot the 95\% confidence interval with robust clustered standard errors. }
	\label{fig: QRFEgroup_small}
\end{figure}

\begin{figure}[!tbp]
	\captionsetup{font=scriptsize}
	\includegraphics[scale=0.65]{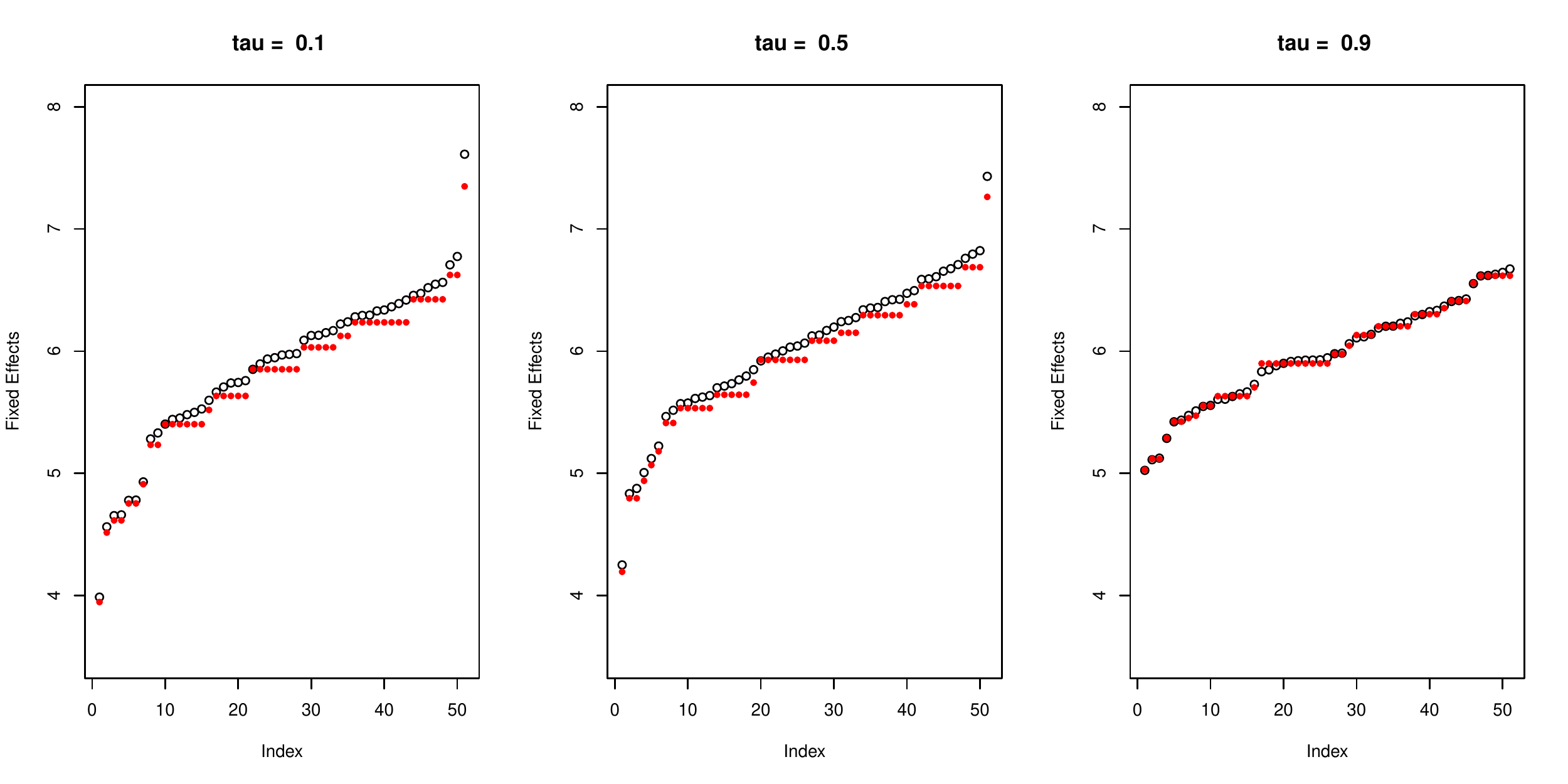}
	\caption{The estimated state fixed effect and the corresponding estimated group structure for various $\tau$ based on model specification (\ref{eq: model1}) excluding the incarceration rate as a control variable: For $\tau \in \{0.1, 0.5, 0.9\}$, the hollow black points plot the ordered panel data quantile regression estimates for individual state fixed effect. The red solid points mark out the estimated grouping and the corresponding group fixed effect estimates using the proposed methodology.}
	\label{fig: fixedeff_small}
\end{figure}

As a robustness check, we report here the results of the empirical analysis for model (\ref{eq: model1}) without the incarcerating rate as a control covariate. Figure \ref{fig: QRFE_small} presents the corresponding common parameters estimation using the fixed effect approach while Figure \ref{fig: QRFEgroup_small} for the results using our proposed grouped fixed effect estimator. Figure \ref{fig: fixedeff_small}  reports the raw and the grouped fixed effect estimates.

\end{document}